\documentclass[english,british]{aa}
\usepackage[T1]{fontenc}
\usepackage[latin9]{inputenc}
\usepackage{color}
\usepackage{babel}
\usepackage{array}
\usepackage{longtable}
\usepackage{rotating}
\usepackage{textcomp}
\usepackage{url}
\usepackage{multirow}
\usepackage{amstext}
\usepackage{amssymb}
\usepackage{graphicx}
\usepackage{esint}
\usepackage[authoryear]{natbib}
\usepackage[unicode=true,
 bookmarks=true,bookmarksnumbered=false,bookmarksopen=false,
 breaklinks=true,pdfborder={0 0 0},backref=false,colorlinks=true]
 {hyperref}
\hypersetup{pdftitle={The dusty torus in the Circinus galaxy: a dense disk and the torus funnel},
 pdfauthor={Konrad R. W. Tristram},
 pdfkeywords={galaxies: active, galaxies: nuclei, galaxies: Seyfert, galaxies: structure, galaxies: individual: Circinus, techniques: interferometric},
 linkcolor = blue, citecolor = blue, urlcolor = blue}
\usepackage{breakurl}
\usepackage{txfonts}
\usepackage{colortbl}
\usepackage[]{pdfcomment} % Also requires package soulpos!!
\usepackage[all]{hypcap}

\makeatletter

\providecommand{\tabularnewline}{\\}
\bibpunct{(}{)}{;}{a}{}{,}

\makeatother

\begin{document}

\title{The dusty torus in the Circinus galaxy:\\
a dense disk and the torus funnel\thanks{Based on observations collected at the European Southern Observatory,
Chile, programme numbers 073.A-9002(A), 060.A-9224(A), 074.B-0213(A/B),
075.B-0215(A), 077.B-0026(A), 081.B-0893(A), 081.B-0908(A/B), 383.B-0159(A),
383.B-0993(A), 087.B-0746(C), 087.B-0971(A-C), and 087.B-0266(H).}}

\author{Konrad R. W. Tristram\inst{1}, Leonard Burtscher\inst{2}, Walter
Jaffe\inst{3}, Klaus Meisenheimer\inst{4}, Sebastian F. Hönig\inst{5},
Makoto Kishimoto\inst{1}, Marc Schartmann\inst{6,2}, Gerd Weigelt\inst{1}}

\institute{Max-Planck-Institut für Radioastronomie, Auf dem Hügel 69, 53121
Bonn, Germany \\
email: \href{mailto:tristram@mpifr-bonn.mpg.de}{tristram@mpifr-bonn.mpg.de}\and
Max-Planck-Institut für extraterrestrische Physik, Postfach 1312,
Gießenbachstraße, 85741 Garching, Germany \and Leiden Observatory,
Leiden University, Niels-Bohr-Weg 2, 2300 CA Leiden, The Netherlands
 \and Max-Planck-Institut für Astronomie, Königstuhl 17, 69117 Heidelberg,
Germany\and Dark Cosmology Center, Niels Bohr Institute, University
of Copenhagen, Juliane Maries Vej 30, 2100 Copenhagen, Denmark \and
Universitätssternwarte München, Scheinerstr. 1, 81679 München, Germany}

\titlerunning{The dusty torus in the Circinus galaxy}

\authorrunning{K. R. W. Tristram et al.}

\abstract{With infrared interferometry it is possible to resolve the nuclear
dust distributions that are commonly associated with the dusty torus
in active galactic nuclei (AGN). The Circinus galaxy hosts the closest
Seyfert 2 nucleus and previous interferometric observations have shown
that its nuclear dust emission is particularly well resolved.} {
The aim of the present interferometric investigation is to better
constrain the dust morphology in this active nucleus. } {To this
end, extensive new observations were carried out with the MID-infrared
Interferometric instrument (MIDI) at the Very Large Telescope Interferometer,
leading to a total of 152 correlated flux spectra and differential
phases between $8$ and $13\,\mathrm{\mu m}$. To interpret this data,
we used a model consisting of black-body emitters with a Gaussian
brightness distribution and with dust extinction.} {The direct analysis
of the data and the modelling confirm that the emission is distributed
in two distinct components: a disk-like emission component with a
size (FWHM) of $\sim0.2\times1.1\,\mathrm{pc}$ and an extended component
with a size of $\sim0.8\times1.9\,\mathrm{pc}$. The disk-like component
is elongated along $\mathit{PA}\sim46\text{\textdegree}$ and oriented
perpendicular to the ionisation cone and outflow. The extended component
is responsible for $80\%$ of the mid-infrared emission. It is elongated
along $\mathit{PA}\sim107\text{\textdegree}$, which is roughly perpendicular
to the disk component and thus in polar direction. It is interpreted
as emission from the inner funnel of an extended dust distribution
and shows a strong increase in the extinction towards the south-east.
We find both emission components to be consistent with dust at $T\sim300\,\mathrm{K}$,
that is we find no evidence of an increase in the temperature of the
dust towards the centre. From this we infer that most of the near-infrared
emission probably comes from parsec scales as well. We further argue
that the disk component alone is not sufficient to provide the necessary
obscuration and collimation of the ionising radiation and outflow.
The material responsible for this must instead be located on scales
of $\sim1\,\mathrm{pc}$, surrounding the disk. We associate this
material with the dusty torus. $ $} { The clear separation of the
dust emission into a disk-like emitter and a polar elongated source
will require an adaptation of our current understanding of the dust
emission in AGN. The lack of any evidence of an increase in the dust
temperature towards the centre poses a challenge for the picture of
a centrally heated dust distribution.}


\date{Received 18 September 2013 / Accepted 26 November 2013}

\maketitle

\section{Introduction\label{sec:introduction}}

Active galactic nuclei (AGN) are thought to play a major role in the
formation and the evolution of galaxies. The AGN phase provides mechanisms
for feedback from the supermassive black hole to its hosting galaxy
and the intergalactic medium. Therefore, a thorough knowledge of the
accretion process in AGN is required to understand their influence
on the formation and evolution of galaxies. Especially little is known
about the accretion process on parsec scales.

A toroidal distribution of warm molecular gas and dust surrounding
the central engine, the so-called \textit{dusty torus}, is a key component
of AGN. First of all, the torus plays an important role in fuelling
the AGN activity: it either forms the passive reservoir of material
for the accretion onto the supermassive black hole or, more intriguingly,
it is itself the active driver of the accretion towards the black
hole \citep{2012Hopkins}. Secondly, the dusty torus is held responsible
for the orientation-dependent obscuration of the central engine \citep[e.g.][]{1993Antonucci,1995Urry}:
when oriented face-on, a direct view of the central engine is possible
through the cavity in the torus (Type 1 AGN); when oriented edge-on,
the view towards the centre is blocked by the gas and dust of the
torus (Type 2 AGN). This scenario is supported by multiple observational
evidence, most importantly by the detection of broad emission lines
in the polarised light of several Type 2 nuclei \citep[e.g.][]{1985Antonucci,2004Lumsden},
indicating that Type 2 sources host the same central engine as Type
1 AGN. However, there is also a growing number of observations that
challenge this simple picture, e.g.\ the discovery of true Type 2
sources (without broad emission lines in polarised light) and X-ray
column densities discrepant with the optical classification (for a
recent discussion of AGN obscuration, see e.g.\ \citealt{2012Bianchi}). 

The dust in the torus is heated by the emission from the accretion
disk. It re-emits this energy in the infrared \citep{1969Rees}: the
innermost dust is close to its sublimation temperature and mainly
emits in the near-infrared, while the dust at larger distances is
at lower temperatures and emits in the mid-infrared \citep{1987Barvainis}.
Direct observations of the torus are thus best carried out at infrared
wavelengths. However, the dust distributions are very compact: they
are essentially unresolved by single-dish observations even with the
largest currently available telescopes \citep[e.g.][]{2009Horst,2009RamosAlmeida}.
Only by employing interferometric methods in the infrared is it possible
to resolve the nuclear dust distributions in AGN. 

To date, several interferometric studies of the nuclear dust distributions
of individual galaxies have been carried out. In the Seyfert 2 galaxy
\object{NGC~1068}, the interferometric observations reveal a hot,
parsec-sized disk that is surrounded by warm dust extended in polar
direction \citep{2004Wittkowski,2004Jaffe1,2004Weigelt,2006Poncelet,2009Raban}.
A two-component structure was also found in the nucleus of the \object{Circinus galaxy}
(\citealt{2007Tristram2}, see below). In \object{NGC~424} (Sy~2)
and \object{NGC~3783} (Sy~1) the thermal dust emission appears extended
along the polar axis of the system \citep{2008Beckert,2012Hoenig,2013Hoenig}.
In \object{NGC~4151}, a Seyfert 1.5 galaxy, interferometric measurements
in the near- and mid-infrared have provided evidence of both hot dust
at the inner rim of the torus as well as warm dust farther out \citep{2003Swain,2009Burtscher,2009Kishimoto2,2010Pott}.
In the radio galaxy \object{NGC~5128} (Centaurus~A) on the other
hand, only half of the mid-infrared emission appears to be of thermal
origin \citep{2007Meisenheimer,2010Burtscher,2011Burtscher}. Near-infrared
reverberation measurements and interferometry of several Type 1 AGN
have shown that the hot inner rim of the torus scales with the square
root of the AGN luminosity \citep{2006Suganuma,2011Kishimoto1}. Whether
this is also true for the cooler dust in the body of the torus remains
unclear after interferometric studies of small samples of galaxies
in the mid-infrared \citep{2009Tristram,2011Tristram,2011Kishimoto2}.
Although a possible common radial structure for AGN tori has been
proposed \citep{2009Kishimoto1}, the first study with a statistically
significant sample of AGN shows a rather diverse picture of the dust
distributions with quite large differences between the dust distributions
in individual galaxies \citep{2013Burtscher}.

Because of computational limitations, initial radiative transfer calculations
of geometrical torus models were carried out for smooth dust distributions
\citep[e.g.][]{1988Krolik,1994Granato,2005Schartmann}. However, it
was realised early on that the distribution of gas and dust is most
likely clumpy \citep{1988Krolik}. For this reason, radiative transfer
calculations of clumpy dust distributions have been carried out more
recently \citep[e.g. ][]{2002Nenkova,2008Schartmann,2010Hoenig2}.
First attempts to address the physics of the accreting nuclear material
in AGN have been undertaken using hydrodynamical simulations \citep{2002Wada,2008Dorodnitsyn,2009Schartmann,2009Wada,2012Wada}.
Nevertheless, the physical picture of the torus remains unclear. Most
torus models were designed to correctly reproduce the infrared spectral
energy distributions (SEDs) of AGN and the silicate feature at $10\,\mathrm{\mu m}$.
Very little can be learned about the torus itself from such comparisons:
very diverse models using different assumptions and parameters can
produce similar SEDs \citep{2012Feltre}. The degeneracies in the
SEDs can, at least partially, be broken by resolving the dust distributions.
This is, as stated above, currently only possible with infrared interferometry.

\begin{table}
\caption{List of the MIDI observation epochs for the Circinus galaxy including
observing date, baseline parameters (telescope combination, range
of baseline lengths and position angles) and number of (un)successful
measurements.\label{tab:observing-epochs}}

\centering

\begin{tabular}{rrr@{ to }rr@{ to }rrr}
\hline 
\hline date & baseline & \multicolumn{2}{c}{$\mathit{BL}$ {[}m{]}} & \multicolumn{2}{c}{$\mathit{PA}$ {[}\textdegree{}{]}} & good & bad\tabularnewline
\hline 
2004-02-12 & U3-U2 &  43 &  43 &  19 & 21 & 2 & 0\tabularnewline
2004-06-03 & U3-U2 &  20 &  29 &  92 & 129 & 2 & 0\tabularnewline
2005-02-21 & U2-U4 & \multicolumn{2}{r}{ 87} & \multicolumn{2}{r}{35} & 0 & 1\tabularnewline
2005-03-01 & U3-U4 &  49 &  62 &  44 & 130 & 6 & 0\tabularnewline
2005-04-18 & U2-U4 & \multicolumn{2}{r}{ 89} & \multicolumn{2}{r}{60} & 1 & 0\tabularnewline
2005-05-26 & U2-U3 &  35 &  43 &  12 & 69 & 6 & 0\tabularnewline
2006-05-18 & U2-U3 &  23 &  31 &  84 & 114 & 4 & 0\tabularnewline
2008-04-17 & U1-U3 &  75 &  84 &  42 & 60 & 0 & 5\tabularnewline
2008-04-18 & U2-U4 &  76 &  89 &  33 & 156 & 26 & 2\tabularnewline
2008-04-26 & E0-G0 &  12 &  15 &  14 & 155 & 23 & 2\tabularnewline
2009-04-15 & U1-U3 &  54 &  92 &  8 & 90 & 17 & 1\tabularnewline
2009-04-27 & E0-G0 &  14 &  16 &  16 & 113 & 16 & 1\tabularnewline
 & H0-G0 &  24 &  25 & 138 & 164 & 3 & 1\tabularnewline
2011-04-14 & U1-U2 & \multicolumn{2}{r}{ 35} & \multicolumn{2}{r}{65} & 1 & 0\tabularnewline
2011-04-17 & U3-U4 &  47 &  62 &  33 & 124 & 10 & 0\tabularnewline
2011-04-18 & U2-U4 &  87 &  87 &  32 & 41 & 7 & 1\tabularnewline
2011-04-19 & U1-U3 &  86 &  86 &  35 & 37 & 2 & 0\tabularnewline
2011-04-20 & U2-U4 &  77 &  81 &  127 & 150 & 16 & 0\tabularnewline
2011-05-06 & C1-A1 &  14 &  15 &  26 & 104 & 10 &  1\tabularnewline
 & D0-B2 &  26 &  27 &  34 & 38 & 0 & 2\tabularnewline
\hline 
totals: &  & \multicolumn{2}{r}{} & \multicolumn{2}{r}{} & 152 & 17\tabularnewline
\hline 
\end{tabular}

\tablefoot{U in the baseline stands for Unit Telescope (UTs); A1 to H0 denote
the stations of the Auxiliary Telescopes (ATs). Measurements were
considered successful (``good'') if a fringe signal was tracked
and sufficient signal was detected (see also Appendix~\ref{sec:list-of-uvpoints}).}
\end{table}
At a distance of about $4\,\mathrm{Mpc}$ ($1\,\mathrm{arcsec}\sim20\,\mathrm{pc}$,
\citealt{1977Freeman}), the \object{Circinus galaxy} is the closest
Seyfert 2 galaxy. It is the second brightest AGN in the mid-infrared
(after \object{NGC~1068}) and, hence, a prime target for detailed
studies of its nuclear distribution of gas and dust. The galaxy can
be considered to be a prototypical Seyfert 2 galaxy with narrow emission
lines \citep{1994Oliva}, broad emission lines in polarised light
\citep{1998Oliva}, a prominent ionisation cone \citep{1997Veilleux,2000Maiolino,2000Wilson,2004Prieto},
an outflow observed in CO \citep{1998Curran,1999Curran}, bipolar
radio lobes \citep{1998Elmouttie}, a Compton thick nucleus and a
reflection component in X-rays \citep{1996Matt,2001Smith,2005Soldi,2009Yang}.
The galaxy is inclined by $\sim65{^\circ}$ and the nucleus is heavily
obscured by dust lanes in the plane of the galaxy so that it is only
visible longward of $\lambda=1.6\,\mathrm{\mu m}$ \citep{2004Prieto}.

\begin{figure*}
\begin{minipage}[b]{12cm}%
\centering

\includegraphics[bb=127bp 254bp 467bp 595bp]{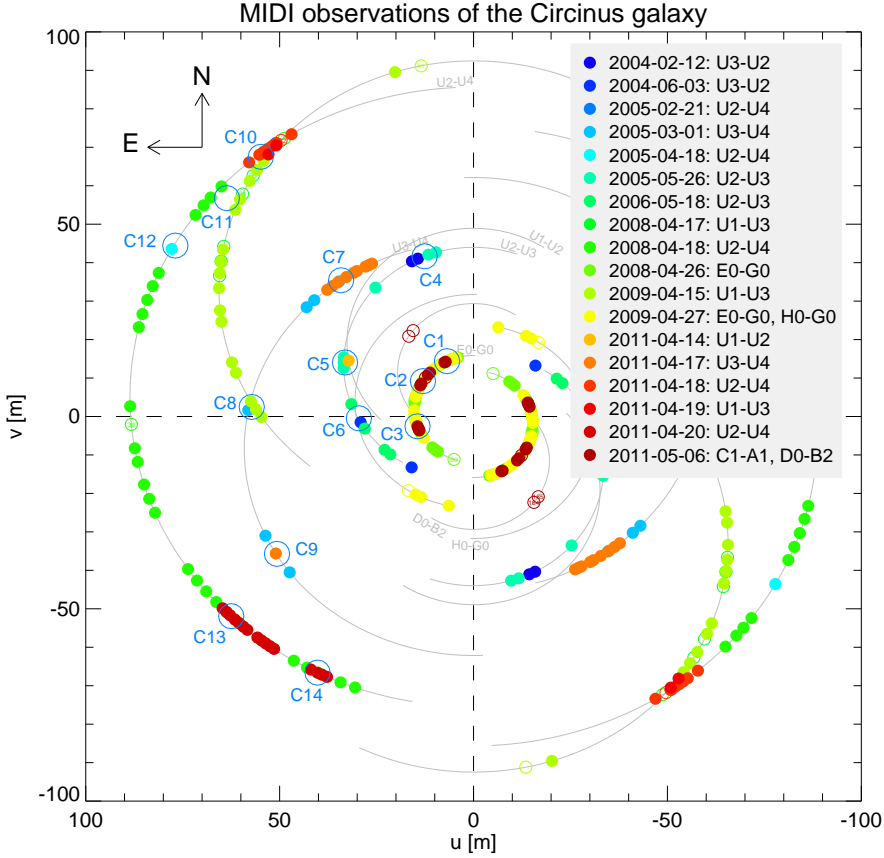}%
\end{minipage}\hfill{}%
\begin{minipage}[b]{6cm}%
\caption{$uv$ plane with all MIDI measurements of the Circinus galaxy. The
individual $uv$ points are colour coded according to the different
observing epochs. Successful measurements are shown by filled circles,
failed measurements by open circles. The full baseline tracks are
plotted in grey for all the telescope combinations used and for a
minimum elevation of the Circinus galaxy of $25\text{\textdegree}$.
In the baseline names, U stands for the UTs, A1 to G0 for the stations
of the ATs. Regions used for the comparison of measurements at different
epochs are encircled and labelled in blue (see Sect.~\ref{sub:data-consistency}).
Every measurement appears twice, symmetric to the centre of the $uv$
plane, because the Fourier transform of a real valued function (such
as the intensity distribution on sky) is hermitian. By consequence,
the measurements which seem covered by the legend are identical to
those on the other side of the $uv$ plane. \label{fig:fig1_uvplane-epochs}}
\end{minipage}
\end{figure*}
A first set of interferometric measurements with the MID-infrared
Interferometric instrument (MIDI) was presented in \citet{2007Tristram2}.
The modelling of the interferometric data showed that an extended,
almost round emission region with $T\lesssim300\,\mathrm{K}$ and
a size of $\sim2.0\,\mathrm{pc}$ surrounds a highly elongated, only
slightly warmer ($T\sim330\,\mathrm{K}$) emission region with a size
of $\sim0.4\,\textrm{pc}$. The latter component was found to have
roughly the same orientation and size as a rotating molecular disk
traced by $\mathrm{H}_{2}\mathrm{O}$ masers \citep{2003Greenhill}.
The observations were interpreted as a geometrically thick dust distribution
with an embedded disk component. This ``torus'' was found to be
oriented perpendicular to the ionisation cone and outflow. The fact
that the model did not reproduce all details of the observations was
attributed to a more complex dust distribution. However, the properties
of the dust components were poorly constrained and no evidence of
hot dust was found.

In this paper, new interferometric observations of the \object{Circinus galaxy}
with MIDI in the mid-infrared are presented in order to shed more
light on both the small scale structure of the dust distribution as
well as on the properties of the extended dust component. The paper
is organised as follows: the observations and data reduction are described
in Sect.~\ref{sec:observations}. The results are discussed in Sect.~\ref{sec:results}.
The description of our modelling of the brightness distribution and
the discussion of the torus properties are given in Sect.~\ref{sec:modelling}
and Sect.~\ref{sec:discussion}, respectively. A summary of the results
is given in Sect.~\ref{sec:conclusions}.

\section{Observations and data reduction\label{sec:observations}}

\subsection{Observations\label{sub:observations}}

The new interferometric observations of the \object{Circinus galaxy}
with MIDI were carried out in April 2008 and 2009 as well as in April
and May 2011. MIDI is the mid-infrared interferometer of the Very
Large Telescope Interferometer (VLTI) on Cerro Paranal in northern
Chile \citep{2003Leinert2}. It combines the light of two telescopes
and produces spectrally dispersed interferograms in the $N$-band
between $8$ and $13\,\mathrm{\mu m}$. All observations of the \object{Circinus galaxy}
were carried out in high sensitivity (\texttt{HIGH\_SENSE}) mode and
with the prism ($\lambda/\delta\lambda\approx30$) as the dispersive
element. With a few exceptions, the calibrator stars \object{HD~120404},
\object{HD~125687} and \object{HD~150798} were observed with the
\object{Circinus galaxy}. Other calibrators observed in the same
night and with the same instrumental setup were only used to determine
the uncertainties of the transfer function (see Sect.~\ref{sub:data-reduction}).

The primary quantities measured by MIDI are the correlated flux spectra
$F_{\mathrm{cor}}(\lambda)$ and the differential phases $\phi_{\mathrm{diff}}(\lambda)$
(differential in wavelength, therefore also called ``chromatic phases'')
from the interferometric measurements, as well as the masked total
flux spectra $F_{\mathrm{tot}}(\lambda)$ from the single-dish measurements.
$F_{\mathrm{cor}}$ and $\phi_{\mathrm{diff}}$ contain information
on the source morphology and depend on the baseline vector (projected
baseline length $\mathit{BL}$ and position angle $\mathit{PA}$),
that is on the projected separation and orientation of the two telescopes.
The visibilities $V(\lambda)$ commonly used in optical/IR interferometry
are in principle obtained by $V(\lambda)=F_{\mathrm{cor}}(\lambda)/F_{\mathrm{tot}}(\lambda)$
(for details on the relation of the quantities for MIDI we refer to
\citealt{2007Tristram1}). Each interferometric measurement determines
one point of the Fourier transform of the brightness distribution
of the source in the so-called $uv$ plane. The actual point measured
depends on the projected baseline of the interferometer, that is on
$\mathit{BL}$ and $\mathit{PA}$. The basic goal of interferometry
is to sample the $uv$ plane as completely as possible in order to
draw inferences on the brightness distribution. To specifically probe
the small scale structures and the more extended dust in the \object{Circinus galaxy},
the new observations were carried out mainly with two long baselines
using pairs of the $8.2\,\mathrm{m}$ Unit Telescopes (UTs) as well
as with the shortest baselines of the VLTI using the $1.8\,\mathrm{m}$
Auxiliary Telescopes (ATs). 

Including the data from \citet{2007Tristram2}, the Circinus galaxy
was observed in a total of 18 epochs between 2004 and 2011. A summary
of the observation epochs, including observing date, baseline properties
and the numbers of good and bad measurements, is given in Table~\ref{tab:observing-epochs}.
Figure~\ref{fig:fig1_uvplane-epochs} shows the measurements in the
$uv$ plane, colour coded according to their observing epoch. A detailed
list of the individual measurements can be found in Table~\ref{tab:list-of-uvpoints}
in Appendix~\ref{sec:list-of-uvpoints}.

To increase the sampling of the $uv$ plane and the observing efficiency
in the new observations, we continuously observed the Circinus galaxy
by directly repeating the interferometry and photometry exposures.
In doing so, the time-consuming acquisition and setup procedures normally
performed when changing sources with MIDI are omitted. Calibrators
were observed at intervals of typically 1.5 hours in order to determine
the variations of the instrumental and atmospheric transfer function
during the night and to estimate the calibration error (see Sect.~\ref{sub:data-reduction}).
Using this observing method, an almost continuous measurement of the
correlated fluxes can be obtained along a $uv$ track, while the projected
baseline vector moves through the $uv$ plane due to the rotation
of the Earth.

The large time span over which the observations were carried out leads
to some inhomogeneity in the data. The instrumental capabilities and
observing procedures evolved with time and both the UTs and ATs were
used for the observations; differing integration times and chopping
frequencies were used for the single-dish spectra; observations were
carried out with different states of the adaptive optics and field
stabilisation systems. The respective quantities and system settings
are included in Table~\ref{tab:list-of-uvpoints} for each $uv$
point. Any ramifications for the data resulting from these differences
should be eliminated during the calibration process (see Sect.~\ref{sub:data-reduction})
because the calibrators were observed with the same set-up. For this
reason, we combine all data into one data set in order to achieve
the best possible $uv$ coverage.

\subsection{Data reduction\label{sub:data-reduction}}

The data was reduced using the data reduction package EWS (Expert
Work Station%
\footnote{The software package ``MIA+EWS'' can be downloaded at:\\
 \url{http://home.strw.leidenuniv.nl/~jaffe/ews/index.html}%
}, \citealt{2004Jaffe2}) Version 2.0. Additional software%
\footnote{This additional software is publicly available at:\\
\url{http://www.blackholes.de/downloads.html}%
} written in IDL was used to diagnose and analyse the data consistently.
EWS implements the coherent integration method (e.g.~\citealt{2004Meisner})
for the reduction of the interferometric data. In this method, the
group delay and phase drifts caused by the atmosphere are determined
from the data itself. Note that because of these drifts only the differential
phases can be obtained with MIDI, not the ``absolute'' Fourier phases.
After the group delay and phase drifts have been removed and the bad
frames have been flagged, the interferometric signal is averaged coherently.
Version 2.0 of EWS includes several improvements especially for the
treatment of fainter sources (i.e.~sources with correlated fluxes
significantly below $1\,\mathrm{Jy}$ on the UTs, or $20\,\mathrm{Jy}$
on the ATs). For example, the group delay and water vapour phase estimation
are significantly more robust and less biased than previously. A description
of the improvements to the data reduction software can be found in
\citet{2012Burtscher}.

\begin{table}
\caption{Values for the parameters in EWS used for the reduction of the MIDI
data of the Circinus galaxy.\label{tab:data-reduction-paramaters}}

\renewcommand{\footnoterule}{}\centering

\begin{minipage}[t]{1\columnwidth}%
\begin{tabular}{r@{ }r@{ }ll}
\hline 
\hline\centering parameter & \multicolumn{2}{c}{value} & description\tabularnewline
\hline 
\texttt{smooth =} & \multicolumn{2}{@{ }r}{10 frames} & width of boxcar for high pass filtering\tabularnewline
\texttt{gsmooth1 =} & 0.36 & sec & 1st pass: coherent smoothing \tabularnewline
\texttt{asmooth1 =} & 2.00 & sec & 1st pass: amplitude smoothing %
\footnote{Using a Gaussian instead of the default boxcar smoothing.%
}\tabularnewline
\texttt{msmooth1 =} & 4.00 & sec & 1st pass: median smoothing\tabularnewline
\texttt{gsmooth2 =} & 0.36 & sec & 2nd pass: coherent integration length\tabularnewline
\texttt{msmooth2 =} & 1.00 & sec & 2nd pass: median smoothing \tabularnewline
\texttt{ngrad =} & 2 &  & 2nd pass: $2\,\mathtt{ngrad}\!+\!1\!=\!5$ delay rate fits\tabularnewline
\texttt{maxopd1 =} & 40 & \textmu{}m & 2nd pass: maximum search OPD\tabularnewline
\texttt{psmooth =} & 0.18 & sec & smoothing for phase estimation\tabularnewline
\texttt{pgrad =} & 0 &  & $2\,\mathtt{pgrad}\!+\!1\!=\!1$ linear phase rate fit\tabularnewline
\texttt{maxopd2 =} & 100 & \textmu{}m & flagging: maximum allowed OPD\tabularnewline
\texttt{minopd =} & 0 & \textmu{}m & flagging: minimum allowed OPD\tabularnewline
\texttt{jumpopd =} & 10 & \textmu{}m & flagging: maximum jump in delays\tabularnewline
\texttt{jitteropd =} & 1.5 & \textmu{}m & flagging: maximum jitter in delays\tabularnewline
\hline 
\end{tabular}%
\end{minipage}
\end{table}
The parameter settings of EWS used to reduce the data of the Circinus
galaxy are listed in Table~\ref{tab:data-reduction-paramaters}.
To extract the spectra, an optimised mask was determined for each
observing epoch. The width of these masks scales with the width of
the PSF and increases from $5.3\pm1.5\,\mathrm{pixels}$ ($0.45\,\mathrm{arcsec}$
for the UTs) at $8\,\mathrm{\mu m}$ to $7.4\pm1.0\,\mathrm{pixels}$
($0.63\,\mathrm{arcsec}$ for the UTs) at $13\,\mathrm{\mu m}$. To
reduce the total flux spectra, the bands for estimating the sky background
were adjusted so that they are always located symmetric with respect
to the source spectra, $15$ or $16\,\mathrm{pixels}$ ($\sim1.33\,\mathrm{arcsec}$)
apart for the UTs and $11\,\mathrm{pixels}$ ($\sim4.2\,\mathrm{arcsec}$)
apart for the ATs. The calibrator database of EWS, which is based
on the database of calibrator spectra by R. van Boekel \citep{2004vanBoekel,2005Verhoelst},
was used to calibrate the data.

In addition to the statistical errors provided by EWS, the uncertainties
due to the variation of the transfer function of the atmosphere and
instrument were estimated from up to five calibrators and added to
the statistical errors in quadrature. In most cases, these calibration
uncertainties dominate over the statistical errors. They are the main
source of uncertainty in our MIDI measurements, with errors from as
low as $5\,\%$ to more than $20\,\%$, depending on the atmospheric
conditions of the night. Similar results were obtained by e.g.\ \citet{2012Burtscher}.
The errors used in \citet{2007Tristram2} are much smaller. There,
only the statistical errors from EWS were used and the uncertainties
were, therefore, underestimated.

\begin{table*}
\caption{Comparison of measurements at similar locations in the $uv$ plane
but from different observing epochs or baselines.\label{tab:comparison-of-measurements}}

\renewcommand{\footnoterule}{}

\begin{minipage}[t]{1\textwidth}%
\centering

\begin{tabular}{rcccccccc}
\hline 
\hline \centering\# & A & B & \multicolumn{2}{c}{$F_{\mathrm{cor}}(\mathrm{B})/F_{\mathrm{cor}}(\mathrm{A})$} & \multicolumn{2}{c}{$F_{\mathrm{tot}}(\mathrm{B})/F_{\mathrm{tot}}(\mathrm{A})$} & \multicolumn{2}{c}{$V(B)/V(A)$}\tabularnewline
(1) & (2) & (3) & \multicolumn{2}{c}{(4)} & \multicolumn{2}{c}{(5)} & \multicolumn{2}{c}{(6)}\tabularnewline
\hline 
C1 & \pdftooltip{2008-04-26 00:55, E0-G0}{\#59: 2008-04-26 00:55:33} & \pdftooltip{2009-04-27 00:42, E0-G0}{\#101: 2009-04-27 00:42:38} & $0.8\pm0.3$ &  & -- &  & -- & \tabularnewline
 & \pdftooltip{2008-04-26 00:55, E0-G0}{\#59: 2008-04-26 00:55:33} & \pdftooltip{2011-05-06 00:25, C1-A1}{\#157: 2011-05-06 00:25:36} & $0.9\pm0.3$ &  & -- &  & -- & \tabularnewline
 & \pdftooltip{2009-04-27 00:42, E0-G0}{\#101: 2009-04-27 00:42:38} & \pdftooltip{2011-05-06 00:25, C1-A1}{\#157: 2011-05-06 00:25:36} & $1.2\pm0.3$ &  & -- &  & -- & \tabularnewline
C2 & \pdftooltip{2008-04-26 02:55, E0-G0}{\#62, 63: 2008-04-26 02:55:34 \& 2008-04-26 03:09:22} & \pdftooltip{2009-04-27 02:52, E0-G0}{\#106: 2009-04-27 02:52:22} & $0.8\pm0.3$ &  & -- &  & -- & \tabularnewline
 & \pdftooltip{2008-04-26 02:55, E0-G0}{\#62, 63: 2008-04-26 02:55:34 \& 2008-04-26 03:09:22} & \pdftooltip{2011-05-06 02:39, C1-A1}{\#162: 2011-05-06 02:39:33} & $0.7\pm0.3$ &  & -- &  & -- & \tabularnewline
 & \pdftooltip{2009-04-27 02:52, E0-G0}{\#106: 2009-04-27 02:52:22} & \pdftooltip{2011-05-06 02:39, C1-A1}{\#162: 2011-05-06 02:39:33} & $0.8\pm0.3$ &  & -- &  & -- & \tabularnewline
C3 & \pdftooltip{2008-04-26 06:31, E0-G0}{\#74: 2008-04-26 06:31:48} & \pdftooltip{2009-04-27 06:15, E0-G0}{\#114: 2009-04-27 06:15:47} & $1.2\pm0.3$ &  & -- &  & -- & \tabularnewline
 & \pdftooltip{2008-04-26 06:31, E0-G0}{\#74: 2008-04-26 06:31:48} & \pdftooltip{2011-05-06 05:48, C1-A1}{\#166: 2011-05-06 05:48:14} & $0.9\pm0.5$ &  & -- &  & -- & \tabularnewline
 & \pdftooltip{2009-04-27 06:15, E0-G0}{\#114: 2009-04-27 06:15:47} & \pdftooltip{2011-05-06 05:48, C1-A1}{\#166: 2011-05-06 05:48:14} & $0.7\pm0.4$ &  & -- &  & -- & \tabularnewline
C4 & \pdftooltip{2004-02-12 06:55, U3-U2}{\#1, 2: 2004-02-12 06:55:11 \& 2004-02-12 07:06:46} & \pdftooltip{2005-05-26 23:29, U2-U3}{\#13, 14: 2005-05-26 23:29:03 \& 2005-05-26 23:43:07} & \cellcolor[rgb]{1.0,0.7,0.3}$1.9\pm0.4$ &  & $0.8\pm0.6$ & b & \cellcolor[rgb]{1.0,1.0,0.3}$2.5\pm0.9$ & \tabularnewline
C5 & \pdftooltip{2005-05-27 04:07, U2-U3}{\#16: 2005-05-27 04:07:02} & \pdftooltip{2011-04-14 08:44, U1-U2}{\#120: 2011-04-14 08:44:53} & \cellcolor[rgb]{1.0,1.0,0.3}$1.3\pm0.2$ &  & $1.1\pm0.4$ & b & $1.1\pm0.3$ & \tabularnewline
C6 & \pdftooltip{2004-06-03 05:50, U3-U2}{\#3: 2004-06-03 05:50:45} & \pdftooltip{2006-05-18 06:16, U2-U3}{\#19, 20: 2006-05-18 06:16:21 \& 2006-05-18 07:09:48} & $1.0\pm0.3$ &  & $1.1\pm0.4$ &  & $0.9\pm0.4$ & \tabularnewline
C7 & \pdftooltip{2005-03-01 04:06, U3-U4}{\#6: 2005-03-01 04:06:04} & \pdftooltip{2011-04-17 00:54, U3-U4}{\#126: 2011-04-17 00:54:28} & \cellcolor[rgb]{1.0,1.0,0.3}$1.5\pm0.3$ & b & -- &  & -- & \tabularnewline
C8 & \pdftooltip{2005-03-01 06:58, U3-U4}{\#9: 2005-03-01 06:58:52} & \pdftooltip{2009-04-15 09:29, U1-U3}{\#96: 2009-04-15 09:29:24} & $1.1\pm0.2$ &  & $1.3\pm0.6$ &  & $0.9\pm0.3$ & \tabularnewline
C9 & \pdftooltip{2005-03-01 09:21, U3-U4}{\#10, 11: 2005-03-01 09:21:08 \& 2005-03-01 10:08:22} & \pdftooltip{2011-04-17 06:40, U3-U4}{\#130: 2011-04-17 06:40:15} & \cellcolor[rgb]{1.0,0.7,0.3}$1.7\pm0.2$ & b & $1.2\pm0.3$ &  & \cellcolor[rgb]{1.0,1.0,0.3}$1.4\pm0.2$ & b\tabularnewline
C10 & \pdftooltip{2009-04-15 05:16, U1-U3}{\#83: 2009-04-15 05:16:40} & \pdftooltip{2011-04-18 02:00, U2-U4}{\#137: 2011-04-18 02:00:18} & \cellcolor[rgb]{1.0,1.0,0.3}$0.5\pm0.3$ & r & \cellcolor[rgb]{1.0,1.0,0.3}$0.6\pm0.3$ &  & $0.8\pm0.4$ & r\tabularnewline
 & \pdftooltip{2009-04-15 05:16, U1-U3}{\#83: 2009-04-15 05:16:40} & \pdftooltip{2011-04-19 04:55, U1-U3}{\#145: 2011-04-19 04:55:06} & $0.8\pm0.3$ &  & $0.9\pm0.3$ &  & $0.9\pm0.4$ & \tabularnewline
 & \pdftooltip{2011-04-18 02:00, U2-U4}{\#137: 2011-04-18 02:00:18} & \pdftooltip{2011-04-19 04:55, U1-U3}{\#140: 2011-04-19 04:55:06} & \cellcolor[rgb]{1.0,1.0,0.3}$1.4\pm0.3$ & b & $1.3\pm0.4$ &  & $1.0\pm0.4$ & b\tabularnewline
C11 & \pdftooltip{2008-04-18 02:42, U2-U4}{\#30: 2008-04-18 02:42:44} & \pdftooltip{2009-04-15 06:00, U1-U3}{\#86: 2009-04-15 06:00:23} & $1.2\pm0.4$ & b & $1.1\pm0.3$ &  & $1.0\pm0.4$ & b\tabularnewline
C12 & \pdftooltip{2005-04-18 03:29, U2-U4}{\#12: 2005-04-18 03:29:53} & \pdftooltip{2008-04-18 02:59, U2-U4}{\#32, 33: 2008-04-18 02:59:04 \& 2008-04-18 03:48:32} & $1.5\pm0.5$ & b & $1.3\pm0.5$ &  & $1.2\pm0.4$ & \tabularnewline
C13 & \pdftooltip{2008-04-18 08:18, U2-U4}{\#49: 2008-04-18 08:18:52} & \pdftooltip{2011-04-20 08:16, U2-U4}{\#143: 2011-04-20 08:16:35} & \cellcolor[rgb]{1.0,1.0,0.3}$1.5\pm0.3$ & b & -- &  & -- & \tabularnewline
C14 & \pdftooltip{2008-04-18 09:25, U2-U4}{\#52, 53: 2008-04-18 09:25:56 \& 2008-04-18 09:36:06} & \pdftooltip{2011-04-20 09:28, U2-U4}{\#154: 2011-04-20 09:28:50} & \cellcolor[rgb]{1.0,1.0,0.3}$1.5\pm0.3$ & b & -- &  & -- & \tabularnewline
\hline 
\end{tabular}

\tablefoot{Column (1) lists the identifiers of the comparison area (cf.\ blue
numbers in Fig.~\ref{fig:fig1_uvplane-epochs}); columns (2) and
(3) list the time and baseline of the two measurements \textquoteleft{}A'
and \textquoteleft{}B' that are compared; the ratios of the correlated
flux spectra, the total flux spectra and the visibilities, averaged
over the entire $N$-band, are given in columns (4), (5) and (6) respectively.
The ratio of the more recent measurement \textquoteleft{}B' over the
earlier measurement \textquoteleft{}A' is calculated. Measurements
which disagree by more than $1\sigma$ are highlighted in yellow,
those that disagree by more than $2\sigma$ are highlighted in orange.
If there is a change in the spectral shape, this is indicated by a
\textquoteleft{}b' (more recent measurement is bluer) or \textquoteleft{}r'
(more recent measurement is redder) after the corresponding value.}%
\end{minipage}
\end{table*}
\begin{figure*}
\centering

\includegraphics[bb=42bp 198bp 552bp 650bp]{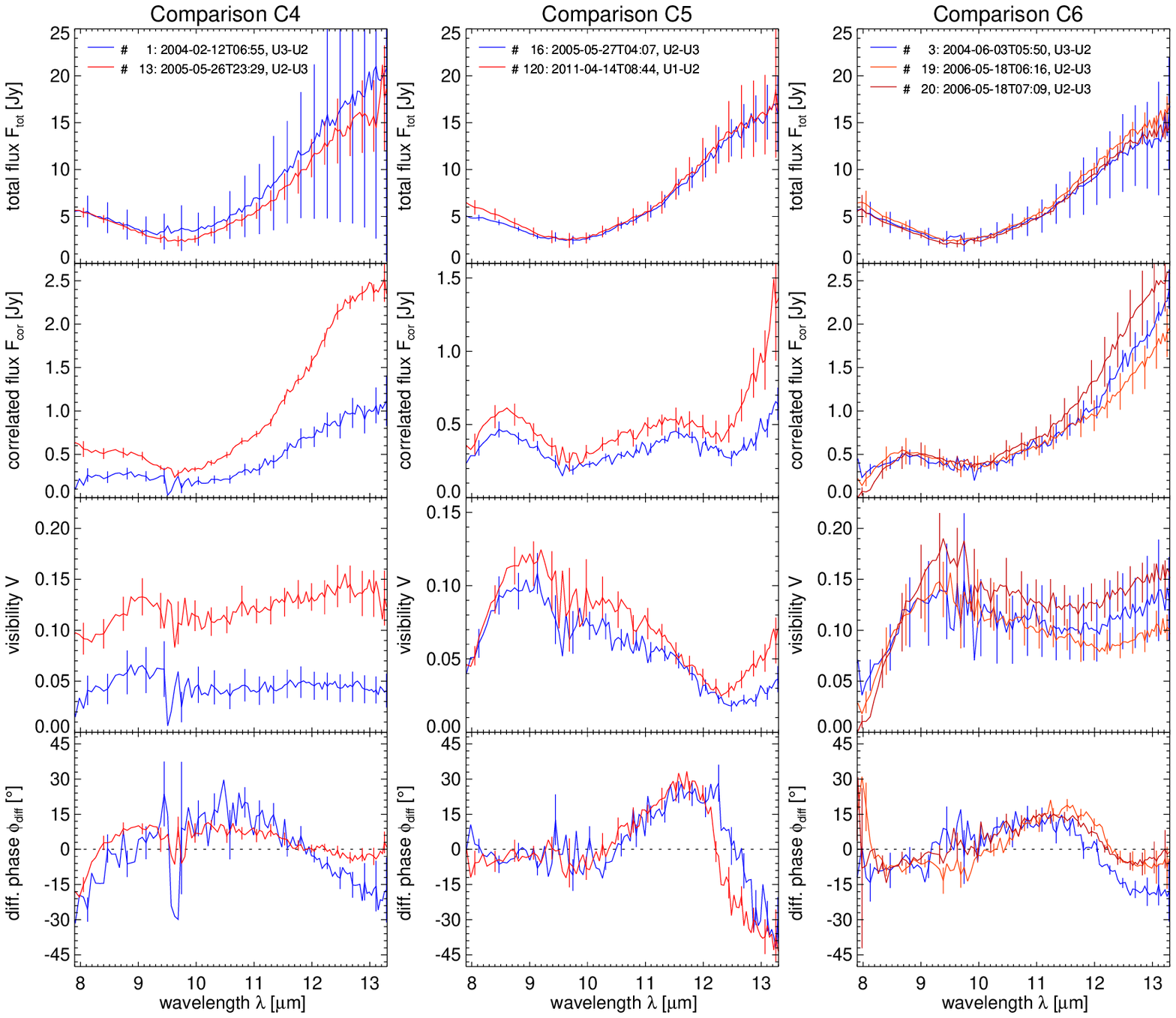}

\caption{Comparison of the total and correlated flux spectra (first two rows),
visibilities (third row) and differential phases (bottom row) at locations
C4 (left column), C5 (central column) and C6 (right column). For clarity,
error bars are only plotted every fifth wavelength bin. Note that
for C4 and C6 the telescopes at the two epochs considered were interchanged
(UT3-UT2 versus UT2-UT3). The phases of measurements \#1 and \#3 were
corrected for this interchange in order to allow an easier comparison
(see also discussion in Sect.~\ref{sub:differential-phases}).\label{fig:fig2_comparison-of-measurements}}
\end{figure*}
Due to the imperfect background subtraction by chopping, the uncertainties
in the total flux measurements are significantly larger than those
of the correlated fluxes. These uncertainties also propagate into
the visibilities. Furthermore, the total flux spectra of the Circinus
galaxy observed with the ATs turn out to be entirely useless. Therefore
most of the following analysis is focused on the correlated fluxes
and the differential phases.

\subsection{Data consistency\label{sub:data-consistency}}

\subsubsection{Individual points in the $uv$ plane\label{sub:consistency-uvpoints}}

Unless the brightness distribution of the object has changed in intensity
or shape, MIDI should always measure the same correlated flux and
differential phase at the same location in the $uv$ plane. Furthermore,
we expect measurements close in $uv$ space to be similar or to change
continuously for measurements at distances less than the telescope
diameters of $8.2\,\mathrm{m}$ for the UTs and $1.8\,\mathrm{m}$
for the ATs \citep[c.f.][]{2014Lopez-Gonzaga}.

For the Circinus galaxy, several locations in the $uv$ plane have
been measured more than once in different observing epochs or using
different telescope combinations. For 14 such locations we check the
consistency of our measurements. The locations are marked in blue
in Fig.~\ref{fig:fig1_uvplane-epochs} and named C1 to C14. The ratios
(averaged over the $N$-band) of the correlated flux spectra, the
total flux spectra and the visibility spectra (where available) for
the 22 possible comparisons at these locations are listed in Table~\ref{tab:comparison-of-measurements}.
Also indicated is whether there is an apparent change in the spectral
slope of the measurements: \textquoteleft{}b' indicates that the more
recent measurement shows a bluer spectrum ($F_{\mathrm{cor}}(13\,\mathrm{\mu m})/F_{\mathrm{cor}}(8\,\mathrm{\mu m})$
decreased), \textquoteleft{}r' indicates that the more recent measurement
shows a redder spectrum ($F_{\mathrm{cor}}(13\,\mathrm{\mu m})/F_{\mathrm{cor}}(8\,\mathrm{\mu m})$
increased).

For the correlated fluxes, 14 comparisons (i.e.\ 64\%) agree within
$1\sigma$, and 20 (i.e.\ 91\%) agree within $2\sigma$. This is
consistent with the statistical expectation and with the results for
other sources, where measurements at approximately the same location
in the $uv$ plane have been repeated (e.g.\ \object{NGC~424}, \object{NGC~3783}
and \object{NGC~4593}, \citealt{2012Hoenig,2013Hoenig,2013Burtscher}).
Due to the larger errors in the total flux measurements, the total
fluxes and visibilities all agree within $2\sigma$. The differential
phases all agree within $1\sigma$, when taking into account the orientation
of the baseline, because the sign of the differential phase switches
for an interchange of the two telescopes, i.e.\ $\phi_{\mathrm{diff}}(\textrm{U2-U1})=-\phi_{\mathrm{diff}}(\textrm{U1-U2})$.
The total and correlated flux spectra, the visibilities and differential
phases for C4, C5 and C6 are plotted in Fig.~\ref{fig:fig2_comparison-of-measurements}.

C4 has the strongest discrepancy: both the correlated flux and the
visibility spectra measured in 2005 are about a factor of two higher
than the ones measured in 2004 (see left column in Fig.~\ref{fig:fig2_comparison-of-measurements}).
These early measurements were carried out without the VLTI infrared
field-stabiliser IRIS (InfraRed Image Sensor, \citealt{2004Gitton}).
As a consequence, the beam overlap in the 2004 measurement was not
optimal, which could be responsible for the lower correlated flux.
In the case of C5 (middle column in Fig.~\ref{fig:fig2_comparison-of-measurements}),
there is a slight increase in the correlated fluxes from 2005 to 2011.
More interestingly, the (unusual) spectral shape with a dip at $12.5\,\mathrm{\mu m}$
remains essentially unchanged. Therefore, we consider this feature
to be a true signal from the source (see discussion in Sect.~\ref{subsub:phases-smallscale}).
In the case of C6 (right column in Fig.~\ref{fig:fig2_comparison-of-measurements}),
one measurement obtained in 2004 lies between two measurements that
were obtained in 2006. This is in perfect agreement with a continuous
drop of the correlated fluxes or visibilities along this $uv$ track.

For most of the comparisons (C4 to C14), the correlated fluxes measured
at later epochs are higher than those measured earlier. This is especially
the case at the short wavelength end of the $N$-band; the more recent
spectra generally appear bluer. All of the observations showing this
possible increase in flux were carried out with the UTs on baselines
longer than $30\,\mathrm{m}$, which essentially probe spatial scales
of $\lesssim80\,\mathrm{mas}$. Additionally, most of the total flux
spectra observed since 2009 appear slightly (although not significantly)
higher (see below). This could be interpreted as evidence of an increase
in the source flux. On the other hand, the correlated fluxes obtained
with the ATs, probing spatial scales of $\lesssim170\,\mathrm{mas}$,
rather suggest the opposite: the correlated fluxes observed in 2009
and 2011 are slightly lower than those observed in 2008 (cf.\ C1,
C2 and C3 in Table~\ref{tab:comparison-of-measurements}). The decrease
is not significant, but an overall increase in the flux on spatial
scales below $170\,\mathrm{mas}$ can be ruled out.

\subsubsection{Total flux\label{sub:consistency-totalflux}}

\begin{figure*}
\centering

\includegraphics[bb=170bp 325bp 420bp 512bp]{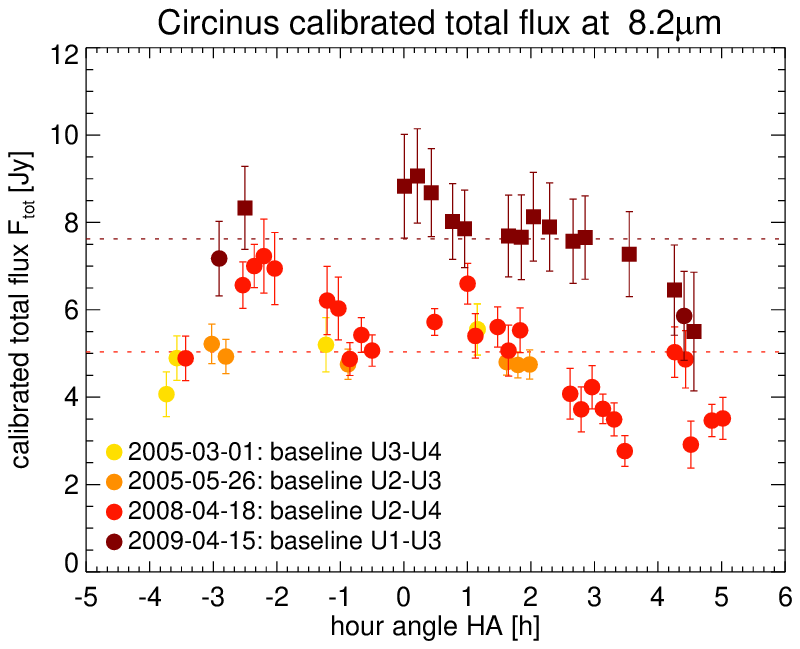}\includegraphics[bb=170bp 325bp 420bp 512bp]{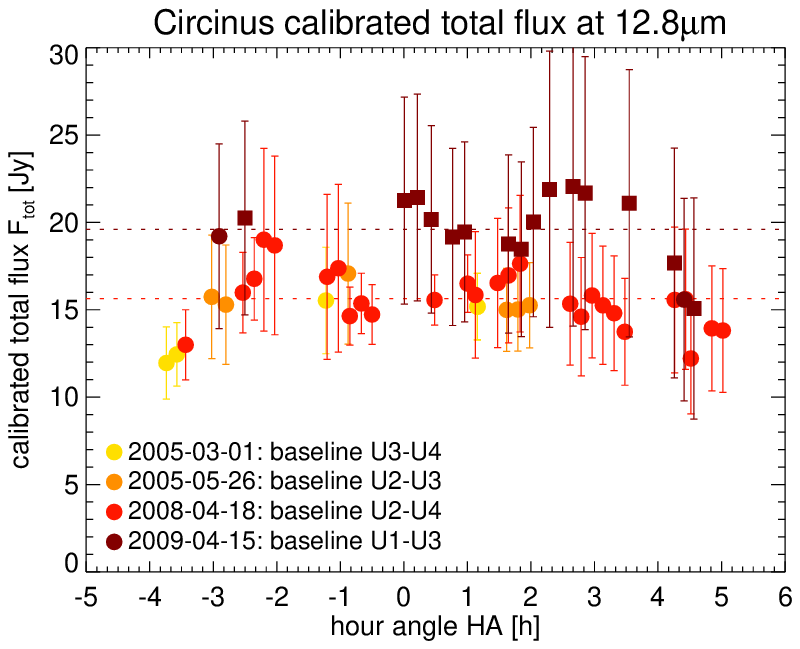}

\caption{Total flux of the Circinus galaxy at $8.2\,\mathrm{\mu m}$ (left)
and $12.8\,\mathrm{\mu m}$ (right) as a function of the hour angle
for four different epochs. The average fluxes at the respective wavelengths
on 2008-04-18 and 2009-04-15 are indicated by the dashed lines.\label{fig:fig3_photometry-variability}}
\end{figure*}
While there seem to be no significant discrepancies between the total
flux spectra of the individual measurements, there are however clear
trends when considering multiple measurements. In principle, MIDI
should always measure the same calibrated total flux spectrum for
a specific source, independent of the instrument settings and the
baseline geometry. However, this is not the case for the Circinus
galaxy, where significant changes in the total flux spectrum appear.
Figure~\ref{fig:fig3_photometry-variability} shows the total fluxes
at $8.2\,\mathrm{\mu m}$ and $12.8\,\mathrm{\mu m}$ as a function
of the hour angle for four different epochs. In each of these epochs,
the total flux spectrum was measured multiple times. The measurements
obtained in 2005 are all consistent with each other. For the measurements
obtained on 2008-04-18 and 2009-04-15, on the other hand, we find
a clear decrease in the total flux with increasing hour angle at wavelengths
shorter than $11.0\,\mathrm{\mu m}$ (see left panel of Fig.~\ref{fig:fig3_photometry-variability}):
on 2008-04-18, the flux at $8.2\,\mathrm{\mu m}$ apparently decreased
from $\sim7\,\mathrm{Jy}$ at the start of the night to $\sim3\,\mathrm{Jy}$
at the end of the night, that is by more than a factor of two. At
longer wavelengths, the flux levels remain more or less constant during
the course of all of the observing runs (see right panel of Fig.~\ref{fig:fig3_photometry-variability}).
In addition, there seems to be a general increase in the total flux
between 2008 and 2009. At $8.2\,\mathrm{\mu m}$ the average flux
(indicated by the dashed lines in Fig.~\ref{fig:fig3_photometry-variability})
increased from $\sim5.0\,\mathrm{Jy}$ to about $7.6\,\mathrm{Jy}$.
At $12.8\,\mathrm{\mu m}$ the increase is from $\sim15.6\,\mathrm{Jy}$
to $\sim19.6\,\mathrm{Jy}$.

We have carried out a thorough analysis to determine the cause for
these changes in the total flux measurements. We checked if (1) possible
changes in the instrumental setup, (2) the different airmasses of
the observation, (3) variations of the atmospheric conditions, (4)
changes in the performance of the adaptive optics system or (5) variability
of the calibrator could have caused the observed flux changes, but
we find none of these explanations to be conclusive (for details,
see \citealt{2013Tristram1}).

Slit losses might be responsible for the continuous flux decreases
on 2008-04-18 and 2009-04-15. Especially before 2009, the Circinus
nucleus was not always perfectly centred in the slit of MIDI due to
an error in the reference position. For all our observations, the
$200\,\mathrm{\mu m}$ wide slit was used, corresponding to widths
of $0.52\,\mathrm{arcsec}$ and $2.29\,\mathrm{arcsec}$ for the UTs
and ATs respectively. Because the mid-infrared emission of the galaxy
is slightly extended already in single-dish observations \citep[e.g.][]{2005Packham}
and the field of view of MIDI rotates on the sky over the course of
a night, this could have lead to a gradual decrease in the measured
total fluxes. On the other hand, the resolved emission constitutes
only about $20\%$ of the total $N$-band emission of the Circinus
nucleus \citep{2005Packham,2010Reunanen}. By consequence, slit losses
due to the orientation of the extended emission cannot account for
a change of the flux by up to a factor of two. Furthermore, it is
unclear why such slit losses should not have played a role in 2005
and why they should only have an effect at short wavelengths, where
potential slit losses should instead be reduced. A more accurate positioning
of the source in the slit in 2009 could also explain the increase
in the total flux for these measurements. The effect is, however,
not sufficient to explain all of the observed increase.

If the light of the Circinus galaxy were significantly polarised,
the rotation of the field of view together with the MIDI and VLTI
optics could lead to a smooth change of the flux over the night. To
obtain the observed change by a factor of 2 at $8.2\,\mathrm{\mu m}$
on 2008-04-18, this would require a degree of polarisation of 50\%
at the short wavelength end of the $N$-band. There are no polarisation
measurements for the Circinus galaxy in the $N$-band, but in the
$K$-band the nucleus of the Circinus galaxy has a polarisation of
the order of $3$ to $4\,\%$ \citep{2000Alexander}. The degree of
polarisation of NGC~1068 in the mid-infrared is less than $3\,\%$
\citep{2000Smith,2007Packham}, that of Mrk~231, a Seyfert 1 galaxy,
$8\,\%$ \citep{2001Siebenmorgen}. It therefore seems unlikely that
the mid-infrared emission of the nucleus of the Circinus galaxy is
much higher polarised and that the degree of polarisation is strongly
wavelength dependent. 

A further possible explanation is that the emission from the Circinus
galaxy itself has changed. As the mid-infrared emission is dominated
by the emission from warm dust, variations on timescales of hours
are not plausible and cannot be held responsible for the flux decrease
observed in the course of a night. An increase in the flux over the
period of one year, on the other hand, is very well possible physically.
Indeed there is further evidence that the total flux of the Circinus
galaxy has increased intrinsically between 2008 and 2009: a flux increase
is also seen in single-dish photometry at $11.9\,\mathrm{\mu m}$
with VISIR and the acquisition images obtained with MIDI. There are,
however, also inconsistencies with an increase in the intrinsic mid-infrared
flux of the nucleus of the Circinus galaxy. Because the increase took
place over a period of less than a year, the variable emission should
come from a region smaller than $1\,\mathrm{ly}\approx0.3\,\mathrm{pc}$,
i.e.\ $15\,\mathrm{mas}$ on the sky. Therefore the increase in the
flux should be mainly seen in the correlated fluxes on longer baselines,
which probe exactly these spatial scales. Although there seems to
be a slight increase in the correlated flux measurements since 2009,
it is only by a few hundred mJy (see previous Sect.). This is by far
not enough to explain the increase in the total flux by more than
$2\:\mathrm{Jy}$. Furthermore, the AT measurements do not show an
increase in the correlated fluxes but rather a decrease. In summary
this means we have (1) a flux \emph{increase} by up to a few hundred
mJy within $\sim80\,\mathrm{mas}$, (2) a possible flux \emph{decrease}
within $\sim170\,\mathrm{mas}$ and (3) an \emph{increase} by more
than $2\:\mathrm{Jy}$ within $\sim500\,\mathrm{mas}$. This is hard
to explain physically. At least two ``bursts'' would have to be
travelling outward through the dust distribution.

\begin{figure*}
\begin{minipage}[b]{12cm}%
\centering

\includegraphics[bb=127bp 255bp 467bp 565bp]{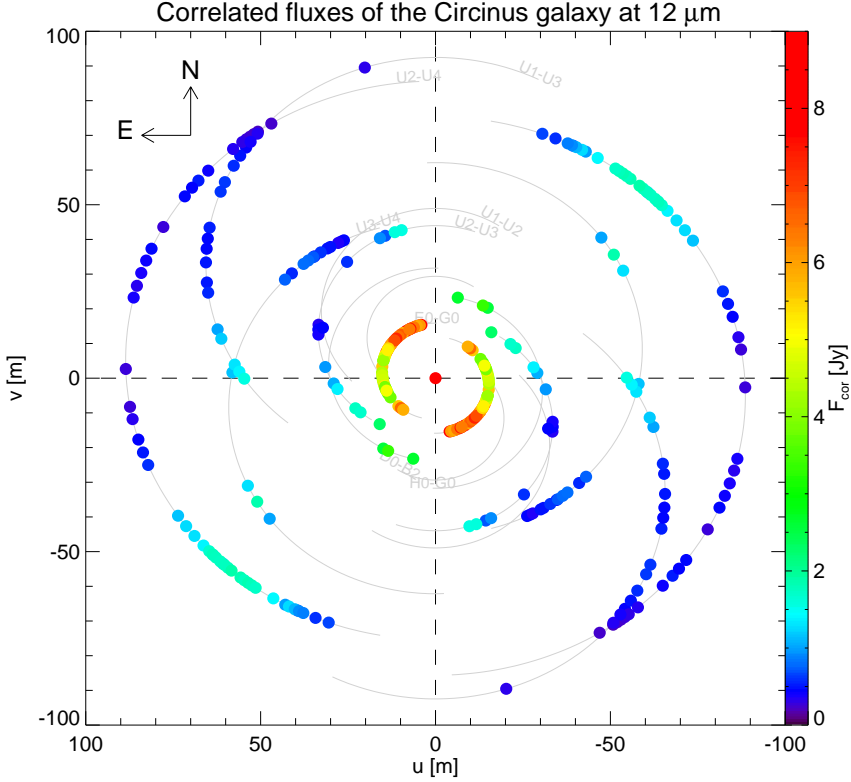}%
\end{minipage}\hfill{}%
\begin{minipage}[b]{6cm}%
\caption{Correlated fluxes at $12\,\mathrm{\mu m}$ for all $uv$ points containing
useful interferometric data. The points are colour coded according
to their correlated flux, $F_{\mathrm{cor}}(12\,\mathrm{\mu m})$,
using a square root colour scaling as indicated in the colour bar
on the right. The $uv$ point at the origin represents the averaged
total flux of the source, which is outside the plotted range of colours:
$\overline{F}_{\mathrm{tot}}(12\,\mathrm{\mu m})=10.7\,\mathrm{Jy}$.
\label{fig:fig4_uvplane-correlated}}
\end{minipage}
\end{figure*}
So far, no studies of the infrared variability of the Circinus galaxy
have been published. Therefore, we started a more detailed investigation
ourselves, including monitoring observations. A detailed discussion
of the variability in the nucleus of the Circinus galaxy goes beyond
the scope of this paper and will be presented in a future publication.

\subsubsection{Conclusion from the data consistency checks\label{sub:consistency-conclusion}}

We conclude that the differences between the individual measurements
of the correlated fluxes at the same position in the $uv$ plane are
in general consistent with the statistical expectations. On the other
hand, there is no clear picture that can explain (1) the continuous
decrease in the total flux during the observations on 2008-04-18 and
2009-04-15 as well as (2) the increase in the total flux between 2008
and 2009. For the following analysis, we will simply assume that the
emission intrinsic to the Circinus galaxy has not changed between
the interferometric measurements. We base this assumption on the fact
that the correlated fluxes on the shortest baseline, i.e.\ within
spatial scales of $\sim170\,\mathrm{mas}$ essentially remained constant.
Because our further analysis and modelling is mainly based on the
correlated fluxes and the differential phases, we are confident that
the general results do not depend on a full discussion of possible
variability. Furthermore, we find no solid basis on which to reject
individual measurements that do not agree with other measurements.
Therefore, we will retain all measured $uv$ points for the following
analysis.

\section{Results\label{sec:results}}

In total we obtained 152 useful measurements of the correlated flux
spectra and differential phases and 74 useful measurements of the
total flux spectra. This includes 20 correlated flux measurements
already published in \citet{2007Tristram2}.

The new reduction of the previously published data in general increases
the data quality. The positive bias of the correlated fluxes and visibilities
in the ozone feature between $9.5$ and $10.0\,\mathrm{\mu m}$ and
at the edges of the $N$-band is reduced, especially for the data
observed in 2004 and 2005. The more accurate group delay estimation
leads to a slight increase in the correlated fluxes at the long wavelength
end, but the overall spectral shape and flux levels remain unchanged.
With correlated flux levels of more than $0.4\,\mathrm{Jy}$ at $12.0\,\mathrm{\mu m}$
in most cases, the result is robust with respect to the data reduction,
and we obtain no contradictions to the values published in 2007. Due
to the improved masking and sky residual estimation, the scatter of
the total flux spectra is reduced significantly. The wavelength calibration
of the MIDI spectra in EWS was also corrected slightly, resulting
in a shift of the spectra to shorter wavelengths by about $0.1\,\mathrm{\mu m}$.
All spectra were corrected for the peculiar redshift of the Circinus
galaxy of $z=0.00145$ ($v_{\mathrm{sys}}=434\pm3\,\mathrm{km}\,\mathrm{s}^{-1}$,
\citealt{2004Koribalski}).

All 74 useful measurements of the total flux spectra were combined
by a weighted average to obtain a single estimate for the total flux
spectrum of the Circinus nucleus: $\overline{F}_{\mathrm{tot}}$.
The spectrum agrees with the one published in 2007. It is shown as
part of Fig.~\ref{fig:fig10_sed}. The spectrum rises from $\sim6\,\mathrm{Jy}$
at $8\,\mathrm{\mu m}$ to $\sim16\,\mathrm{Jy}$ at $13\,\mathrm{\mu m}$,
which is quite ``red'' ($\overline{F}_{\mathrm{tot}}(8\,\mathrm{\mu m})<\overline{F}_{\mathrm{tot}}(13\,\mathrm{\mu m})$)
and indicative of emission from warm ($T\sim290\,\mathrm{K}$) dust.
The spectrum is dominated by a deep silicate absorption feature over
almost the entire $N$-band. For the following, we will consider the
total flux spectrum as a measurement with a projected baseline length
of $\mathit{BL}=0\,\mathrm{m}$.

We use all measurements of the correlated fluxes and phases individually
and do not average measurements close in $uv$ space. All useful correlated
fluxes at $12\,\mathrm{\mu m}$ are listed in Table~\ref{tab:list-of-uvpoints}
in the Appendix and plotted in Fig.~\ref{fig:fig4_uvplane-correlated}.
With correlated fluxes at $12\,\mathrm{\mu m}$ ranging from $\sim8\,\mathrm{Jy}$
(corresponding to $V\sim0.8$) on the shortest baselines to $\sim0.4\,\mathrm{Jy}$
($V\sim0.04$) on certain long baselines, we clearly resolve the mid-infrared
emission in the nucleus of the Circinus galaxy. The $uv$ plane also
shows that along certain position angles, the correlated flux is higher
than along others. A very prominent example is the increase at the
end of the baseline UT2-UT4, leading to cyan and green colours (corresponding
to $F_{\mathrm{cor}}(12\,\mathrm{\mu m})>1.5\,\mathrm{Jy}$) in Fig.~\ref{fig:fig4_uvplane-correlated}.
This increase will be discussed in Sect.~\ref{sub:angular_dependency}.

Many of the correlated flux spectra (see Fig.~\ref{fig:alldata-correlated}
in the Appendix) have a shape similar to the total flux spectrum.
Especially on short baselines, the correlated flux spectra are similar
to the total flux spectrum when the spectral change due to the resolution
effect at different wavelengths is taken into account.

On longer baselines, however, this is not always the case. Most noticeably,
the short wavelength emission often either disappears completely or
there is a downturn in the correlated flux shortward of $8.7\,\mathrm{\mu m}$
without any significant signal in the differential phases (see e.g.\ C6
in Fig.~\ref{fig:fig2_comparison-of-measurements}). Interestingly,
a similar decrease, albeit mainly in the visibilities and for $\lambda<9.0\,\mathrm{\mu m}$,
might also be present in certain $uv$ points for \object{NGC~424}
\citep{2012Hoenig} and \object{NGC~3783} \citep{2013Hoenig}. Contamination
of the total flux by the wings of a spatially extended polycyclic
aromatic hydrocarbon (PAH) feature at $7.7\,\mathrm{\mu m}$ was discussed
as a possible reason in \citet{2012Hoenig}. This would, however,
only affect the visibilities and not the correlated fluxes we are
considering here. Furthermore, the $8.6$ and $11.3\,\mathrm{\mu m}$
PAH features are completely absent in our MIDI total flux spectrum
(see inset of Fig.~\ref{fig:fig10_sed}) or the nuclear spectra obtained
by \citet{2006Roche}%
\footnote{On scales of tens of arcseconds ($>200\,\mathrm{pc}$) however, the
Circinus galaxy indeed shows significant PAH emission at $7.7$, $8.6$
and $11.3\,\mathrm{\mu m}$ \citep{1996Moorwood,2008Galliano}, most
likely from the circumnuclear starburst.%
}. Because the $7.7\,\mathrm{\mu m}$ feature is roughly correlated
to the $8.6\,\mathrm{\mu m}$ feature \citep{2008Galliano} and, in
AGN environments, suppressed with respect to the $11.7\,\mathrm{\mu m}$
feature \citep{2007Smith}, we conclude that any contribution from
the wing of the $7.7\,\mathrm{\mu m}$ feature to our nuclear fluxes
is negligible. For these reasons we rule out any PAH contamination
to be responsible for the downturn of the correlated flux shortward
of $8.7\,\mathrm{\mu m}$. A further possible explanation could be
instrumental/calibration effects, especially correlation losses \citep{2012Burtscher}.
However, this would not explain why the downturn appears so abruptly
for $\lambda<8.7\,\mathrm{\mu m}$. The correlation losses should
instead be a smooth function of wavelength and only become significant
for $F_{\mathrm{cor}}(12\,\mathrm{\mu m})<150\,\mathrm{mJy}$$ $
\citep{2013Burtscher}. This is not the case for the measurements
at hand. It seems that the emission at the shortest wavelengths is
almost fully resolved out and thus comes from an extended emission
region (see discussion in Sect.~\ref{sub:hot-dust}).

In a few measurements, also a decrease in the correlated fluxes at
longer wavelengths is present (see e.g.\ C5 in Fig.~\ref{fig:fig2_comparison-of-measurements}).
In these cases, strong gradients in the differential phases also appear.
We interpret these signatures (which are also present in the visibilities)
as evidence of a more complex brightness distribution with small scale
structure (see discussion in Sect.~\ref{subsub:phases-smallscale}).

\subsection{Radial dependency of the correlated fluxes\label{sub:radial_dependency}}

\begin{figure}
\centering

\includegraphics[bb=170bp 325bp 420bp 512bp]{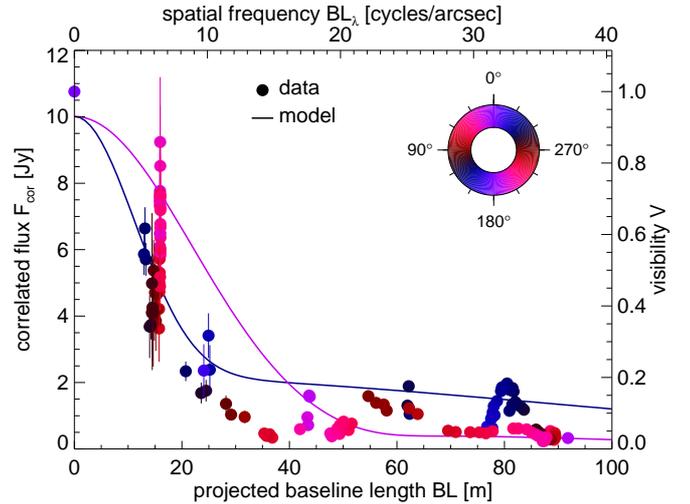}

\caption{Correlated fluxes ($F_{\mathrm{cor}}$, left ordinate) or visibilities
($V=F_{\mathrm{cor}}/\overline{F}_{\mathrm{tot}}$, right ordinate)
of the Circinus galaxy at $12\,\mathrm{\mu m}$ as a function of the
projected baseline length $\mathit{BL}$ (bottom axis) or spatial
frequency (top axis). The data are colour coded with the position
angle (see compass on the top right). Overplotted by two thick continuous
lines are the correlated fluxes of the three-component model discussed
in Sect.~\ref{sec:modelling}. The correlated fluxes of the model
are along $\mathit{PA}=17\text{\textdegree}$ (violet) and $\mathit{PA}=137\text{\textdegree}$
(dark blue). Note that the errors in the measurements with UT baselines
($\mathit{BL}>30\,\mathrm{m}$) are smaller than the plot symbol and
that the model does not fully reproduce the total flux (plotted at
$\mathit{BL}=0\,\mathrm{m}$) for $\lambda>11.5\,\mathrm{\mu m}$
(see Sect.~\ref{sub:spectral-energy-distribution})$ $.\label{fig:fig5_correlated-flux-radial}}
\end{figure}

The correlated fluxes (and visibilities) at $12\,\mathrm{\mu m}$
as a function of the projected baseline length $\mathit{BL}$ are
shown in Fig.~\ref{fig:fig5_correlated-flux-radial}. The correlated
fluxes quickly drop from the total flux of $10.7\,\mathrm{Jy}$ to
less than $2\,\mathrm{Jy}$ ($V\lesssim20\%$) at $\mathit{BL}\sim30\,\mathrm{m}$.
On longer baselines, the correlated fluxes remain on more or less
the same level between $0.2\,\mathrm{Jy}$ and $2.0\,\mathrm{Jy}$.
Note that the apparent scatter in the measurements at a certain baseline
length is mainly due to measurements at different position angles
(see next Sect.).

The rapid drop at short baseline lengths implies that the mid-infrared
emission is mostly resolved out by the interferometer. More precisely,
only about $20\%$ of the mid-infrared emission comes from structures
smaller than about $70\,\mathrm{mas}$ in diameter. $80\%$ of the
emission is located on spatial scales between $\sim70\,\mathrm{mas}$
and $\sim500\,\mathrm{mas}$. The new observations with the ATs (at
baseline lengths of $\mathit{BL}\sim15\,\mathrm{m}$) now also probe
the majority of the flux in this extended emission region.

Because the correlated fluxes do not decrease much farther for $30\,\mathrm{m}\lesssim\mathit{BL}\lesssim95\,\mathrm{m}$
(a change of the baseline length and, thus, spatial resolution by
a factor of three), a single or several unresolved structures (``clumps'')
with sizes below the resolution limit of the interferometer of about
$15\,\mathrm{mas}$ must be present. The two regimes of the visibility
function suggest that the corresponding brightness distribution has
two distinct spatial scales: a large scale that is quickly resolved
by the interferometer when increasing the baseline length and a small
scale that essentially remains unresolved even at the longest baselines
(i.e.\ at the smallest spatial scales probed by the interferometer).

\subsection{Angular dependency of the correlated fluxes\label{sub:angular_dependency}}

\begin{figure*}
\centering

\includegraphics[bb=170bp 325bp 420bp 512bp]{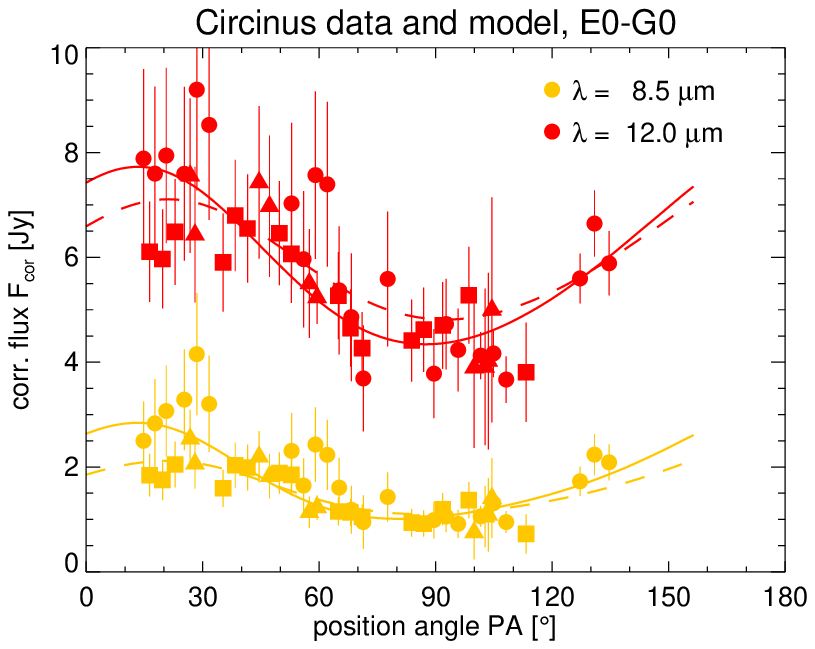}\includegraphics[bb=170bp 325bp 420bp 512bp]{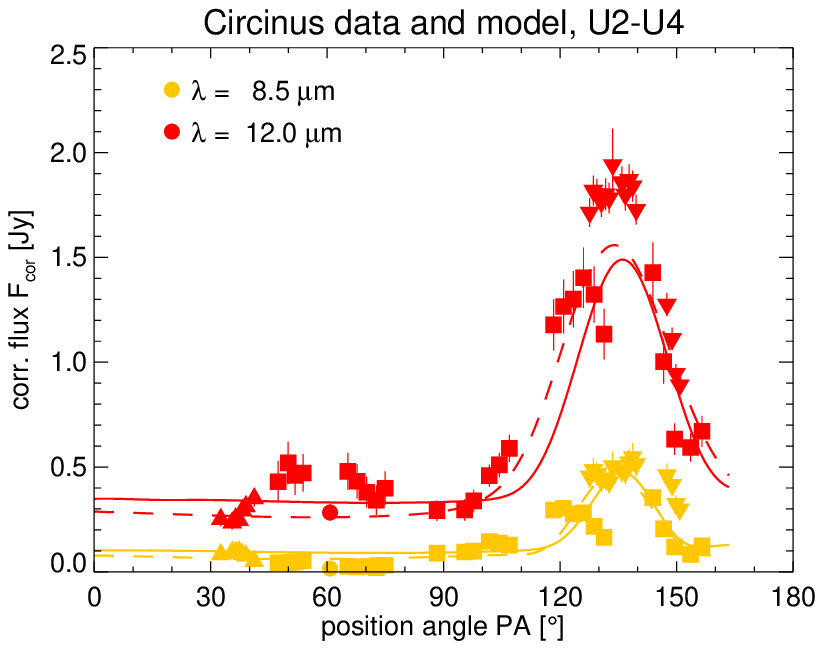}

\caption{Correlated fluxes of the Circinus galaxy at $8.5$ and $12.0\,\mathrm{\mu m}$
as a function of the position angle for the baselines E0-G0 (with
$13\,\mathrm{m}<\mathit{BL}<16\,\mathrm{m}$, left) and UT2-UT4 (with
$76\,\mathrm{m}<\mathit{BL}<90\,\mathrm{m}$, right). Also plotted
are the correlated fluxes of the three-component model discussed in
Sect.~\ref{sec:modelling}: fit 3 by continuous lines, fit 2 and
1 by dashed lines in the left and right panels, respectively.\label{fig:fig6_correlated-flux-pangle}}
\end{figure*}

There are two telescope combinations, E0-G0%
\footnote{This telescope combination is identical to the combination C1-A1 measured
in 2011 modulo an interchange of the telescopes.%
} (using the ATs) and UT2-UT4 (using the UTs), where the projected
baseline lengths roughly remain the same while the position angle
changes over a wide range due to the rotation of the Earth. This means
the same spatial scales are probed in different directions. The correlated
fluxes can therefore be directly compared to infer information on
the source size in different directions under the assumption of a
smooth, centrally peaked brightness distribution. The correlated fluxes
at $8.5$ and $12.0\,\mathrm{\mu m}$ for the two mentioned baselines
are shown in Fig.~\ref{fig:fig6_correlated-flux-pangle} as a function
of the position angle. On both baselines and at both wavelengths,
a clear dependency of the correlated fluxes on the position angle
is present.

With the E0-G0 baseline, $13\,\mathrm{m}<\mathit{BL}<16\,\mathrm{m}$,
spatial scales of the order of $150\,\mathrm{mas}$ are probed. The
uncertainties and scatter in the data points are relatively large
because these measurements were obtained with the ATs. Nevertheless,
the correlated fluxes show a clear trend with a broad minimum at $\mathit{PA}\sim90\text{\textdegree}$
at both $8.5$ and $12.0\,\mathrm{\mu m}$. This essentially means
that the emission is better resolved in this direction, i.e. it is
more extended along this position angle.

The UT2-UT4 baseline has projected baseline lengths between $76\,\mathrm{m}$
and $90\,\mathrm{m}$, corresponding to spatial scales of about $25\,\mathrm{mas}$.
The correlated fluxes show a very pronounced peak at $\mathit{PA}\sim135\text{\textdegree}$
with correlated fluxes more than three times higher than at other
position angles. This means that the source appears significantly
less resolved in this direction, while it is better resolved in all
other directions. Such a dependency on the position angle can be explained
by an extremely elongated emission component. Furthermore there is
a second, much weaker peak at $\mathit{PA}\sim60\text{\textdegree}$
in the correlated fluxes at $12.0\,\mathrm{\mu m}$. This peak is
not present at $8.5\,\mathrm{\mu m}$. We interpret this behaviour
as evidence of further small scale structure. Finally, the $12.0\,\mathrm{\mu m}$
fluxes are always above $\sim0.3\,\mathrm{Jy}$, indicating that there
is still unresolved flux at these wavelengths. The fluxes at $8.5\,\mathrm{\mu m}$
are consistent with zero at $\mathit{PA}\sim60\text{\textdegree}$.
Thus the $8.5\,\mathrm{\mu m}$ emission is completely resolved out
by the interferometer.

In summary, we conclude from the direct and completely model free
analysis of the data that there are two different orientations in
the mid-infrared brightness distribution of the Circinus galaxy: on
spatial scales of $\sim150\,\mathrm{mas}$ the emission is moderately
elongated along $\mathit{PA}\sim90\text{\textdegree}$, while on smaller
spatial scales, the emission is highly elongated along $\mathit{PA}\sim45\text{\textdegree}$.

\section{Modelling\label{sec:modelling}}

To get a better understanding of the overall structure of the emission,
we model the data. Motivated by the evidence of (1) an essentially
unresolved emission component, (2) a small, highly elongated component
and (3) an extended, only slightly elongated component, we expand
the two-component model from \citet{2007Tristram2} to three components.
In the following, we will refer to these three components as $i=1$
the ``unresolved'', $i=2$ the ``disk-like'' and $i=3$ the ``extended''
components of the emission. Despite being ``unresolved'', component
1 has a non-zero size because its brightness depends on its surface
and temperature.

The model is not intended to directly represent any physical structure
for the emission. To begin with, it is intended to capture the general
morphology and spectral properties of the surface brightness distribution
of the source, by fitting the interferometric data and the total flux
spectrum in the wavelength range between $\lambda=8.0$ and $13.0\,\mathrm{\mu m}$.

The three components are modelled as black-body emitters with a Gaussian
brightness distribution. The Gaussian emitters can be elliptical and
each emitter is behind an absorption screen responsible for the silicate
absorption. For each component, the elliptical Gaussian is a function
of the position on the sky $\left(\alpha,\delta\right)$ and takes
the functional form 

\begin{equation}
G_{i}(\alpha,\delta)=f_{i}\cdot\exp\left(-4\ln2\cdot\left[\left(\frac{\alpha_{i}'}{r_{i}\cdot\Delta_{i}}\right)^{2}+\left(\frac{\delta_{i}'}{\Delta_{i}}\right)^{2}\right]\right),
\end{equation}
where $\alpha_{i}'=\left(\alpha-\alpha_{i}\right)\cdot\cos\psi_{i}+\left(\delta-\delta_{i}\right)\cdot\sin\psi_{i}$
and $\delta_{i}'=\left(\alpha-\alpha_{i}\right)\cdot\sin\psi_{i}-\left(\delta-\delta_{i}\right)\cdot\cos\psi_{i}$
are the positional coordinates of the Gaussians, which are offset
by $\left(\alpha_{i},\delta_{i}\right)$ from the centre and rotated
by the position angle $\psi_{i}$. $f_{i}$ specifies the maximum
surface filling and emissivity factor of this component at the position
$\left(\alpha_{i},\delta_{i}\right)$, $\Delta_{i}$ is the full width
half maximum (FWHM) of the Gaussian along its major axis (oriented
along $\psi_{i}$), and $r_{i}$ is the ratio of the minor to major
FWHM. A maximum surface filling factor of 1 is possible for the sum
of all three components. This was not handled entirely correctly in
the modelling in \citet{2007Tristram2}. There, the emission of the
two components was simply added one to another, which could effectively
result in a total filling factor greater than one. This physical inconsistency,
however, did not have any consequences on the overall results. Now
we take this additional, physical constraint into account, and use
modified Gaussians calculated as

\begin{equation}
\tilde{G}_{i}(\alpha,\delta)=\left[G_{i}(\alpha,\delta)-\left(\left(\sum_{j=1}^{i}G_{j}(\alpha,\delta)-1\right)>0\right)\right]>0.
\end{equation}
The final flux density of the model at a certain wavelength $\lambda$
is given by

\begin{equation}
F\left(\lambda,\alpha,\delta\right)=\sum_{i=1}^{3}\tilde{G}_{i}(\alpha,\delta)\cdot F_{\mathrm{BB}}\left(T_{i},\lambda\right)\cdot\mathrm{e}^{-\tau_{i}\cdot\tau(\lambda)},
\end{equation}
where $F_{\mathrm{BB}}$ is the black-body intensity depending on
the temperature $T_{i}$, and $\tau_{i}$ is the optical depth of
the silicate feature. For the extended component ($i=3$), the optical
depth can linearly change along the major axis, i.e.\ $\tau_{3}(\alpha,\delta)=\tau_{3}+\xi_{3}\cdot d$,
where $d$ is the distance from $(\alpha,\delta)=(0,0)$ projected
onto the major axis. This gradient is a first approximation for the
overall change of the silicate absorption depth and is motivated by
the gradient seen in the mid-infrared spectra by \citet{2006Roche}.
It will be discussed in more detail in Sect.~\ref{subsub:phases-largescale}.
The template absorption profile $\tau(\lambda)$ is the same as in
\citet{2007Tristram2}, i.e.\ it is derived from the extinction curves
of \citet{2005Schartmann} and \citet{2004Kemper}.

In total, the model has 25 parameters (8 for each component plus $\xi_{3}$).
However, because the innermost component is assumed to be essentially
unresolved, it is not elongated, and thus $\psi_{1}=0$ and $r_{1}=1$.
Furthermore, because MIDI only measures differential phases and not
the full Fourier phases (see Sect.~\ref{sub:data-reduction}), the
absolute position on the sky is undetermined. We therefore define
the disk-like component to lie at the origin of our coordinates, that
is $\left(\alpha_{2},\delta_{2}\right)\equiv\left(0,0\right)$. Furthermore,
the offset of the extended component is not well constrained and it
does not lead to a significant improvement of the fit (see below).
We therefore also assume the extended component to be located at the
centre, $\left(\alpha_{3},\delta_{3}\right)\equiv\left(0,0\right)$.
The coordinates of the unresolved component are therefore to be interpreted
as offsets with respect to the other components. We are then left
with up to 19 free parameters in the model. To carry out the fit,
the brightness distribution of the model is calculated for $8.0\,\mathrm{\mu m}<\lambda<13.0\,\mathrm{\mu m}$,
then Fourier transformed to obtain the complex correlated fluxes:
\begin{equation}
\mathcal{F}\left(\lambda,u,v\right)=\iint d\alpha\, d\delta\, F\left(\lambda,\alpha,\delta\right)e^{-2\pi\, i\,(u\alpha+v\delta)/\lambda}.
\end{equation}
The moduli of the correlated fluxes, $F_{\mathrm{corr}}=\left|\mathcal{F}\right|$
, and the wavelength differential phases, $\phi_{\mathrm{diff}}$,
are extracted at the measured $uv$ points and compared to the measurements.
All calculations are carried out in IDL using custom programmes and
employ the programme \textsc{\large mpfit} \citep{2009Markwardt}
for non-linear least squares fitting.

Because the components are self-similar and can become degenerate,
we first fit certain parts of the $uv$ plane separately. This approach
allows us to understand the source morphology on the different spatial
scales (cf.\ Sect.~\ref{sec:results}) first, before fitting for
the entire brightness distribution. In fit 1, we fit the data from
baseline U2-U4 only using the two inner components ($i=1,2$); in
fit 2, we fit the data with $\mathit{BL}<20\,\mathrm{m}$ (essentially
baseline E0-G0) using the extended component ($i=3$). The parameters
of the remaining components are kept fixed. Only in fit 3, is all
the data fitted by all three components, using the parameters of the
previous fits as first guesses. The best fit parameters for all three
fits are listed in Table~\ref{tab:model-parameters-1}.
\begin{table}
\caption{Parameters of the three-component Gaussian fits. Parameters held fixed
at certain values are enclosed in brackets.\label{tab:model-parameters-1}}

\begin{tabular}{ccccc}
\hline 
\hline  &  & fit 1 & fit 2 & fit 3\tabularnewline
$i$ & parameter & baseline U2-U4 & $\mathit{BL}<20\,\mathrm{m}$ & all baselines\tabularnewline
\hline 
\multirow{6}{*}{\begin{sideways}
1: unresolved\enskip{}
\end{sideways}} & $\alpha_{1}$ {[}mas{]} & $2.9_{-18}^{+15}$ & $\left(3.0\right)$ & $-8.5_{-2.8}^{+7.0}$\tabularnewline
 & $\delta_{1}$ {[}mas{]} & $6.5_{-24}^{+10}$ & $\left(6.0\right)$ & $11.6_{-4.4}^{+2.6}$\tabularnewline
 & $\Delta_{1}$ {[}mas{]} & $10_{-1.2}^{+14}$ & $\left(10\right)$ & $12\pm2$\tabularnewline
 & $\tau_{1}$ & $0.80_{-0.5}^{+1.0}$ & $\left(0.8\right)$ & $1.23_{-0.24}^{+0.15}$\tabularnewline
 & $T_{1}$ {[}K{]} & $264_{-22}^{+35}$ & $\left(260\right)$ & $317\pm22$\tabularnewline
 & $f_{1}$ & $\left(1.00\right)$ & $\left(1.0\right)$ & $0.49_{-0.17}^{+0.29}$\tabularnewline
\hline 
\multirow{6}{*}{\begin{sideways}
2: disk-like\enskip{}
\end{sideways}} & $\Delta_{2}$ {[}mas{]} & $45\pm11$ & $\left(45\right)$ & $57\pm15$\tabularnewline
 & $r_{2}$ & $0.17_{-0.06}^{+0.20}$ & $\left(0.17\right)$ & $0.16\pm0.04$\tabularnewline
 & $\psi_{2}$ {[}\textdegree{}{]} & $44\pm6$ & $\left(44\right)$ & $46\pm3$\tabularnewline
 & $\tau_{2}$ & $1.92{}_{-0.43}^{+0.19}$ & $\left(2.00\right)$ & $1.88\pm0.40$\tabularnewline
 & $T_{2}$ {[}K{]} & $288_{-11}^{+32}$ & $\left(290\right)$ & $290_{-13}^{+22}$\tabularnewline
 & $f_{2}$ & $\left(1.00\right)$ & $\left(1.00\right)$ & $0.55_{-0.14}^{+0.24}$\tabularnewline
\hline 
\multirow{7}{*}{\begin{sideways}
3: extended\enskip{}
\end{sideways}} & $\Delta_{3}$ {[}mas{]} & -- & $92_{-4}^{+10}$ & $93_{-12}^{+6}$\tabularnewline
 & $r_{3}$ & -- & $0.60\pm0.12$ & $0.45_{-0.05}^{+0.07}$\tabularnewline
 & $\psi_{3}$ {[}\textdegree{}{]} & -- & $114\pm3$ & $107\pm8$\tabularnewline
 & $\tau_{3}$ & -- & $2.71_{-0.08}^{+0.36}$ & $2.40_{-0.6}^{+0.1}$\tabularnewline
 & $\xi_{3}$ {[}$\mathrm{arcsec}^{-1}${]} & -- & $30_{-4}^{+3}$ & $27\pm6$\tabularnewline
 & $T_{3}$ {[}K{]} & -- & $297_{-3}^{+145}$ & $304_{-8}^{+62}$\tabularnewline
 & $f_{3}$ & $\left(0.0\right)$ & $0.33_{-0.26}^{+0.08}$ & $0.34_{-0.22}^{+0.03}$\tabularnewline
\hline 
 & $\chi_{\mathrm{red}}^{2}$ & 5.22 & 1.06 & 6.32\tabularnewline
\hline 
\end{tabular}

\tablefoot{The parameters for each of the three components $i=\left[1,2,3\right]$
are: $\left(\alpha_{i},\delta_{i}\right)$ -- position; $\Delta_{i}$
-- FWHM; $r_{i}$ -- axis ratio; $\psi_{i}$ -- position angle; $\tau_{i}$
-- silicate optical depth; $\xi_{i}$ -- gradient of the silicate
optical depth; $T_{i}$ -- temperature; $f_{i}$ -- surface filling
factor. See text for details.}
\end{table}

For fit 1, the extended component is suppressed by setting $f_{3}\equiv0$.
It can be omitted entirely because the flux from this component is
almost completely resolved out ($V<2\%$) for $\mathit{BL}>70\,\mathrm{m}$.
Therefore, the correlated fluxes on the baseline U2-U4 primarily depend
on the inner two components. The two components can reproduce the
general trends of the data ($\chi_{\mathrm{red}}^{2}=5.2$, dashed
lines in the right panel of Fig.~\ref{fig:fig6_correlated-flux-pangle}).
The strong increase in the correlated flux at $\mathit{PA}\sim130\text{\textdegree}$
is well reproduced by the strong elongation of component $i=2$. For
this reason, it is called ``disk-like'', which will be discussed
in more detail in Sect.~\ref{subsub:morphology-disk}. Due to the
well measured position angle dependence of this peak, the position
angle of this elongated component is well determined. On the other
hand, second order variations of the correlated flux, such as the
second peak at $\mathit{PA}\sim60\text{\textdegree}$ (see Sect.~\ref{sub:angular_dependency})
or changes in the spectral shape, are not reproduced by our simple
model. They require a more complex brightness distribution. The differential
phases can be at least qualitatively reproduced to some degree by
an offset of the unresolved component from the disk-like component.

For fit 2, the two inner components are included in the fit because
they contribute significantly to the correlated flux. However, their
parameters are held fixed at the (approximate) values of fit 1. Because
the AT data changes more smoothly as a function of position in the
$uv$ plane and because it has larger uncertainties, both the correlated
fluxes as well as the differential phases can be well modelled by
the extended component ($\chi_{\mathrm{red}}^{2}=1.1$, dashed lines
in the left panel of Fig.~\ref{fig:fig6_correlated-flux-pangle}).
In fact, the extended component alone is sufficient to achieve the
good fit; the inner two components are not required to model the AT
data.

Finally, in fit 3, all free parameters are fitted using the 152 correlated
flux spectra and differential phases as well as the averaged total
flux spectrum. For this fit, we obtain a relatively high value of
the reduced chi-squared, $\chi_{\mathrm{red}}^{2}=6.3$. This means
that our simple model is not a ``good'' fit to the full data set;
it only traces the general structure of the emission. Nonetheless
the quality of our fit is comparable to the similar fit (also using
three components) of the correlated fluxes and differential phases
in NGC~1068, where $\chi_{\mathrm{red}}^{2}(\textrm{NGC 1068})=6.2$
\citep{2014Lopez-Gonzaga}. To determine a possible offset of the
extended component like in NGC~1068, we carried out a parameter scan
for $\alpha_{3}$ and $\delta_{3}$. Although we find a best fit for
$\left(\alpha_{3},\delta_{3}\right)=\left(-45,3\right)\,\mathrm{mas}$
with $\chi_{\mathrm{red}}^{2}=6.0$, we also find very different offsets
with similarly good values of $\chi_{\mathrm{red}}^{2}$. Due to this
ambiguity and the only small improvement of the fit, we see no compelling
evidence for a particular offset of the extended component and therefore
set $\left(\alpha_{3},\delta_{3}\right)\equiv\left(0,0\right)$. The
correlated fluxes of fit 3 are compared to the data as a function
of the position angle in Fig.~\ref{fig:fig6_correlated-flux-pangle}
(continuous curves). The dispersed correlated fluxes and differential
phases of this fit for all $uv$-points are plotted in red in Figs.~\ref{fig:alldata-correlated}
and \ref{fig:alldata-phases}.

The formal fit errors calculated from the covariance matrix are very
small and not representative of our fit of a smooth model to a more
complex emission distribution. To determine more realistic uncertainties
for the fit parameters of all three fits, we use the non-parametric
bootstrap with replacement \citep{1979Efron}. We employ a block bootstrap
\citep{1985Hall} due to the correlation of consecutively observed
$uv$ points, i.e.\ we resample the data using small sequences of
$uv$ points instead of individual $uv$ points before fitting the
resampled data. The errors listed in Table~\ref{tab:model-parameters-1}
are directly estimated from the bootstrap distribution of the respective
parameter and mark the $68.3\%$ (1$\sigma$) confidence intervals.

Our model can fit the data on the shortest baselines very well, which
means that it reproduces the low spatial frequencies of the source
adequately. On longer baselines, however, the data is not well reproduced
by our model. This is predominantly due to small scale variations
of the correlated fluxes and differential phases at longer baselines
(cf.\ Fig.\ref{fig:fig4_uvplane-correlated}), which cannot be reproduced
by our smooth model. We interpret these variations as signatures for
small scale structures that our model obviously cannot replicate.

Finally, a few remarks on degeneracies: several parameters of our
model are not independent. The clearest example is the degeneracy
between the temperature $T_{i}$ and the surface filling factor $f_{i}$.
Because we are fitting a narrow wavelength range ($8\,\mathrm{\mu m}<\lambda<13\,\mathrm{\mu m}$),
the temperatures of our dust components are not well constrained.
A small change in temperature has a direct influence on the brightness
of the source, which can be compensated by changing the surface filling
factor. Similar degeneracies are present between the size and the
axis ratio of the source, which all change the emitted flux density.
Depending on how well these parameters are constrained by the interferometric
measurements, these parameters can become degenerate.

\section{Discussion\label{sec:discussion}}

\subsection{Morphology\label{sub:morphology}}

The direct analysis of the data (Sect.~\ref{sec:results}) and our
modelling (Sect.~\ref{sec:modelling}) confirm that the mid-infrared
emission in the nucleus of the Circinus galaxy comes from at least
two distinct components: a highly elongated, compact ``disk-like''
component and a moderately elongated, extended component. To some
degree, the distinction between the two components is suggested by
the two different regimes of the correlated fluxes as a function of
the projected baseline length (see Sect.~\ref{sub:radial_dependency}).
Primarily, however, the distinction is suggested by the different
orientations of the two components: the two components are elongated
roughly perpendicular to one another. Two clearly separated emission
components have also been found in NGC~1068 and NGC~3783 \citep{2009Raban,2013Hoenig},
and a two-component morphology in the infrared appears to be common
to a large number of AGN \citep{2011Kishimoto2,2013Burtscher}.

We interpret the mid-infrared emission as emission from warm dust
in the context of the hydrodynamic models of dusty tori in AGN by
\citet{2009Schartmann}, \citet{2009Wada} and \citet{2012Wada}.
These models find a relatively cold, geometrically thin and very turbulent
disk in the mid-plane of the torus, surrounded by a filamentary structure.
The latter consists of long radial filaments with a hot tenuous medium
in between. We associate the central, highly elongated component in
the Circinus nucleus with the dense disk in these simulations, and
we interpret the extended mid-infrared emission in the context of
the filamentary torus structure seen in these models.

\begin{figure}
\includegraphics[bb=0bp 0bp 512bp 512bp,width=1\columnwidth]{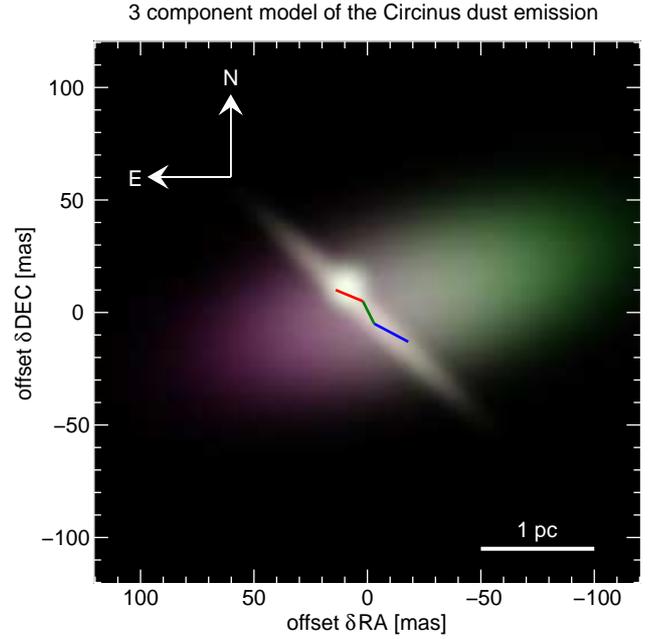}

\caption{False-colour image of the three-component model for the mid-infrared
emission of the nucleus of the Circinus galaxy (fit 3). The colours
red, green and blue correspond to the model at $13.0\,\mathrm{\mu m}$,
$10.5\,\mathrm{\mu m}$ and $8.0\,\mathrm{\mu m}$, respectively.
The colour scaling is logarithmic in order to show both bright as
well as faint features. Clearly the colour gradient of the extended
component due to the increase in the silicate depth towards the south-east
is visible. This colour gradient leads to a chromatic photocentre
shift towards the north-west. Despite the lower surface brightness,
$80\%$ of the emission comes from the extended component. Also plotted
is the trace of the water maser disk: the blue and red parts trace
the approaching and receding sides of the maser disk respectively.
Note that the relative offset of the mid-infrared emission with respect
to the maser disk is not known (see text for details). \label{fig:fig07_model-image}}

\end{figure}
A false-colour image of our best fitting model (fit 3) is shown in
Fig.~\ref{fig:fig07_model-image}, with the model images at $13.0\,\mathrm{\mu m}$,
$10.5\,\mathrm{\mu m}$ and $8.0\,\mathrm{\mu m}$ mapped to the red,
green and blue channels of the image, respectively.

When interpreting our observations, we have to take into account that
the emission is dominated by the warmest dust at a certain location,
which normally comes from the dust clouds directly illuminated by
the central UV source. There are probably also considerable amounts
of cooler dust. However, the cooler material only contributes insignificantly
to the infrared emission (see also Sect.~\ref{sub:spectral-energy-distribution}).

\subsubsection{The disk-like component\label{subsub:morphology-disk}}

The disk-like component is highly elongated and has a major axis FWHM
of $\Delta_{2}\sim1.1\,\mathrm{pc}$. Due to the strong position angle
dependency of the correlated fluxes for the longest baselines, the
position angle of the major axis is very well constrained: $ $$\psi_{2}=46\pm3\text{\textdegree}$.
The strong elongation of this component with an axis ratio of more
than $6:1$ at first suggests an interpretation as a highly inclined
disk, as in \citet{2007Tristram2}. This interpretation is supported
by the close agreement in orientation and size of this component with
the warped maser disk from \citet{2003Greenhill}. The masers were
modelled by a thin disk extending from $r_{\mathrm{in}}\sim0.1\,\mathrm{pc}$
to $r_{\mathrm{out}}\sim0.4\,\mathrm{pc}$. The maser disk is warped
with the position angle changing from $29\text{\textdegree}\pm3\text{\textdegree}$
at $r_{\mathrm{in}}$ to $56\text{\textdegree}\pm6\text{\textdegree}$
at $r_{\mathrm{out}}$. With a position angle of $\psi_{2}\sim46\text{\textdegree}$,
our disk-like component now matches this orientation much better than
previously. The larger size of the mid-infrared disk as compared to
the maser disk could be evidence of the disk extending out to larger
radii than is probed by the maser emission. We emphasise that the
agreement is only in orientation and size, not in the absolute position.
With MIDI alone, no absolute astrometry is possible because the absolute
phase signal is destroyed by the atmosphere (see Sect.~\ref{sub:data-reduction}).
By consequence, the relative position between the maser disk and our
disk-like component cannot be determined. In Fig.\ \ref{fig:fig07_model-image},
the disk-like component and the maser disk are plotted centred to
allow a good comparison, but the two structures might well be offset
with respect to each other.

Due to the strong elongation of the disk-like component, the associated
disk would have to be close to edge-on and cannot be very thick. If
we assume that the emission comes directly from an infinitesimally
thin and flat disk, we can use the observed axis ratio to constrain
its inclination: $i>75\text{\textdegree}$. Or, if seen edge-on ($i=90\text{\textdegree}$),
as suggested by the maser emission, we can derive a rough estimate
for the maximum thickness of the disk: $r_{2}\cdot\Delta_{2}<0.2\,\mathrm{pc}$.

An interpretation as emission directly from the disk is, however,
problematic. Due to the high densities required for maser emission
($n(\mathrm{H}_{2})\gtrsim10^{8}\,\mathrm{cm}^{-3}$, \citealt{1991Kylafis}),
we expect the disk to be optically thick and to thus appear in absorption
rather than in emission when seen close to edge-on. This is corroborated
by the dense disks in the hydrodynamical torus models, which appear
as dark lanes in the mid-infrared (see Fig.~8 in \citealt{2009Schartmann}
for an example). The effect is further enhanced by the anisotropic
radiation characteristic of the accretion disk, which emits less energy
in the direction of the disk plane. A dense disk, by consequence,
appears dark. Our disk-like component, on the contrary, appears in
emission.

Instead of the disk itself, we could be seeing the directly illuminated
inner rim of a highly inclined disk, similar to what is seen for circumstellar
disks (e.g.\ in \object{KK Ophiuchi}, \citealt{2013Kreplin}). However,
the inner rim of the disk, where the dust sublimates, is expected
to lie at $r_{\mathrm{sub}}=r_{\mathrm{in}}\sim0.03\,\mathrm{pc}$
for the Circinus galaxy. This is much smaller than the parsec-sized
elongation we observe. In addition, the inner rim is expected to consist
of hot dust close to the dust sublimation temperature, which is in
disagreement with the warm dust at $T\sim300\,\mathrm{K}$ suggested
by our observations. Only if the inner region of the disk were completely
dust free and the inner rim were located at a much larger distances
($r_{\mathrm{sub}}\ll r_{\mathrm{in}}\sim1\,\mathrm{pc}$), would
the size and temperature of the emission be in agreement with our
interferometric observations (see also discussion in Sect.~\ref{sub:hot-dust}).

Therefore, two alternative explanations seem more likely: First, the
emission could come from directly illuminated dusty material slightly
above or below the dense disk, such as filaments of dense material
swept up from the disk by turbulence. Or, the warp of the disk allows
us to see a part of the disk that is directly illuminated by the central
source. From the maser observations, \citet{2003Greenhill} deduce
that the western side of the warped disk is oriented such that we
can see its surface, which is directly illuminated by the accretion
disk (see their Fig.~8). Depending on the exact geometry of the warp
and the illumination characteristics of the accretion disk, this could
lead to the observed highly elongated mid-infrared emission. More
sophisticated radiative transfer calculations will have to be carried
out to constrain the possible geometries.

\subsubsection{Unresolved component\label{subsub:morphology-unresolved}}

The additional unresolved emission component is required to explain
the measurable correlated fluxes when the disk-like component is otherwise
fully resolved out. If, in fit 3, this unresolved component is located
north-east of the disk-like component, the fit quality is improved.
This is mainly because of a better agreement between the differential
phases of the model and the observations, especially for the strong
phase signal at $(u,v)\approx(-35,35)\,\mathrm{m}$ (dark blue points
at the beginning of the U3-U4 baseline in Fig.~\ref{fig:fig09_uvplane-diffphases}).
We interpret this as a sign that we are tracing smaller scale structure,
e.g.\ an enhanced part of the disk-like component, possibly a region
with less obscuration leading to the comparatively low silicate optical
depth of this emission component. For the following discussion we
will therefore consider the unresolved emission as a part of the disk-like
emission.

\subsubsection{Polar elongated dust\label{subsub:morphology-extended}}

The extended emission component is responsible for $\sim80\%$ of
the emission in the $N$-band, independent of the wavelength. This
can already be seen in Fig.\ \ref{fig:fig5_correlated-flux-radial},
where the visibility quickly drops to values of $\lesssim20\%$ with
increasing baseline length. Similarly high contributions to the total
emission from extended dust have been found for NGC~1068 and NGC~3783
\citep{2009Raban,2013Hoenig}. There is, however, one significant
difference: in these galaxies, the contribution of the extended component
depends strongly on the wavelength (e.g.\  from $55\%$ to $90\%$
between $\lambda=8.0$ and $13.0\,\mathrm{\mu m}$ for NGC~3783).
This is not the case in the nucleus of the Circinus galaxy due to
the lack of any significant temperature difference between the components
(see discussion in Sect.~\ref{sub:hot-dust}).

The extended emission is significantly elongated along $\mathit{PA}\sim107\text{\textdegree}$,
which is roughly (but not exactly) perpendicular to the disk-like
component ($\mathit{PA}\sim46\text{\textdegree}$) and, thus, in polar
direction. In fact, there is mounting evidence that this is a common
characteristic of Seyfert galaxies: also in NGC~1068, NGC~424 and
NGC~3783 the majority of the emission is extended in polar direction
\citep{2007Poncelet,2008Raban,2012Hoenig,2013Hoenig}. This obviously
raises the question: Where is the dust responsible for this emission
located?

The polar dust emission could originate from dust \emph{within} the
ionisation cone. Dust clouds inside the ionisation cone or, more generally,
the Narrow Line Region may contribute significantly to the mid-infrared
emission of AGN \citep[e.g.][]{2003Radomski,2007Poncelet,2008Schweitzer}.
A contribution of $80\%$ in our case nevertheless appears very high
for the following reason. The Circinus galaxy has a sharply delimited
ionisation cone, and the central engine is fully obscured along other
lines of sight. While the projected half opening angle of the visible
ionisation cone onto the plane of the sky is $\vartheta_{\mathrm{proj}}\sim45\text{\textdegree}$
\citep{2000Wilson}, the physical half opening angle has been estimated
to be $\vartheta_{\mathrm{cone}}\sim41{^\circ}$ from modelling of
the narrow line kinematics \citep{2013Fischer}. This means that more
than 70\% of the sky are covered by the equatorial obscurer. Furthermore,
the dusty material in the ionisation cone must have a low covering
factor, typically $<30\%$. Therefore, the total covering fraction
of the dust in the ionisation cone is most likely $<10\%$, which
is consistent with the estimate by e.g.\ \citet{2009Mor}. It is
then difficult to explain why this material is responsible for 80\%
of the mid-infrared emission, while all the other dust intercepting
most of the energy from the central engine radiates much less. Furthermore,
the dust emission in the ionisation cone is expected to be optically
thin and to show the silicate feature in emission. While this may
be the case in NGC~424 \citep{2012Hoenig}, this is certainly not
the case for the Circinus galaxy: the extended emission has, on average,
the highest silicate optical depth.

Deep single-dish images of the Circinus nucleus \citep{2005Packham,2010Reunanen}
already show a faint halo surrounding the nuclear point source. This
halo is extended $\sim2\,\mathrm{arcsec}$ along $\mathit{PA}\sim100\text{\textdegree}$
and $280\text{\textdegree}$. It is interpreted as emission from dense
dusty material which enters the ionisation cone preferentially from
one direction due to a nuclear gaseous bar and is entrained outward
in an outflow possibly driven by radiation pressure \citep{2000Maiolino,2005Packham}.
Our extended emission component naturally appears as a continuation
of this outer halo towards the nucleus. We therefore interpret the
extended component as enhanced emission from the southern edge of
the ionisation cone, that is from the inner funnel of the torus. The
emission from the funnel most likely has smaller scale structure,
such as bright regions along the cone edge or along filaments. We
consider such smaller scale structures to be responsible for the irregular
behaviour of the visibilities and differential phases on longer baselines
(cf.\ Sect.~\ref{subsub:phases-smallscale}).

In addition to an enhanced supply of material to one side of the ionisation
cone, a preferential illumination of the cone edge along $\mathit{PA}\sim-90\text{\textdegree}$
(and the counter cone along $\mathit{PA}\sim90\text{\textdegree}$)
may be responsible for the enhanced emission in this direction: the
best estimate for the orientation of the accretion disk comes from
the innermost disk masers (see Sect.~\ref{subsub:morphology-disk}),
suggesting a disk axis along $-60\text{\textdegree}$. This value
also agrees well with the position angle of the radio lobes, $\mathit{PA}\sim115\text{\textdegree}$
and $295\text{\textdegree}=-65\text{\textdegree}$ \citep{1998Elmouttie},
which are most probably launched in the innermost region of the accretion
disk. The flux from a thin, optically thick accretion disk depends
on the polar angle $\vartheta$ as $F\propto\cos\vartheta\left(1+2\cos\vartheta\right)$,
i.e.\ it emits anisotropically with the strongest emission in direction
of the disk axis \citep{1987Netzer}. This implies that the radiation
towards the edges of the ionisation cone at $\mathit{PA}\sim-90\text{\textdegree}$
and $+90\text{\textdegree}$ ($\vartheta\sim30\text{\textdegree}$)
are more illuminated than the opposite cone edges at $\mathit{PA}\sim0\text{\textdegree}$
and $180\text{\textdegree}$ ($\vartheta\sim60\text{\textdegree}$)
by a factor of about two$ $. The unobscured, western cone edge along
at $\mathit{PA}\sim-90\text{\textdegree}$ thus appears brighter.

\subsection{Differential phases\label{sub:differential-phases}}

\subsubsection{Characteristics of the differential phase signal\label{subsub:phases-characteristics}}

In our MIDI data, we find relatively strong differential phases with
amplitudes (peak-to-valley) of up to $\sim80\text{\textdegree}$.
The phase signal may be caused either by atmospheric phase residuals
(mainly chromatic dispersion due to water vapour) or it may be intrinsic
to the source. We are convinced that the signal is intrinsic to the
source for the following reasons: (1) In several cases (especially
for the measurements using the UTs), the phase measurements have a
high signal to noise ratio, $S/N>5$. (2) The differential phases
agree or are very similar when measured twice at different epochs
(cf.\ Sect.~\ref{sub:data-consistency}). This is further corroborated
by a flip of the differential phase for an interchange of the telescopes.
Two examples of such a flip are shown in Fig.\ \ref{fig:fig2_comparison-of-measurements}
(comparisons C4 and C6: bottom row, first and last column). (3) We
see no significant differential phase signal in any other AGN \citep[see their Fig. 63]{2013Burtscher}.
An exception is NGC~1068, which shows strong differential phases
similar to those of the Circinus galaxy \citep{2014Lopez-Gonzaga}.
It is unlikely that strong phase residuals remain after calibration
only for the two brightest AGN observed with MIDI. It is more likely
that we are able to detect their intrinsic phase signal more easily
due to their brightness. We thus conclude that the observed differential
phase signal is dominated by the intrinsic structure of the brightness
distribution.

\begin{figure}
\centering

\includegraphics[bb=170bp 284bp 419bp 532bp,clip]{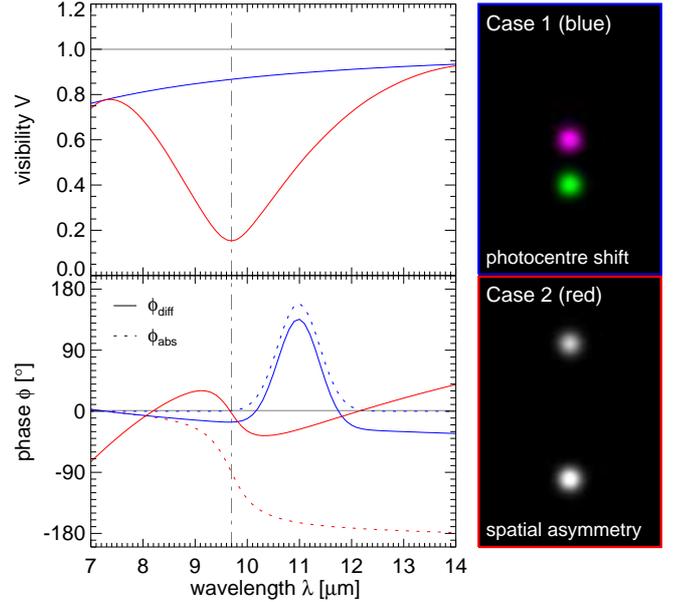}

\caption{Examples for the two ways to produce a smooth differential phase signal:
(1) a chromatic photocentre shift (blue) and (2) a purely spatial
asymmetry of the brightness distribution (red). In case 1, the source
is shifted to the bottom between $\lambda\sim10\,\mathrm{\mu m}$
and $12\,\mathrm{\mu m}$; in case 2, the ``binary'' has a brightness
ratio of $0.7:1.0$. The resulting visibilities and phases are shown
on the left, with the absolute Fourier phases $\phi_{\mathrm{abs}}$
and the differential phases $\phi_{\mathrm{diff}}$ (as would be measured
with MIDI) plotted by dotted and continuous lines, respectively. On
the right, the two brightness distributions are shown as false-colour
images, with the brightness distribution at $13.0\,\mathrm{\mu m}$,
$10.5\,\mathrm{\mu m}$ and $8.0\,\mathrm{\mu m}$ mapped to the red,
green and blue channels, respectively. \label{fig:fig08_phase-example}}
\end{figure}
\begin{figure*}
\begin{minipage}[b]{12cm}%
\centering

\includegraphics[bb=127bp 255bp 467bp 565bp]{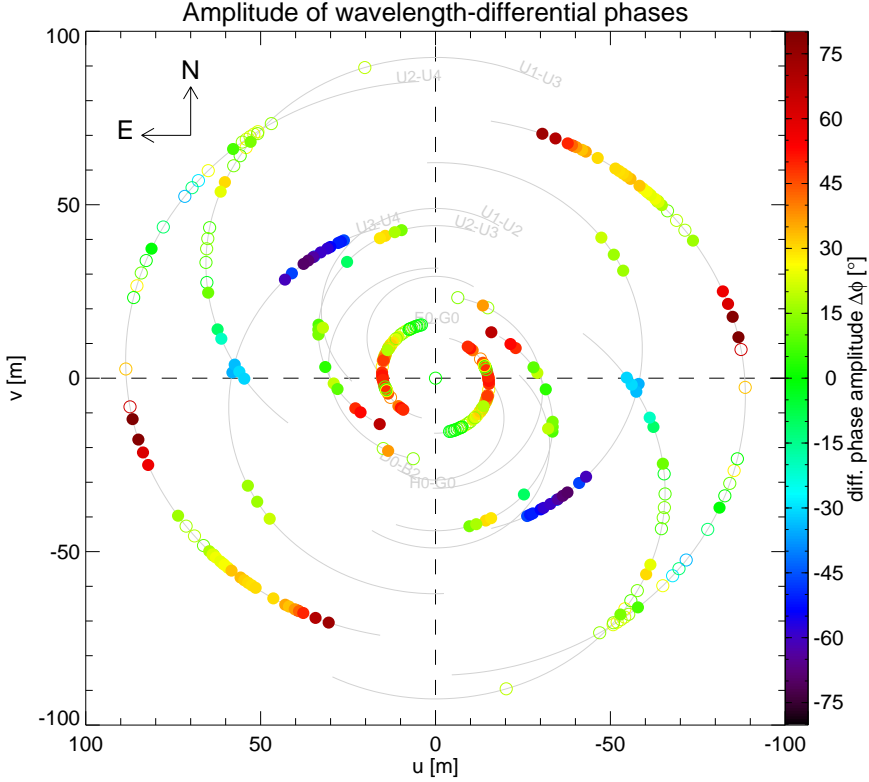}%
\end{minipage}\hfill{}%
\begin{minipage}[b]{6cm}%
\caption{Amplitude of the differential phases in the $uv$ plane: each $uv$
point is colour coded with the value of $\Delta\phi$, the difference
of the differential phase at $10\,\mathrm{\mu m}$ to those at the
edges of the $N$-band ($\lambda=8.2\,\mathrm{\mu m}$ \& $13.0\,\mathrm{\mu m}$).
Filled dots denote a phase signal with $>3\sigma$ significance averaged
over the $N$-band, empty circles a phase signal with $<3\sigma$
significance, i.e.\ a differential phase consistent with 0. \label{fig:fig09_uvplane-diffphases}}
\end{minipage}
\end{figure*}
Differential phases appear if there is an asymmetry in the brightness
distribution. This asymmetry does not necessarily have to be chromatic:
due to the different spatial frequencies probed by different wavelengths,
a phase signal is also introduced for non-chromatic, yet spatially
asymmetric sources%
\footnote{In fact, even symmetric, non-chromatic brightness distributions can
have a phase signal, as the examples of a ring or a uniform disk show.
In these cases, however, there is a sharp phase jump by exactly 180\textdegree{}
at the point where $V=0$. This has been measured, e.g.\ for the
star \object{Vega} \citep{2006Schmitt}. Because we do not see such
sharp phase jumps in the data of the Circinus galaxy, we do not further
consider these special cases here.%
}. We thus distinguish two ways to produce a differential phase signal
in our MIDI data (Fig.~\ref{fig:fig08_phase-example}, see also \citealt{2007Deroo}):
\begin{enumerate}
\item A chromatic photocentre shift of a brightness distribution that is
otherwise point-symmetric at each wavelength (plotted blue in Fig.~\ref{fig:fig08_phase-example}).
The photocentre shift has to be nonlinear in wavenumber to produce
a differential phase signal measurable with MIDI. An example of this
is a point-symmetric brightness distribution that is shifted in one
direction in an emission line. In this first case, the correlated
fluxes (and visibilities) of the source remain \textit{unchanged}
with respect to a source with no photocentre shift. 
\item A brightness distribution that is asymmetric at each wavelength but
otherwise has the same spectrum everywhere. In this case, the differential
phases are caused by phase gradients (``jumps'') of the complex
visibilities as they are probed at different spatial frequencies for
different wavelengths. An example is a binary with unequal brightness
but identical spectrum (plotted red in Fig.~\ref{fig:fig08_phase-example}).
For such a binary, the differential phases appear as a smoothed step.
The amplitude and the smoothing of the step depends on the brightness
ratio of the two components %
\footnote{In the case of an equal binary, the jump is exactly 180\textdegree{}
and sharp at the spatial frequency where the complex visibility changes
sign, similar to the case of a ring or uniform disk. %
}. For this second case, the phase jump occurs at a \textit{minimum}
of the visibilities (cf.\ top left panel in Fig.~\ref{fig:fig08_phase-example}).
This is in contrast to case one, where the visibilities remain unchanged. 
\end{enumerate}
Typically, in interferometry, a mixture of both cases occurs and it
is hard to separate the two effects. For example in a binary, the
two companions usually have different spectral types. Short of being
able to fully reconstruct the true wavelength-dependent brightness
distribution of the nuclear dust in the Circinus galaxy, we will use
the discussed properties of the differential phases to derive some
information on the asymmetries of the brightness distribution.

In general, the differential phases of the Circinus galaxy change
smoothly over the entire $N$-band, without any sharp phase jumps.
Figure~\ref{fig:fig09_uvplane-diffphases} shows the amplitude of
the differential phase signal, $\Delta\phi=\phi_{\mathrm{diff}}(10.0\,\mathrm{\mu m})-\frac{1}{2}\left(\phi_{\mathrm{diff}}(8.2\,\mathrm{\mu m})+\phi_{\mathrm{diff}}(13.0\,\mathrm{\mu m})\right)$,
in the $uv$ plane. Differential phases with more than $3\sigma$
significance (averaged over $8.2\,\mathrm{\mu m}<\lambda<13.0\,\mathrm{\mu m}$)
are plotted by filled dots. Phases which are consistent with $\Delta\phi=0$
are plotted by open circles. A negative value means the phase has
a concave ($\cup$-shaped) dependency, while positive values mean
the phase has a convex shape (i.e.\ $\cap$-shaped). To make the
plot easier to read, $\Delta\phi$ was corrected for an interchange
of the telescopes.

Clearly there are distinct areas in the $uv$ plane with strong differential
phases. Furthermore, we find equally strong differential phases on
large spatial scales (short baselines) as on small spatial scales
(long baselines). This directly implies that not only small scale
structure (``clumps'') can be made responsible for the phase signal.
These would only produce a phase signal on the longest baselines.

\subsubsection{The phase signal of the large scale structure\label{subsub:phases-largescale}}

On the shortest baseline (E0-G0), we measure a phase signal of up
to $\Delta\phi=70\text{\textdegree}$, with a more or less smooth
dependency of $\Delta\phi$ on the position angle. For $ $$\mathit{PA}\sim15\text{\textdegree}$
to $ $$30\text{\textdegree}$ the differential phase is consistent
with zero, then increases to reach its maximum between $\mathit{PA}\sim80\text{\textdegree}$
and $ $$140\text{\textdegree}$. Furthermore we find that the phase
signal is somewhat $\vee$-shaped, reminiscent of the shape of the
silicate absorption feature. This and the smooth dependency with position
angle strongly suggest that the phases are dominated by large scale
variations of the silicate absorption strength, consistent with case
1 discussed above. Indeed, the differential phases on the shortest
baselines can be very well explained by a linear gradient of the optical
depth of the silicate feature over the source. In our modelling, we
applied a linear gradient in the optical depth to the extended component
of our three-component model (see Sect.~\ref{sec:modelling}): $\tau_{3}(\alpha,\delta)=\tau_{3}+\xi_{3}\cdot\left(\delta\cdot\cos\psi_{3}+\alpha\cdot\sin\psi_{3}\right)$.
Consistent with the position angle dependence of the phase signal,
the gradient is oriented in direction of the major axis of this component.
A gradient in the optical depth effectively leads to a shift of the
emission in the direction of lower absorption (see also Fig.~\ref{fig:fig07_model-image}).
The shift is towards the north-west, and it is non-linear in wavelength
due to the specific wavelength dependency of the silicate optical
depth. In fact, an enhancement of the emission towards the north-west
was already suggested in \citet{2007Tristram2}.

From our model fit, we find a gradient of the optical depth of $\xi_{3}\sim27\,\mathrm{arcsec}^{-1}$.
The positive value means that the silicate absorption depth increases
towards the south-east, i.e. this side of the nucleus is more heavily
obscured. This is consistent with observations at larger scales, where
the south-eastern sides of the inclined galactic disk \citep{1977Freeman}
and a circumnuclear molecular ring \citep{1998Curran,1999Curran}
are located closer to us. Furthermore, the ionisation cone is one-sided
towards the north-west \citep[e.g.][]{1994Marconi,2000Wilson}, and
colour maps \citep[e.g.][]{2004Prieto} show redder colours towards
the south-east -- all implying stronger obscuration towards the south-east.
The highly inclined nuclear maser disk is the only structure with
a size directly comparable to our dust distribution. Our results would
be most consistent if the south-eastern side of the warped disk is
located closer to us, similar to the galactic scale structures and
as somewhat implied by \citet{2003Greenhill}. It appears that the
Circinus galaxy is more heavily obscured towards the south-east all
the way down to the very nucleus. The disk-like obscurers on different
spatial scales (the galactic disk, the circumnuclear ring and the
nuclear disk) all have the north-western part of their axes slightly
directed towards us.

The rather large gradient in the silicate absorption depth is necessary
to obtain a significant colour change on scales of less than $100\,\mathrm{mas}$.
Using long-slit spectroscopy in the mid-infrared, \citet{2006Roche}
mapped the strength of the silicate feature along $\mathit{PA}=10\text{\textdegree}$
and $\mathit{PA}=100\text{\textdegree}$. They found a general trend
of the optical depth along $\mathit{PA}=100\text{\textdegree}$, increasing
from $\tau_{9.7\mathrm{\mu m}}\sim1.0$ at $1.3\,\mathrm{arcsec}$
west of the nucleus to $\tau_{9.7\mathrm{\mu m}}\sim2.6$ at $1.3\,\mathrm{arcsec}$
east of the nucleus. A similar trend is not found along $\mathit{PA}=10\text{\textdegree}$.
These measurements further support that the south-eastern side of
the nucleus is more obscured. The observed large scale trend corresponds
to a gradient of $\xi_{\mathrm{Roche}}\sim0.6\,\mathrm{arcsec}^{-1}$.
This is more than one order of magnitude less than our result. Such
a low gradient would not produce a significant phase signal in the
interferometric measurements by far. Therefore, our gradient cannot
be directly connected to the larger scale obscuration on scales of
tens of parsecs. We instead interpret our strong colour gradient as
evidence of more dense dusty material on scales of less than an arcsecond
(i.e.\ less than $20\,\mathrm{pc}$). This is consistent with evidence
for much of the obscuration in the nucleus taking place on parsec
scales \citep{2004Prieto}.

\subsubsection{The phase signal for high spatial frequencies\label{subsub:phases-smallscale}}

For higher spatial frequencies (smaller scales, measured with longer
baselines) the interpretation of the phase signal is more difficult.
The amplitude and the shape of the differential phases change significantly
between different locations in the $uv$ plane. While some areas have
a strong phase signal of up to $\Delta\phi=90\text{\textdegree}$,
the measurements are consistent with no significant phase signal in
other areas. In contrast to the phases on the shortest baselines,
there is no preferential direction, and we find no general correlation
(or anti-correlation) of $\Delta\phi$ with $F_{\mathrm{cor}}$. There
seem to be ``patches'' of stronger phases in different areas of
the $uv$ plane. In particular, we see no general trend for the phase
signal to be strongest at minima in the correlated fluxes. This argues
against the phase signal being caused by purely achromatic asymmetries
(case 2, see Sect.~\ref{subsub:phases-characteristics}). We therefore
suspect colour gradients and especially changes in the optical depth
on intermediate scales to be responsible for these phases.

\begin{figure*}
\begin{minipage}[b]{12cm}%
\centering

\includegraphics[bb=127bp 296bp 467bp 552bp]{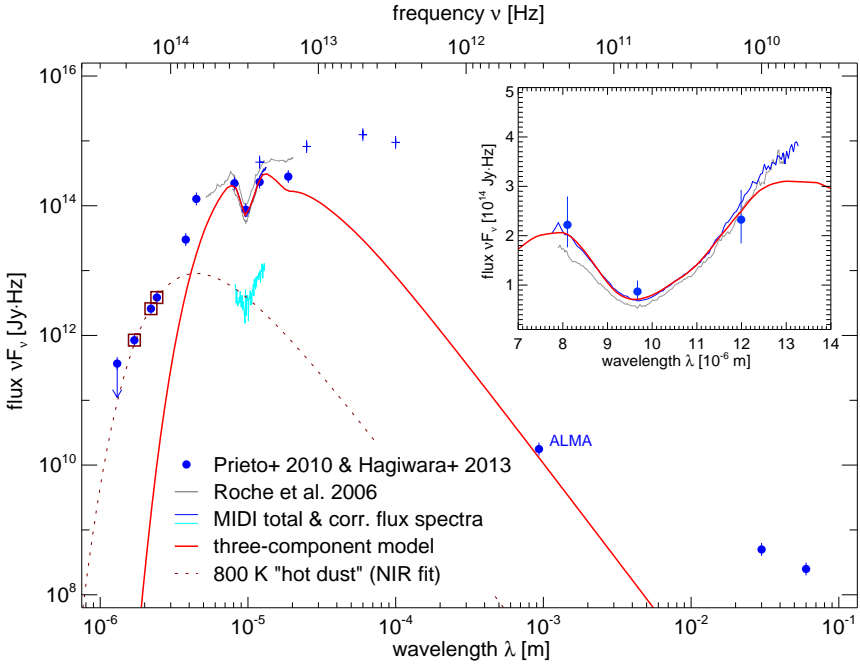}%
\end{minipage}\hfill{}%
\begin{minipage}[b]{6cm}%
\caption{Nuclear Spectral Energy Distribution (SED) of the Circinus galaxy.
Blue dots are high resolution photometry compiled by \citet{2010Prieto}
plus an additional measurement with ALMA at $ $$\lambda=9.3\cdot10^{-4}\,\mathrm{m}$
($321\,\mathrm{GHz}$) by \citet{2013Hagiwara}. The measurement in
the $J$-band, which is only an upper limit, is marked with an arrow.
Also shown are low resolution IRAS photometry (blue crosses), a MIDI
correlated flux spectrum (cyan line), the averaged MIDI total flux
spectrum (blue line) and the high resolution spectrum from \citet[grey line]{2006Roche}.
The SED of our best fitting model is plotted in red. A possible compact
hot dust component, fitted to the SED data marked by the dark red
boxes, is shown by the dotted, dark red line. A zoom on the $N$-band
region with linear scaling of the axes is shown as an inset. See text
for further details. \label{fig:fig10_sed}}
\end{minipage}
\end{figure*}
An exception are the $uv$ points at $(u,v)\approx(33,14)\,\mathrm{m}$.
These correspond to the comparison C5 (see Sect.~\ref{sub:data-consistency}).
The correlated fluxes and differential phases for this location in
the $uv$ plane are shown in the middle column of Fig.~\ref{fig:fig2_comparison-of-measurements}.
Clearly, the correlated flux (and the visibility) has a dip at $\lambda\sim12.4\,\mathrm{\mu m}$
which strongly deviates from the normal spectral shape of the Circinus
galaxy. The dip is at the same wavelength as a strong gradient and
zero-crossing of the differential phase, suggestive of case 2 discussed
in Sect.~\ref{subsub:phases-characteristics}. This means a discrete
asymmetry is probed by these measurements (and only by these measurements!).
A closer inspection of the data reveals that the wavelength at which
the minimum occurs varies slightly from $\lambda=12.3\,\mathrm{\mu m}$
to $\lambda=12.5\,\mathrm{\mu m}$ for the individual measurements.
This is roughly followed by corresponding changes of the zero-crossing
of the differential phases, as we would expect for a phase jump at
a visibility minimum. Assuming an unequal binary, we can directly
calculate the projected separation $d_{\mathrm{proj}}$ of the two
components: as can be derived easily, the separation is given by $d_{\mathrm{proj}}=\lambda_{0}/\mathit{BL}_{0}\cdot(2n+1)/2$,
where $\lambda_{0}$ and $\mathit{BL}_{0}$ are the wavelength and
projected baseline length at which the minimum in the visibilities
and the phase jump occur, and $n\in\mathbb{N}=\left[0,1,2,3,\ldots\right]$.
In our case, we obtain $d_{\mathrm{proj}}=(2n+1)\cdot35.5\,\mathrm{mas}$.
Note that only $n=0$ appears likely, because separations of more
than $100\,\mathrm{mas}$ are unlikely considering the overall size
of the emission. Also, for higher values of $n$, we would quickly
start observing more than one step in the phases within $ $$8.0\,\mathrm{\mu m}<\lambda<13.0\,\mathrm{\mu m}$.
In any case, the separation is intermediate between the sizes of the
inner disk-like component and the extended dust component. It is certainly
too large to be attributed to individual clumps of the torus. It instead
indicates intermediate scale asymmetries in the brightness distribution,
as could be caused by warps in the inner disk-like component or enhanced
emission regions along the inner edge of the dust distribution (cf.\ Sect.~\ref{subsub:morphology-extended}).

We conclude that the strong differential phases observed on longer
baselines provide evidence of intermediate scale asymmetries in the
dust emission, both chromatic and achromatic, caused by the specific
illumination and geometry of the dust distribution. Interferometric
observations enabling image reconstruction on the one hand and more
sophisticated modelling on the other hand will be required to reveal
the details of the underlying brightness distribution.

\subsection{Spectral Energy Distribution\label{sub:spectral-energy-distribution}}

To be meaningful, our model has to not only reproduce the interferometric
data in the $N$-band, but also be consistent with the entire spectral
energy distribution (SED) of the Circinus galaxy. The SED of the Circinus
galaxy from \citet{2010Prieto} is shown in Fig.~\ref{fig:fig10_sed}.
With an angular resolution of $<1.5\,\mathrm{arcsec}$, these measurements
represent the true nuclear SED. We add to this SED a new measurement
of the continuum emission at $ $$\lambda=9.3\cdot10^{-4}\,\mathrm{m}$
($321\,\mathrm{GHz}$) obtained by \citet{2013Hagiwara} with ALMA:
$F_{\mathrm{ALMA}}=55\,\mathrm{mJy}$. Low resolution measurements
with IRAS \citep{1990Moshir} are plotted as blue crosses for reference.
The error bars correspond to an assumed uncertainty of $0.1\,\mathrm{dex}$
for all photometric data points (e.g.\ due to variability). A MIDI
correlated flux spectrum (\#41, $\mathit{BL}=87\,\mathrm{m}$, selected
for its low correlated flux and good $S/N$) is plotted in cyan, and
the averaged total flux spectrum from MIDI is plotted in blue. The
SED of our model is shown in red. For comparison, we also reproduce
the high spatial resolution spectrum of the inner $0.81\,\mathrm{arcsec}$
from \citet{2006Roche} in grey.

Not surprisingly, the SED of our model closely resembles the Planck
curve for a black body of $T\sim300\,\mathrm{K}$, with some modifications
due to absorption, most notably the silicate features at $\lambda=9.7\,\mathrm{\mu m}$
and $20\,\mathrm{\mu m}$. Although we only fitted our model to the
data obtained with MIDI, it is consistent with the entire high resolution
SED: it represents the flux in the mid-infrared well while producing
less or equal flux at other wavelength ranges.

At the long wavelength end of the N-band (see inset of Fig.~\ref{fig:fig10_sed})
our model under-predicts the total flux spectrum for $\lambda\gtrsim12\,\mathrm{\mu m}$.
We interpret this as evidence of additional, cooler dust, which is
responsible for this emission and which is not fully captured by our
model. On the other hand, the measurement with ALMA sets a strong
constraint on the emission from cooler dust in the innermost $1.3\,\mathrm{arcsec}$
($26\,\mathrm{pc}$) of the Circinus galaxy. While the SED of our
model is consistent with the ALMA measurement, it rules out any major
contribution to the SED from additional dust in the nucleus with $T<200\,\mathrm{K}$.
This is consistent with the results of radiative transfer calculations
of dusty tori, which in general show a steep decrease in the torus
SED towards longer wavelengths. 

Towards shorter wavelengths, the model SED rapidly decreases following
the (absorbed) Wien tail of the black-body emission. The observed
SED in the near-infrared is much higher. The observed fluxes in the
$J$, $\mathit{H}$ and $K$-bands are compatible with an additional
compact component of hot dust ($T\sim800\,\mathrm{K}$, dotted curve
in Fig.~\ref{fig:fig10_sed}). Such a component would not contribute
significantly to the emission in the mid-infrared. It would still
be (marginally) consistent with the lowest correlated fluxes measured
with MIDI, which can be considered as upper limits for any contribution
by a compact hot emission component (e.g. the cyan curve in Fig.~\ref{fig:fig10_sed}).
However, our interferometric measurements rule out a compact emitter
that produces sufficient flux to explain the photometric measurements
in the $L$ and $M$-band. We tested this by fixing the temperature
of the unresolved component to $T_{1}=600\,\mathrm{K}$ and adjusting
$f_{1}$ so that the near-infrared part of the SED is reasonably well
reproduced. We then obtain much higher correlated fluxes than observed
at the short wavelength end, unless $\Delta_{1}\gtrsim15\,\mathrm{mas}$.
This means that most of the emission at these wavelengths must be
extended by at least $0.3\,\mathrm{pc}$ and less than $3.8\,\mathrm{pc}$
($190\,\mathrm{mas}$, unresolved by single dish observations). This
is discussed further in Sect.~\ref{sub:hot-dust}.

In conclusion, we find that our model is consistent with the overall
SED of the Circinus galaxy. We further conclude that the nuclear SED
is dominated by emission from dust close to $T\sim300\,\mathrm{K}$
and that the contribution from dust at much hotter ($T>800\,\mathrm{K}$)
or cooler ($T<200\,\mathrm{K}$) temperatures is very low on parsec
scales.

\subsection{Lack of a compact hot dust component\label{sub:hot-dust}}

In \citet{2007Tristram2}, no significant temperature gradient was
found in the dust distribution of the Circinus galaxy. One of the
main goals of the new observations was to search for hotter dust using
longer baselines, or generally for an increase in the dust temperature
towards the centre, similar to what is observed in NGC~1068 \citep{2004Jaffe1,2008Raban,2014Lopez-Gonzaga}.
NGC~1068 is similarly well resolved (both in terms of visibility
and scales of the supposed sublimation radius) as the Circinus galaxy.
In NGC~1068, the correlated fluxes are on average bluer than the
total flux ($ $$F_{\mathrm{cor}}(13\,\mathrm{\mu m})/F_{\mathrm{cor}}(8\,\mathrm{\mu m})<F_{\mathrm{tot}}(13\,\mathrm{\mu m})/F_{\mathrm{tot}}(8\,\mathrm{\mu m})$).
This clearly indicates an increase in the dust temperature towards
the centre. In the Circinus galaxy, however, the emission at the short
wavelength end of the $N$-band is almost completely resolved out
even on the longest baselines, and we see no significant difference
between the temperatures of our model components (for all three components,
$T\sim300\,\mathrm{K}$). The new measurements thus confirm the earlier
results: we find no evidence of a temperature increase towards the
centre. This is unexpected, because the resolution limit of our interferometric
observations,  $\sim0.1\,\mathrm{pc}$, corresponds to a few times
the sublimation radius ($r_{\mathrm{sub}}\sim0.03\,\mathrm{pc}$).
We should start seeing hotter dust on these scales.

There are two possible explanations for the lack of hot dust at the
centre: either we (still) cannot see it, or it is truly absent. In
the first case, the hot dust component could be so highly obscured
on scales of a few $r_{\mathrm{sub}}$ that we simply do not see it.
However, this is somewhat in contradiction with the decrease in the
silicate absorption depth towards the centre unless the innermost
dust contains less silicates. Full radiative transfer calculations
will have to be carried out to verify if a hot dust component can
be fully obscured while reproducing the other characteristics of Circinus
nucleus. In the second, more speculative case, there might simply
be no hot dust at the inner rim of the torus at all. In Sect.~\ref{sub:consistency-totalflux},
we discussed a possible variability in the Circinus galaxy. It is
therefore conceivable that the active nucleus was previously in a
more active state or even that an explosive event occurred some time
ago \citep{1997Bland-Hawthorn}. The increased activity would lead
to an increased sublimation radius. If the inner rim were to contract
only slowly after the activity decreased, the innermost dust would
be at much cooler temperatures for some time. Such a large and relatively
cool inner rim would not only explain the lack of hot dust but also
the highly elongated disk-like component (cf. Sect.~\ref{subsub:morphology-disk}).
On the other hand, this would require very long timescales for dust
formation, preventing the inner rim from quickly moving back inwards.
The amplitude of variability would have to be three to four orders
of magnitude to result in a significantly cooler inner rim. Moreover,
the near-infrared excess of the SED with respect to our model indicates
that dust at $T>500\,\mathrm{K}$ must exist (see Sect.~\ref{sub:spectral-energy-distribution}).

The explanations are not very satisfactory. Furthermore, the problem
is not the lack of a hot dust component alone, but rather the absence
of any evidence of an increase in the dust temperature towards the
centre. The interferometric measurements roughly probe spatial scales
from $20$ to $200\,\mathrm{mas}$ ($0.4$ to $4.0\,\mathrm{pc}$).
If we include our single dish measurements or the spectra observed
by \citet{2006Roche}, the spatial scales explored are up to $500\,\mathrm{mas}$
($\sim10\,\mathrm{pc}$) or $2\,\mathrm{arcsec}$ ($\sim40\,\mathrm{pc}$),
respectively. The temperature of directly illuminated dust%
\footnote{The dust directly illuminated by the central engine generally dominates
the emission, see discussion in Sect.~\ref{sub:morphology}.%
} depends on the distance as $T\propto(L/r^{2})^{1/5.6}\propto r^{-1/2.8}$
\citep{1987Barvainis}. Consequently, we would expect a temperature
difference by a factor of $\left(25/170\right)^{-1/2.8}\approx2$
in the range of the interferometric measurements, or a temperature
change by a factor of more than 3 when including the single dish data.
This is clearly not observed. This lack of a temperature gradient
poses a challenge for the picture of a centrally heated dust distribution
in the nucleus of the Circinus galaxy.

\subsection{What is the ``dusty torus'', and where is it?\label{sub:obscuration}}

A central question is, of course, which of the observed components
can be associated with the dusty torus in unified schemes of AGN?
Here we consider the ``torus'' defined by its role in classical
unification schemes: it obscures the central engine for certain (edge-on)
viewing angles while at the same time collimating the ionising radiation
and outflows. Additionally, the dust in the inner funnel of the torus
scatters emission from the central engine towards an observer. The
Circinus galaxy certainly shows all these characteristics of a dusty
torus (see Sect.~\ref{sec:introduction}).

One explanation for the obscuration and collimation could be a constant
scale height of the inner disk, or even a puffed-up inner rim within
a few $r_{\mathrm{sub}}$, similar to what has been suggested for
disks around young stellar objects \citep{2001Natta,2001Dullemond}.
The latter scenario has recently also been proposed for AGN \citep{2012Hoenig}.
However, if the obscuration and collimation already take place within
a few $r_{\mathrm{sub}}$ (i.e. within $\sim0.1\,\mathrm{pc}$ for
the Circinus galaxy), then the material in the disk at larger radii
should be shadowed and significantly colder than $300\,\mathrm{K}$.
This contradicts the parsec-sized, equatorially extended emission
of our disk-like component as well as the requirements for the pumping
of the disk masers, which require a direct line of sight to the central
engine \citep{2003Greenhill,2005Lo}. Also, such a geometry is hard
to reconcile with the outflow ($120\text{\textdegree}$ opening angle)
traced by water masers on sub-parsec scales, which implies that the
material is still not fully collimated on these spatial scales.

We therefore argue that most of the collimation (and with it obscuration)
most likely take place on scales of $\sim1\,\mathrm{pc}$, surrounding
the disk. This is consistent with the disk masers and the inner wide
angle outflow not extending beyond this scale and the large scale
structures having been collimated within a few parsec. In this picture,
a significant part of the polar elongated emission originates from
the inner funnel of the obscuring material, which is equivalent to
the outer edge of the ionisation cone (see Sect.~\ref{subsub:morphology-extended}).
Most of the obscuration and collimation is not due to the disk, but
due to the material located above or below it. We therefore consider
that it is this material which is commonly associated with the ``obscuring
torus'' in AGN.

\section{Summary and conclusions\label{sec:conclusions}}

We have obtained an extensive new interferometric dataset in the mid-infrared
in order to study the nuclear dust distribution in the Circinus galaxy.
Due to our optimised observing strategy we could increase the $uv$
coverage from 21 $uv$ points to 152. This allows us to more precisely
determine the properties of the dust distribution in the Circinus
galaxy. Using both the correlated fluxes and the differential phases
measured by the interferometer, we come to the following conclusions:
\begin{itemize}
\item We confirm the two-component structure previously found in the dust
distribution of the Circinus galaxy, consisting of an inner dense
disk-like component and an an extended emission region. We interpret
this as emission from warm dust in the context of an inner dense disk
surrounded by a less dense, geometrically thick and filamentary dust
distribution.
\item The disk-like component is highly elongated with a size (FWHM) of
$\sim0.2\,\mathrm{pc}\times1.1\,\mathrm{pc}$. Its major axis is along
$\mathit{PA}\sim46\text{\textdegree}$, which is in perfect agreement
to the orientation of the nuclear maser disk and oriented perpendicular
to the ionisation cone and outflow along $\mathit{PA}\sim-44\text{\textdegree}$.
Because a flat, dense disk should appear in absorption rather than
emission when seen close to edge-on, we interpret this component either
as emission from material associated with the inner funnel of the
torus directly above or below the disk or from the directly illuminated
portion of a warped disk slightly oriented towards us.
\item The extended dust emission is responsible for $80\%$ of the mid-infrared
emission. It has a FWHM of $\sim0.8\,\mathrm{pc}\times1.9\,\mathrm{pc}$
and is elongated along $\mathit{PA}\sim107\text{\textdegree}$, that
is, roughly in polar direction. We see this component as the inner
part of the extended emission already seen by single dish imaging.
It is interpreted as the emission from the inner funnel of a more
extended dust distribution and especially as emission from the funnel
edge along $\mathit{PA}\sim-90\text{\textdegree}$. Dense dusty material
enters the ionisation cone primarily on this side of the funnel, which
is also preferentially illuminated by the inclined accretion disk.
\item We detect significant differential phases, indicating both chromatic
and spatial asymmetries in the dust emission. On scales of $\sim1.0\,\mathrm{pc}$,
the differential phases are well explained by a strong increase in
the silicate absorption depth towards the south-east. This strong
gradient in the silicate absorption implies that a significant amount
of obscuration takes place on parsec scales. It also shows that the
galaxy is more obscured towards the south-east all the way to the
nucleus. The more complex differential phase signal on longer baselines
is most likely caused by intermediate scale asymmetries in the brightness
distribution, both chromatic and achromatic. These are probably not
individual clumps of a clumpy torus, but rather intermediate scale
structures such as the warp of the inner disk, the sharp cone edges
or large filamentary structures which are directly illuminated by
the central engine.
\item We find no indications for a temperature difference in the dust distribution
between $\sim0.1\,\mathrm{pc}$ to $\sim40\,\mathrm{pc}$. On all
of these scales, the mid-infrared spectrum has roughly the same slope,
indicative of dust at $T\sim300\,\mathrm{K}$. Because the SED shows
excess emission in the near-infrared with respect to emission by $300\,\mathrm{K}$
dust, we argue that this emission must also be extended on scales
$>0.3\,\mathrm{pc}$. Otherwise, the Rayleigh-Jeans part of the compact
hot dust component would appear as an unresolved source in our interferometric
data in the mid-infrared. This is in contrast to the results for other
well studied AGN such as NGC~1068 and NGC~424, where a clear temperature
increase towards the centre is observed. This result also has a consequence
for MATISSE \citep{2008Lopez}, the successor to MIDI at the VLTI:
we predict relatively low visibilities also in the $L$ and $M$-band
for the Circinus galaxy.
\item We further argue that the dense inner disk is not sufficient to provide
the obscuration and collimation observed on larger scales, that is,
the function normally attributed to the dusty torus in unified schemes
of AGN. These must be caused by additional material above and below
the disk on scales of $\sim1\,\mathrm{pc}$. The emission from the
polar component is substantially from the inner funnel of this dust
distribution.
\end{itemize}
Our interferometric observations show a quite complex picture of the
dust emission in the nucleus of the Circinus galaxy. To fully and
unambiguously reveal the underlying brightness distribution, interferometric
observations enabling image reconstruction will be necessary. Because
MIDI is only a two beam interferometer, it measures no closure phases
and a reliable image reconstruction is therefore not readily possible.
This will, however, soon be possible with the upcoming second generation
mid-infrared beam combiner for the VLTI, MATISSE, which will combine
four telescopes at a time and thus provide three closure phases.

The presence of a bright disk-like component, polar elongated dust
emission and the lack of a temperature difference are not expected
for typical models of the centrally heated dust distributions of AGN.
New sets of detailed radiative transfer calculations will be required
to explain our observations and to better understand the three-dimensional
dust morphology in the nuclei of active galaxies.
\begin{acknowledgements}
We thank the ESO staff operating the VLT(I) and the night astronomers
on duty during our observations for their support. Without their assistance
and help, it would not have been possible to collect this unique data
set. We also thank D. Asmus and A. Kreplin for helpful discussions
and their suggestions to improve the manuscript, as well as the anonymous
referee for the detailed comments strengthening the manuscript. We
are very grateful to K. Zelenevskiy for proof-reading the manuscript.

\end{acknowledgements}
\bibliographystyle{aamod}
\bibliography{paper}

\begin{thebibliography}{105}
\expandafter\ifx\csname natexlab\endcsname\relax\def\natexlab#1{#1}\fi

\bibitem[{{Alexander} {et~al.}(2000){Alexander}, {Heisler}, {Young}, {Lumsden},
  {Hough}, \& {Bailey}}]{2000Alexander}
{Alexander}, D.~M., {Heisler}, C.~A., {Young}, S., {et~al.} 2000,
  \begingroup\hypersetup{urlcolor=[rgb]{0.9,0.0,0.9}}\href{http://dx.doi.org/10.1046/j.1365-8711.2000.03285.x}{\mnras}\endgroup,
   \href{http://cdsads.u-strasbg.fr/abs/2000MNRAS.313..815A}{313, 815}

\bibitem[{{Antonucci}(1993)}]{1993Antonucci}
{Antonucci}, R. 1993,
  \begingroup\hypersetup{urlcolor=[rgb]{0.9,0.0,0.9}}\href{http://dx.doi.org/10.1146/annurev.aa.31.090193.002353}{\araa}\endgroup,
   \href{http://cdsads.u-strasbg.fr/abs/1993ARA\%26A..31..473A}{31, 473}

\bibitem[{{Antonucci} \& {Miller}(1985)}]{1985Antonucci}
{Antonucci}, R.~R.~J. \& {Miller}, J.~S. 1985,
  \begingroup\hypersetup{urlcolor=[rgb]{0.9,0.0,0.9}}\href{http://dx.doi.org/10.1086/163559}{\apj}\endgroup,
   \href{http://cdsads.u-strasbg.fr/abs/1985ApJ...297..621A}{297, 621}

\bibitem[{{Barvainis}(1987)}]{1987Barvainis}
{Barvainis}, R. 1987,
  \begingroup\hypersetup{urlcolor=[rgb]{0.9,0.0,0.9}}\href{http://dx.doi.org/10.1086/165571}{\apj}\endgroup,
   \href{http://cdsads.u-strasbg.fr/abs/1987ApJ...320..537B}{320, 537}

\bibitem[{{Beckert} {et~al.}(2008){Beckert}, {Driebe}, {H{\"o}nig}, \&
  {Weigelt}}]{2008Beckert}
{Beckert}, T., {Driebe}, T., {H{\"o}nig}, S.~F., \& {Weigelt}, G. 2008,
  \begingroup\hypersetup{urlcolor=[rgb]{0.9,0.0,0.9}}\href{http://dx.doi.org/10.1051/0004-6361:20078881}{\aap}\endgroup,
   \href{http://cdsads.u-strasbg.fr/abs/2008A%26A...486L..17B}{486, L17}

\bibitem[{{Bianchi} {et~al.}(2012){Bianchi}, {Maiolino}, \&
  {Risaliti}}]{2012Bianchi}
{Bianchi}, S., {Maiolino}, R., \& {Risaliti}, G. 2012,
  \begingroup\hypersetup{urlcolor=[rgb]{0.9,0.0,0.9}}\href{http://dx.doi.org/10.1155/2012/782030}{Advances
  in Astronomy}\endgroup,
  \href{http://cdsads.u-strasbg.fr/abs/2012AdAst2012E..17B}{2012}

\bibitem[{{Bland-Hawthorn} {et~al.}(1997){Bland-Hawthorn}, {Gallimore},
  {Tacconi}, {Brinks}, {Baum}, {Antonucci}, \& {Cecil}}]{1997Bland-Hawthorn}
{Bland-Hawthorn}, J., {Gallimore}, J.~F., {Tacconi}, L.~J., {et~al.} 1997,
  \begingroup\hypersetup{urlcolor=[rgb]{0.9,0.0,0.9}}\href{http://dx.doi.org/10.1023/A:1000567831370}{\apss}\endgroup,
   \href{http://cdsads.u-strasbg.fr/abs/1997Ap%26SS.248....9B}{248, 9}

\bibitem[{{Burtscher}(2011)}]{2011Burtscher}
{Burtscher}, L. 2011, PhD thesis, Max-Planck-Institut f{\"u}r Astronomie,
  K{\"o}nigstuhl 17, 69117 Heidelberg, Germany

\bibitem[{{Burtscher} {et~al.}(2009){Burtscher}, {Jaffe}, {Raban},
  {Meisenheimer}, {Tristram}, \& {R{\"o}ttgering}}]{2009Burtscher}
{Burtscher}, L., {Jaffe}, W., {Raban}, D., {et~al.} 2009,
  \begingroup\hypersetup{urlcolor=[rgb]{0.9,0.0,0.9}}\href{http://dx.doi.org/10.1088/0004-637X/705/1/L53}{\apjl}\endgroup,
   \href{http://cdsads.u-strasbg.fr/abs/2009ApJ...705L..53B}{705, L53}

\bibitem[{{Burtscher} {et~al.}(2010){Burtscher}, {Meisenheimer}, {Jaffe},
  {Tristram}, \& {R{\"o}ttgering}}]{2010Burtscher}
{Burtscher}, L., {Meisenheimer}, K., {Jaffe}, W., {Tristram}, K.~R.~W., \&
  {R{\"o}ttgering}, H.~J.~A. 2010,
  \begingroup\hypersetup{urlcolor=[rgb]{0.9,0.0,0.9}}\href{http://dx.doi.org/10.1071/AS09068}{\pasa}\endgroup,
   \href{http://cdsads.u-strasbg.fr/abs/2010PASA...27..490B}{27, 490}

\bibitem[{{Burtscher} {et~al.}(2013){Burtscher}, {Meisenheimer}, {Tristram},
  {Jaffe}, {H{\"o}nig}, {Davies}, {Kishimoto}, {Pott}, {R{\"o}ttgering},
  {Schartmann}, {Weigelt}, \& {Wolf}}]{2013Burtscher}
{Burtscher}, L., {Meisenheimer}, K., {Tristram}, K.~R.~W., {et~al.} 2013,
  \begingroup\hypersetup{urlcolor=[rgb]{0.9,0.0,0.9}}\href{http://dx.doi.org/10.1051/0004-6361/201321890}{\aap}\endgroup,
   \href{http://cdsads.u-strasbg.fr/abs/2013A%26A...558A.149B}{558, A149}

\bibitem[{{Burtscher} {et~al.}(2012){Burtscher}, {Tristram}, {Jaffe}, \&
  {Meisenheimer}}]{2012Burtscher}
{Burtscher}, L., {Tristram}, K.~R.~W., {Jaffe}, W.~J., \& {Meisenheimer}, K.
  2012, in Society of Photo-Optical Instrumentation Engineers (SPIE) Conference
  Series,\begingroup\hypersetup{urlcolor=[rgb]{0.9,0.0,0.9}}\href{http://dx.doi.org/10.1117/12.925521}{
  Vol. 8445}\endgroup,
  \href{http://cdsads.u-strasbg.fr/abs/2012SPIE.8445E..1GB}{84451G}

\bibitem[{{Curran} {et~al.}(1998){Curran}, {Johansson}, {Rydbeck}, \&
  {Booth}}]{1998Curran}
{Curran}, S.~J., {Johansson}, L.~E.~B., {Rydbeck}, G., \& {Booth}, R.~S. 1998,
  \aap,  \href{http://cdsads.u-strasbg.fr/abs/1998A%26A...338..863C}{338, 863}

\bibitem[{{Curran} {et~al.}(1999){Curran}, {Rydbeck}, {Johansson}, \&
  {Booth}}]{1999Curran}
{Curran}, S.~J., {Rydbeck}, G., {Johansson}, L.~E.~B., \& {Booth}, R.~S. 1999,
  \aap,  \href{http://cdsads.u-strasbg.fr/abs/1999A%26A...344..767C}{344, 767}

\bibitem[{{Deroo} {et~al.}(2007){Deroo}, {van Winckel}, {Verhoelst}, {Min},
  {Reyniers}, \& {Waters}}]{2007Deroo}
{Deroo}, P., {van Winckel}, H., {Verhoelst}, T., {et~al.} 2007,
  \begingroup\hypersetup{urlcolor=[rgb]{0.9,0.0,0.9}}\href{http://dx.doi.org/10.1051/0004-6361:20066516}{\aap}\endgroup,
   \href{http://cdsads.u-strasbg.fr/abs/2007A%26A...467.1093D}{467, 1093}

\bibitem[{{Dorodnitsyn} {et~al.}(2008){Dorodnitsyn}, {Kallman}, \&
  {Proga}}]{2008Dorodnitsyn}
{Dorodnitsyn}, A., {Kallman}, T., \& {Proga}, D. 2008,
  \begingroup\hypersetup{urlcolor=[rgb]{0.9,0.0,0.9}}\href{http://dx.doi.org/10.1086/529374}{\apjl}\endgroup,
   \href{http://cdsads.u-strasbg.fr/abs/2008ApJ...675L...5D}{675, L5}

\bibitem[{{Dullemond} {et~al.}(2001){Dullemond}, {Dominik}, \&
  {Natta}}]{2001Dullemond}
{Dullemond}, C.~P., {Dominik}, C., \& {Natta}, A. 2001,
  \begingroup\hypersetup{urlcolor=[rgb]{0.9,0.0,0.9}}\href{http://dx.doi.org/10.1086/323057}{\apj}\endgroup,
   \href{http://cdsads.u-strasbg.fr/abs/2001ApJ...560..957D}{560, 957}

\bibitem[{{Efron}(1979)}]{1979Efron}
{Efron}, B. 1979,
  \begingroup\hypersetup{urlcolor=[rgb]{0.9,0.0,0.9}}\href{http://dx.doi.org/10.1214/aos/1176344552}{The
  Annals of Statistics}\endgroup, 7, 1

\bibitem[{{Elmouttie} {et~al.}(1998){Elmouttie}, {Haynes}, {Jones}, {Sadler},
  \& {Ehle}}]{1998Elmouttie}
{Elmouttie}, M., {Haynes}, R.~F., {Jones}, K.~L., {Sadler}, E.~M., \& {Ehle},
  M. 1998,
  \begingroup\hypersetup{urlcolor=[rgb]{0.9,0.0,0.9}}\href{http://dx.doi.org/10.1046/j.1365-8711.1998.01592.x}{\mnras}\endgroup,
   \href{http://cdsads.u-strasbg.fr/abs/1998MNRAS.297.1202E}{297, 1202}

\bibitem[{{Feltre} {et~al.}(2012){Feltre}, {Hatziminaoglou}, {Fritz}, \&
  {Franceschini}}]{2012Feltre}
{Feltre}, A., {Hatziminaoglou}, E., {Fritz}, J., \& {Franceschini}, A. 2012,
  \begingroup\hypersetup{urlcolor=[rgb]{0.9,0.0,0.9}}\href{http://dx.doi.org/10.1111/j.1365-2966.2012.21695.x}{\mnras}\endgroup,
   \href{http://cdsads.u-strasbg.fr/abs/2012MNRAS.426..120F}{426, 120}

\bibitem[{{Fischer} {et~al.}(2013){Fischer}, {Crenshaw}, {Kraemer}, \&
  {Schmitt}}]{2013Fischer}
{Fischer}, T.~C., {Crenshaw}, D.~M., {Kraemer}, S.~B., \& {Schmitt}, H.~R.
  2013,
  \begingroup\hypersetup{urlcolor=[rgb]{0.9,0.0,0.9}}\href{http://dx.doi.org/10.1088/0067-0049/209/1/1}{\apjs}\endgroup,
   \href{http://cdsads.u-strasbg.fr/abs/2013ApJS..209....1F}{209, 1}

\bibitem[{{Freeman} {et~al.}(1977){Freeman}, {Karlsson}, {Lynga}, {Burrell},
  {van Woerden}, {Goss}, \& {Mebold}}]{1977Freeman}
{Freeman}, K.~C., {Karlsson}, B., {Lynga}, G., {et~al.} 1977, \aap,
  \href{http://cdsads.u-strasbg.fr/abs/1977A%26A....55..445F}{55, 445}

\bibitem[{{Galliano} {et~al.}(2008){Galliano}, {Madden}, {Tielens}, {Peeters},
  \& {Jones}}]{2008Galliano}
{Galliano}, F., {Madden}, S.~C., {Tielens}, A.~G.~G.~M., {Peeters}, E., \&
  {Jones}, A.~P. 2008,
  \begingroup\hypersetup{urlcolor=[rgb]{0.9,0.0,0.9}}\href{http://dx.doi.org/10.1086/587051}{\apj}\endgroup,
   \href{http://cdsads.u-strasbg.fr/abs/2008ApJ...679..310G}{679, 310}

\bibitem[{{Gitton} {et~al.}(2004){Gitton}, {Leveque}, {Avila}, \& {Phan
  Duc}}]{2004Gitton}
{Gitton}, P.~B., {Leveque}, S.~A., {Avila}, G., \& {Phan Duc}, T. 2004, in New
  Frontiers in Stellar Interferometry, Proceedings of SPIE, ed. W.~A. {Traub},
  Vol. 5491,  \href{http://cdsads.u-strasbg.fr/abs/2004SPIE.5491..944G}{944}

\bibitem[{{Granato} \& {Danese}(1994)}]{1994Granato}
{Granato}, G.~L. \& {Danese}, L. 1994, \mnras,
  \href{http://cdsads.u-strasbg.fr/abs/1994MNRAS.268..235G}{268, 235}

\bibitem[{{Greenhill} {et~al.}(2003){Greenhill}, {Booth}, {Ellingsen},
  {Herrnstein}, {Jauncey}, {McCulloch}, {Moran}, {Norris}, {Reynolds}, \&
  {Tzioumis}}]{2003Greenhill}
{Greenhill}, L.~J., {Booth}, R.~S., {Ellingsen}, S.~P., {et~al.} 2003,
  \begingroup\hypersetup{urlcolor=[rgb]{0.9,0.0,0.9}}\href{http://dx.doi.org/10.1086/374862}{\apj}\endgroup,
   \href{http://cdsads.u-strasbg.fr/abs/2003ApJ...590..162G}{590, 162}

\bibitem[{{Hagiwara} {et~al.}(2013){Hagiwara}, {Miyoshi}, {Doi}, \&
  {Horiuchi}}]{2013Hagiwara}
{Hagiwara}, Y., {Miyoshi}, M., {Doi}, A., \& {Horiuchi}, S. 2013,
  \begingroup\hypersetup{urlcolor=[rgb]{0.9,0.0,0.9}}\href{http://dx.doi.org/10.1088/2041-8205/768/2/L38}{\apjl}\endgroup,
   \href{http://cdsads.u-strasbg.fr/abs/2013ApJ...768L..38H}{768, L38}

\bibitem[{Hall(1985)}]{1985Hall}
Hall, P. 1985,
  \begingroup\hypersetup{urlcolor=[rgb]{0.9,0.0,0.9}}\href{http://dx.doi.org/http://dx.doi.org/10.1016/0304-4149(85)90212-1}{Stochastic
  Processes and their Applications}\endgroup, 20, 231

\bibitem[{{H{\"o}nig} \& {Kishimoto}(2010)}]{2010Hoenig2}
{H{\"o}nig}, S.~F. \& {Kishimoto}, M. 2010,
  \begingroup\hypersetup{urlcolor=[rgb]{0.9,0.0,0.9}}\href{http://dx.doi.org/10.1051/0004-6361/200912676}{\aap}\endgroup,
   \href{http://cdsads.u-strasbg.fr/abs/2010A%26A...523A..27H}{523, A27}

\bibitem[{{H{\"o}nig} {et~al.}(2012){H{\"o}nig}, {Kishimoto}, {Antonucci},
  {Marconi}, {Prieto}, {Tristram}, \& {Weigelt}}]{2012Hoenig}
{H{\"o}nig}, S.~F., {Kishimoto}, M., {Antonucci}, R., {et~al.} 2012,
  \begingroup\hypersetup{urlcolor=[rgb]{0.9,0.0,0.9}}\href{http://dx.doi.org/10.1088/0004-637X/755/2/149}{\apj}\endgroup,
   \href{http://cdsads.u-strasbg.fr/abs/2012ApJ...755..149H}{755, 149}

\bibitem[{{H{\"o}nig} {et~al.}(2013){H{\"o}nig}, {Kishimoto}, {Tristram},
  {Prieto}, {Gandhi}, {Asmus}, {Antonucci}, {Burtscher}, {Duschl}, \&
  {Weigelt}}]{2013Hoenig}
{H{\"o}nig}, S.~F., {Kishimoto}, M., {Tristram}, K.~R.~W., {et~al.} 2013,
  \begingroup\hypersetup{urlcolor=[rgb]{0.9,0.0,0.9}}\href{http://dx.doi.org/10.1088/0004-637X/771/2/87}{\apj}\endgroup,
   \href{http://cdsads.u-strasbg.fr/abs/2013ApJ...771...87H}{771, 87}

\bibitem[{{Hopkins} {et~al.}(2012){Hopkins}, {Hayward}, {Narayanan}, \&
  {Hernquist}}]{2012Hopkins}
{Hopkins}, P.~F., {Hayward}, C.~C., {Narayanan}, D., \& {Hernquist}, L. 2012,
  \begingroup\hypersetup{urlcolor=[rgb]{0.9,0.0,0.9}}\href{http://dx.doi.org/10.1111/j.1365-2966.2011.20035.x}{\mnras}\endgroup,
   \href{http://cdsads.u-strasbg.fr/abs/2012MNRAS.420..320H}{420, 320}

\bibitem[{{Horst} {et~al.}(2009){Horst}, {Duschl}, {Gandhi}, \&
  {Smette}}]{2009Horst}
{Horst}, H., {Duschl}, W.~J., {Gandhi}, P., \& {Smette}, A. 2009,
  \begingroup\hypersetup{urlcolor=[rgb]{0.9,0.0,0.9}}\href{http://dx.doi.org/10.1051/0004-6361:200809878}{\aap}\endgroup,
   \href{http://adsabs.harvard.edu/abs/2009A%26A...495..137H}{495, 137}

\bibitem[{{Jaffe} {et~al.}(2004){Jaffe}, {Meisenheimer}, {R{\"o}ttgering},
  {Leinert}, {Richichi}, {Chesneau}, {Fraix-Burnet}, {Glazenborg-Kluttig},
  {Granato}, {Graser}, {Heijligers}, {K{\"o}hler}, {Malbet}, {Miley},
  {Paresce}, {Pel}, {Perrin}, {Przygodda}, {Schoeller}, {Sol}, {Waters},
  {Weigelt}, {Woillez}, \& {de Zeeuw}}]{2004Jaffe1}
{Jaffe}, W., {Meisenheimer}, K., {R{\"o}ttgering}, H.~J.~A., {et~al.} 2004,
  \begingroup\hypersetup{urlcolor=[rgb]{0.9,0.0,0.9}}\href{http://dx.doi.org/10.1038/nature02531}{\nat}\endgroup,
   \href{http://cdsads.u-strasbg.fr/abs/2004Natur.429...47J}{429, 47}

\bibitem[{{Jaffe}(2004)}]{2004Jaffe2}
{Jaffe}, W.~J. 2004, in New Frontiers in Stellar Interferometry, Proceedings of
  SPIE, ed. W.~A. {Traub}, Vol. 5491,
  \href{http://cdsads.u-strasbg.fr/abs/2004SPIE.5491..715J}{715}

\bibitem[{{Kemper} {et~al.}(2004){Kemper}, {Vriend}, \& {Tielens}}]{2004Kemper}
{Kemper}, F., {Vriend}, W.~J., \& {Tielens}, A.~G.~G.~M. 2004,
  \begingroup\hypersetup{urlcolor=[rgb]{0.9,0.0,0.9}}\href{http://dx.doi.org/10.1086/421339}{\apj}\endgroup,
   \href{http://cdsads.u-strasbg.fr/abs/2004ApJ...609..826K}{609, 826}

\bibitem[{{Kishimoto} {et~al.}(2011{\natexlab{a}}){Kishimoto}, {H{\"o}nig},
  {Antonucci}, {Barvainis}, {Kotani}, {Tristram}, {Weigelt}, \&
  {Levin}}]{2011Kishimoto1}
{Kishimoto}, M., {H{\"o}nig}, S.~F., {Antonucci}, R., {et~al.}
  2011{\natexlab{a}},
  \begingroup\hypersetup{urlcolor=[rgb]{0.9,0.0,0.9}}\href{http://dx.doi.org/10.1051/0004-6361/201016054}{\aap}\endgroup,
   \href{http://cdsads.u-strasbg.fr/abs/2011A%26A...527A.121K}{527, A121}

\bibitem[{{Kishimoto} {et~al.}({2009b}){Kishimoto}, {H{\"o}nig}, {Antonucci},
  {Kotani}, {Barvainis}, {Tristram}, \& {Weigelt}}]{2009Kishimoto2}
{Kishimoto}, M., {H{\"o}nig}, S.~F., {Antonucci}, R., {et~al.} {2009b},
  \begingroup\hypersetup{urlcolor=[rgb]{0.9,0.0,0.9}}\href{http://dx.doi.org/10.1051/0004-6361/200913512}{\aap}\endgroup,
   \href{http://adsabs.harvard.edu/abs/2009A%26A...507L..57K}{507, L57}

\bibitem[{{Kishimoto} {et~al.}(2011{\natexlab{b}}){Kishimoto}, {H{\"o}nig},
  {Antonucci}, {Millour}, {Tristram}, \& {Weigelt}}]{2011Kishimoto2}
{Kishimoto}, M., {H{\"o}nig}, S.~F., {Antonucci}, R., {et~al.}
  2011{\natexlab{b}},
  \begingroup\hypersetup{urlcolor=[rgb]{0.9,0.0,0.9}}\href{http://dx.doi.org/10.1051/0004-6361/201117367}{\aap}\endgroup,
   \href{http://cdsads.u-strasbg.fr/abs/2011A%26A...536A..78K}{536, A78}

\bibitem[{{Kishimoto} {et~al.}({2009a}){Kishimoto}, {H{\"o}nig}, {Tristram}, \&
  {Weigelt}}]{2009Kishimoto1}
{Kishimoto}, M., {H{\"o}nig}, S.~F., {Tristram}, K.~R.~W., \& {Weigelt}, G.
  {2009a},
  \begingroup\hypersetup{urlcolor=[rgb]{0.9,0.0,0.9}}\href{http://dx.doi.org/10.1051/0004-6361:200811062}{\aap}\endgroup,
   \href{http://cdsads.u-strasbg.fr/abs/2009A%26A...493L..57K}{493, L57}

\bibitem[{{Koribalski} {et~al.}(2004){Koribalski}, {Staveley-Smith}, {Kilborn},
  {Ryder}, {Kraan-Korteweg}, {Ryan-Weber}, {Ekers}, {Jerjen}, {Henning},
  {Putman}, {Zwaan}, {de Blok}, {Calabretta}, {Disney}, {Minchin}, {Bhathal},
  {Boyce}, {Drinkwater}, {Freeman}, {Gibson}, {Green}, {Haynes}, {Juraszek},
  {Kesteven}, {Knezek}, {Mader}, {Marquarding}, {Meyer}, {Mould}, {Oosterloo},
  {O'Brien}, {Price}, {Sadler}, {Schr{\"o}der}, {Stewart}, {Stootman}, {Waugh},
  {Warren}, {Webster}, \& {Wright}}]{2004Koribalski}
{Koribalski}, B.~S., {Staveley-Smith}, L., {Kilborn}, V.~A., {et~al.} 2004,
  \begingroup\hypersetup{urlcolor=[rgb]{0.9,0.0,0.9}}\href{http://dx.doi.org/10.1086/421744}{\aj}\endgroup,
   \href{http://adsabs.harvard.edu/abs/2004AJ....128...16K}{128, 16}

\bibitem[{{Kreplin} {et~al.}(2013){Kreplin}, {Weigelt}, {Kraus}, {Grinin},
  {Hofmann}, {Kishimoto}, {Schertl}, {Tambovtseva}, {Clausse}, {Massi},
  {Perraut}, \& {Stee}}]{2013Kreplin}
{Kreplin}, A., {Weigelt}, G., {Kraus}, S., {et~al.} 2013,
  \begingroup\hypersetup{urlcolor=[rgb]{0.9,0.0,0.9}}\href{http://dx.doi.org/10.1051/0004-6361/201220806}{\aap}\endgroup,
   \href{http://cdsads.u-strasbg.fr/abs/2013A%26A...551A..21K}{551, A21}

\bibitem[{{Krolik} \& {Begelman}(1988)}]{1988Krolik}
{Krolik}, J.~H. \& {Begelman}, M.~C. 1988,
  \begingroup\hypersetup{urlcolor=[rgb]{0.9,0.0,0.9}}\href{http://dx.doi.org/10.1086/166414}{\apj}\endgroup,
   \href{http://cdsads.u-strasbg.fr/abs/1988ApJ...329..702K}{329, 702}

\bibitem[{{Kylafis} \& {Norman}(1991)}]{1991Kylafis}
{Kylafis}, N.~D. \& {Norman}, C.~A. 1991,
  \begingroup\hypersetup{urlcolor=[rgb]{0.9,0.0,0.9}}\href{http://dx.doi.org/10.1086/170071}{\apj}\endgroup,
   \href{http://cdsads.u-strasbg.fr/abs/1991ApJ...373..525K}{373, 525}

\bibitem[{{Leinert} {et~al.}(2003){Leinert}, {Graser}, {Przygodda}, {Waters},
  {Perrin}, {Jaffe}, {Lopez}, {Bakker}, {B{\"o}hm}, {Chesneau}, {Cotton},
  {Damstra}, {de Jong}, {Glazenborg-Kluttig}, {Grimm}, {Hanenburg}, {Laun},
  {Lenzen}, {Ligori}, {Mathar}, {Meisner}, {Morel}, {Morr}, {Neumann}, {Pel},
  {Schuller}, {Rohloff}, {Stecklum}, {Storz}, {von der L{\"u}he}, \&
  {Wagner}}]{2003Leinert2}
{Leinert}, C., {Graser}, U., {Przygodda}, F., {et~al.} 2003,
  \begingroup\hypersetup{urlcolor=[rgb]{0.9,0.0,0.9}}\href{http://dx.doi.org/10.1023/A:1026158127732}{\apss}\endgroup,
   \href{http://cdsads.u-strasbg.fr/abs/2003Ap%26SS.286...73L}{286, 73}

\bibitem[{{Lo}(2005)}]{2005Lo}
{Lo}, K.~Y. 2005,
  \begingroup\hypersetup{urlcolor=[rgb]{0.9,0.0,0.9}}\href{http://dx.doi.org/10.1146/annurev.astro.41.011802.094927}{\araa}\endgroup,
   \href{http://cdsads.u-strasbg.fr/abs/2005ARA%26A..43..625L}{43, 625}

\bibitem[{{Lopez} {et~al.}(2008){Lopez}, {Antonelli}, {Wolf}, {Lagarde},
  {Jaffe}, {Navarro}, {Graser}, {Petrov}, {Weigelt}, {Bresson}, {Hofmann},
  {Beckman}, {Henning}, {Laun}, {Leinert}, {Kraus}, {Robbe-Dubois}, {Vakili},
  {Richichi}, {Abraham}, {Augereau}, {Behrend}, {Berio}, {Berruyer},
  {Chesneau}, {Clausse}, {Connot}, {Demyk}, {Danchi}, {Dugu{\'e}}, {Finger},
  {Flament}, {Glazenborg}, {Hannenburg}, {Heininger}, {Hugues}, {Hron},
  {Jankov}, {Kerschbaum}, {Kroes}, {Linz}, {Lizon}, {Mathias}, {Mathar},
  {Matter}, {Menut}, {Meisenheimer}, {Millour}, {Nardetto}, {Neumann},
  {Nussbaum}, {Niedzielski}, {Mosoni}, {Olofsson}, {Rabbia}, {Ratzka}, {Rigal},
  {Roussel}, {Schertl}, {Schmider}, {Stecklum}, {Thiebaut}, {Vannier}, {Valat},
  {Wagner}, \& {Waters}}]{2008Lopez}
{Lopez}, B., {Antonelli}, P., {Wolf}, S., {et~al.} 2008, in Society of
  Photo-Optical Instrumentation Engineers (SPIE) Conference
  Series,\begingroup\hypersetup{urlcolor=[rgb]{0.9,0.0,0.9}}\href{http://dx.doi.org/10.1117/12.789412}{
  Vol. 7013}\endgroup,
  \href{http://cdsads.u-strasbg.fr/abs/2008SPIE.7013E..70L}{70132B}

\bibitem[{{L{\'o}pez-Gonzaga} {et~al.}(2014){L{\'o}pez-Gonzaga}, {Jaffe},
  {Burtscher}, {Tristram}, \& {Meisenheimer}}]{2014Lopez-Gonzaga}
{L{\'o}pez-Gonzaga}, N., {Jaffe}, W., {Burtscher}, L., {Tristram}, K.~R.~W., \&
  {Meisenheimer}, K. 2014, submitted to \aap

\bibitem[{{Lumsden} {et~al.}(2004){Lumsden}, {Alexander}, \&
  {Hough}}]{2004Lumsden}
{Lumsden}, S.~L., {Alexander}, D.~M., \& {Hough}, J.~H. 2004,
  \begingroup\hypersetup{urlcolor=[rgb]{0.9,0.0,0.9}}\href{http://dx.doi.org/10.1111/j.1365-2966.2004.07469.x}{\mnras}\endgroup,
   \href{http://cdsads.u-strasbg.fr/abs/2004MNRAS.348.1451L}{348, 1451}

\bibitem[{{Maiolino} {et~al.}(2000){Maiolino}, {Alonso-Herrero}, {Anders},
  {Quillen}, {Rieke}, {Rieke}, \& {Tacconi-Garman}}]{2000Maiolino}
{Maiolino}, R., {Alonso-Herrero}, A., {Anders}, S., {et~al.} 2000,
  \begingroup\hypersetup{urlcolor=[rgb]{0.9,0.0,0.9}}\href{http://dx.doi.org/10.1086/308444}{\apj}\endgroup,
   \href{http://cdsads.u-strasbg.fr/abs/2000ApJ...531..219M}{531, 219}

\bibitem[{{Marconi} {et~al.}(1994){Marconi}, {Moorwood}, {Origlia}, \&
  {Oliva}}]{1994Marconi}
{Marconi}, A., {Moorwood}, A.~F.~M., {Origlia}, L., \& {Oliva}, E. 1994, The
  Messenger,  \href{http://cdsads.u-strasbg.fr/abs/1994Msngr..78...20M}{78, 20}

\bibitem[{{Markwardt}(2009)}]{2009Markwardt}
{Markwardt}, C.~B. 2009, in Astronomical Society of the Pacific Conference
  Series, Vol. 411, Astronomical Data Analysis Software and Systems XVIII, ed.
  D.~A. {Bohlender}, D.~{Durand}, \& P.~{Dowler},
  \href{http://cdsads.u-strasbg.fr/abs/2009ASPC..411..251M}{251}

\bibitem[{{Matt} {et~al.}(1996){Matt}, {Fiore}, {Perola}, {Piro}, {Fink},
  {Grandi}, {Matsuoka}, {Oliva}, \& {Salvati}}]{1996Matt}
{Matt}, G., {Fiore}, F., {Perola}, G.~C., {et~al.} 1996, \mnras,
  \href{http://cdsads.u-strasbg.fr/abs/1996MNRAS.281L..69M}{281, L69}

\bibitem[{{Meisenheimer} {et~al.}(2007){Meisenheimer}, {Tristram}, {Jaffe},
  {Israel}, {Neumayer}, {Raban}, {R{\"o}ttgering}, {Cotton}, {Graser},
  {Henning}, {Leinert}, {Lopez}, {Perrin}, \& {Prieto}}]{2007Meisenheimer}
{Meisenheimer}, K., {Tristram}, K.~R.~W., {Jaffe}, W., {et~al.} 2007,
  \begingroup\hypersetup{urlcolor=[rgb]{0.9,0.0,0.9}}\href{http://dx.doi.org/10.1051/0004-6361:20066967}{\aap}\endgroup,
   \href{http://cdsads.u-strasbg.fr/abs/2007A%26A...471..453M}{471, 453}

\bibitem[{{Meisner} {et~al.}(2004){Meisner}, {Tubbs}, \& {Jaffe}}]{2004Meisner}
{Meisner}, J.~A., {Tubbs}, R.~N., \& {Jaffe}, W.~J. 2004, in New Frontiers in
  Stellar Interferometry, Proceedings of SPIE, ed. W.~A. {Traub}, Vol. 5491,
  \href{http://cdsads.u-strasbg.fr/abs/2004SPIE.5491..725M}{725}

\bibitem[{{Moorwood} {et~al.}(1996){Moorwood}, {Lutz}, {Oliva}, {Marconi},
  {Netzer}, {Genzel}, {Sturm}, \& {de Graauw}}]{1996Moorwood}
{Moorwood}, A.~F.~M., {Lutz}, D., {Oliva}, E., {et~al.} 1996, \aap,
  \href{http://cdsads.u-strasbg.fr/abs/1996A%26A...315L.109M}{315, L109}

\bibitem[{{Mor} {et~al.}(2009){Mor}, {Netzer}, \& {Elitzur}}]{2009Mor}
{Mor}, R., {Netzer}, H., \& {Elitzur}, M. 2009,
  \begingroup\hypersetup{urlcolor=[rgb]{0.9,0.0,0.9}}\href{http://dx.doi.org/10.1088/0004-637X/705/1/298}{\apj}\endgroup,
   \href{http://cdsads.u-strasbg.fr/abs/2009ApJ...705..298M}{705, 298}

\bibitem[{{Moshir} {et~al.}(1990){Moshir}, {Kopan}, {Conrow}, {McCallon},
  {Hacking}, {Gregorich}, {Rohrbach}, {Melnyk}, {Rice}, {Fullmer}, {White}, \&
  {Chester}}]{1990Moshir}
{Moshir}, M., {Kopan}, G., {Conrow}, T., {et~al.} 1990, in Bulletin of the
  American Astronomical Society, Vol.~22,
  \href{http://cdsads.u-strasbg.fr/abs/1990BAAS...22Q1325M}{1325}

\bibitem[{{Natta} {et~al.}(2001){Natta}, {Prusti}, {Neri}, {Wooden}, {Grinin},
  \& {Mannings}}]{2001Natta}
{Natta}, A., {Prusti}, T., {Neri}, R., {et~al.} 2001,
  \begingroup\hypersetup{urlcolor=[rgb]{0.9,0.0,0.9}}\href{http://dx.doi.org/10.1051/0004-6361:20010334}{\aap}\endgroup,
   \href{http://cdsads.u-strasbg.fr/abs/2001A%26A...371..186N}{371, 186}

\bibitem[{{Nenkova} {et~al.}(2002){Nenkova}, {Ivezi{\'c}}, \&
  {Elitzur}}]{2002Nenkova}
{Nenkova}, M., {Ivezi{\'c}}, {\v Z}., \& {Elitzur}, M. 2002,
  \begingroup\hypersetup{urlcolor=[rgb]{0.9,0.0,0.9}}\href{http://dx.doi.org/10.1086/340857}{\apjl}\endgroup,
   \href{http://cdsads.u-strasbg.fr/abs/2002ApJ...570L...9N}{570, L9}

\bibitem[{{Netzer}(1987)}]{1987Netzer}
{Netzer}, H. 1987, \mnras,
  \href{http://cdsads.u-strasbg.fr/abs/1987MNRAS.225...55N}{225, 55}

\bibitem[{{Oliva} {et~al.}(1998){Oliva}, {Marconi}, {Cimatti}, \&
  {Alighieri}}]{1998Oliva}
{Oliva}, E., {Marconi}, A., {Cimatti}, A., \& {Alighieri}, S.~D.~S. 1998, \aap,
   \href{http://cdsads.u-strasbg.fr/abs/1998A\%26A...329L..21O}{329, L21}

\bibitem[{{Oliva} {et~al.}(1994){Oliva}, {Salvati}, {Moorwood}, \&
  {Marconi}}]{1994Oliva}
{Oliva}, E., {Salvati}, M., {Moorwood}, A.~F.~M., \& {Marconi}, A. 1994, \aap,
  \href{http://cdsads.u-strasbg.fr/abs/1994A\%26A...288..457O}{288, 457}

\bibitem[{{Packham} {et~al.}(2005){Packham}, {Radomski}, {Roche}, {Aitken},
  {Perlman}, {Alonso-Herrero}, {Colina}, \& {Telesco}}]{2005Packham}
{Packham}, C., {Radomski}, J.~T., {Roche}, P.~F., {et~al.} 2005,
  \begingroup\hypersetup{urlcolor=[rgb]{0.9,0.0,0.9}}\href{http://dx.doi.org/10.1086/427691}{\apjl}\endgroup,
   \href{http://adsabs.harvard.edu/abs/2005ApJ...618L..17P}{618, L17}

\bibitem[{{Packham} {et~al.}(2007){Packham}, {Young}, {Fisher}, {Volk},
  {Mason}, {Hough}, {Roche}, {Elitzur}, {Radomski}, \& {Perlman}}]{2007Packham}
{Packham}, C., {Young}, S., {Fisher}, S., {et~al.} 2007,
  \begingroup\hypersetup{urlcolor=[rgb]{0.9,0.0,0.9}}\href{http://dx.doi.org/10.1086/518602}{\apjl}\endgroup,
   \href{http://cdsads.u-strasbg.fr/abs/2007ApJ...661L..29P}{661, L29}

\bibitem[{{Poncelet} {et~al.}(2007){Poncelet}, {Doucet}, {Perrin}, {Sol}, \&
  {Lagage}}]{2007Poncelet}
{Poncelet}, A., {Doucet}, C., {Perrin}, G., {Sol}, H., \& {Lagage}, P.~O. 2007,
  \begingroup\hypersetup{urlcolor=[rgb]{0.9,0.0,0.9}}\href{http://dx.doi.org/10.1051/0004-6361:20067012}{\aap}\endgroup,
   \href{http://cdsads.u-strasbg.fr/abs/2007A%26A...472..823P}{472, 823}

\bibitem[{{Poncelet} {et~al.}(2006){Poncelet}, {Perrin}, \&
  {Sol}}]{2006Poncelet}
{Poncelet}, A., {Perrin}, G., \& {Sol}, H. 2006,
  \begingroup\hypersetup{urlcolor=[rgb]{0.9,0.0,0.9}}\href{http://dx.doi.org/10.1051/0004-6361:20053608}{\aap}\endgroup,
   \href{http://cdsads.u-strasbg.fr/abs/2006A\%26A...450..483P}{450, 483}

\bibitem[{{Pott} {et~al.}(2010){Pott}, {Malkan}, {Elitzur}, {Ghez}, {Herbst},
  {Sch{\"o}del}, \& {Woillez}}]{2010Pott}
{Pott}, J., {Malkan}, M.~A., {Elitzur}, M., {et~al.} 2010,
  \begingroup\hypersetup{urlcolor=[rgb]{0.9,0.0,0.9}}\href{http://dx.doi.org/10.1088/0004-637X/715/2/736}{\apj}\endgroup,
   \href{http://adsabs.harvard.edu/abs/2010ApJ...715..736P}{715, 736}

\bibitem[{{Prieto} {et~al.}(2004){Prieto}, {Meisenheimer}, {Marco}, {Reunanen},
  {Contini}, {Clenet}, {Davies}, {Gratadour}, {Henning}, {Klaas}, {Kotilainen},
  {Leinert}, {Lutz}, {Rouan}, \& {Thatte}}]{2004Prieto}
{Prieto}, M.~A., {Meisenheimer}, K., {Marco}, O., {et~al.} 2004,
  \begingroup\hypersetup{urlcolor=[rgb]{0.9,0.0,0.9}}\href{http://dx.doi.org/10.1086/423422}{\apj}\endgroup,
   \href{http://cdsads.u-strasbg.fr/abs/2004ApJ...614..135P}{614, 135}

\bibitem[{{Prieto} {et~al.}(2010){Prieto}, {Reunanen}, {Tristram}, {Neumayer},
  {Fernandez-Ontiveros}, {Orienti}, \& {Meisenheimer}}]{2010Prieto}
{Prieto}, M.~A., {Reunanen}, J., {Tristram}, K.~R.~W., {et~al.} 2010,
  \begingroup\hypersetup{urlcolor=[rgb]{0.9,0.0,0.9}}\href{http://dx.doi.org/10.1111/j.1365-2966.2009.15897.x}{\mnras}\endgroup,
   \href{http://adsabs.harvard.edu/abs/2010MNRAS.402..724P}{402, 724}

\bibitem[{{Raban} {et~al.}(2008){Raban}, {Heijligers}, {R{\"o}ttgering},
  {Meisenheimer}, {Jaffe}, {K{\"a}ufl}, \& {Henning}}]{2008Raban}
{Raban}, D., {Heijligers}, B., {R{\"o}ttgering}, H., {et~al.} 2008,
  \begingroup\hypersetup{urlcolor=[rgb]{0.9,0.0,0.9}}\href{http://dx.doi.org/10.1051/0004-6361:20077444}{\aap}\endgroup,
   \href{http://adsabs.harvard.edu/abs/2008A%26A...484..341R}{484, 341}

\bibitem[{{Raban} {et~al.}(2009){Raban}, {Jaffe}, {R{\"o}ttgering},
  {Meisenheimer}, \& {Tristram}}]{2009Raban}
{Raban}, D., {Jaffe}, W., {R{\"o}ttgering}, H., {Meisenheimer}, K., \&
  {Tristram}, K.~R.~W. 2009,
  \begingroup\hypersetup{urlcolor=[rgb]{0.9,0.0,0.9}}\href{http://dx.doi.org/10.1111/j.1365-2966.2009.14439.x}{\mnras}\endgroup,
   \href{http://adsabs.harvard.edu/abs/2009MNRAS.394.1325R}{394, 1325}

\bibitem[{{Radomski} {et~al.}(2003){Radomski}, {Pi{\~n}a}, {Packham},
  {Telesco}, {De Buizer}, {Fisher}, \& {Robinson}}]{2003Radomski}
{Radomski}, J.~T., {Pi{\~n}a}, R.~K., {Packham}, C., {et~al.} 2003,
  \begingroup\hypersetup{urlcolor=[rgb]{0.9,0.0,0.9}}\href{http://dx.doi.org/10.1086/367612}{\apj}\endgroup,
   \href{http://cdsads.u-strasbg.fr/abs/2003ApJ...587..117R}{587, 117}

\bibitem[{{Ramos Almeida} {et~al.}(2009){Ramos Almeida}, {Levenson},
  {Rodr{\'{\i}}guez Espinosa}, {Alonso-Herrero}, {Asensio Ramos}, {Radomski},
  {Packham}, {Fisher}, \& {Telesco}}]{2009RamosAlmeida}
{Ramos Almeida}, C., {Levenson}, N.~A., {Rodr{\'{\i}}guez Espinosa}, J.~M.,
  {et~al.} 2009,
  \begingroup\hypersetup{urlcolor=[rgb]{0.9,0.0,0.9}}\href{http://dx.doi.org/10.1088/0004-637X/702/2/1127}{\apj}\endgroup,
   \href{http://adsabs.harvard.edu/abs/2009ApJ...702.1127R}{702, 1127}

\bibitem[{{Rees} {et~al.}(1969){Rees}, {Silk}, {Werner}, \&
  {Wickramasinghe}}]{1969Rees}
{Rees}, M.~J., {Silk}, J.~I., {Werner}, M.~W., \& {Wickramasinghe}, N.~C. 1969,
  \begingroup\hypersetup{urlcolor=[rgb]{0.9,0.0,0.9}}\href{http://dx.doi.org/10.1038/223788a0}{\nat}\endgroup,
   \href{http://cdsads.u-strasbg.fr/abs/1969Natur.223..788R}{223, 788}

\bibitem[{{Reunanen} {et~al.}(2010){Reunanen}, {Prieto}, \&
  {Siebenmorgen}}]{2010Reunanen}
{Reunanen}, J., {Prieto}, M.~A., \& {Siebenmorgen}, R. 2010,
  \begingroup\hypersetup{urlcolor=[rgb]{0.9,0.0,0.9}}\href{http://dx.doi.org/10.1111/j.1365-2966.2009.15997.x}{\mnras}\endgroup,
   \href{http://cdsads.u-strasbg.fr/abs/2010MNRAS.402..879R}{402, 879}

\bibitem[{{Roche} {et~al.}(2006){Roche}, {Packham}, {Telesco}, {Radomski},
  {Alonso-Hererro}, {Aitken}, {Colina}, \& {Perlman}}]{2006Roche}
{Roche}, P.~F., {Packham}, C., {Telesco}, C.~M., {et~al.} 2006,
  \begingroup\hypersetup{urlcolor=[rgb]{0.9,0.0,0.9}}\href{http://dx.doi.org/10.1111/j.1365-2966.2006.10072.x}{\mnras}\endgroup,
   \href{http://cdsads.u-strasbg.fr/abs/2006MNRAS.367.1689R}{367, 1689}

\bibitem[{{Schartmann} {et~al.}(2005){Schartmann}, {Meisenheimer}, {Camenzind},
  {Wolf}, \& {Henning}}]{2005Schartmann}
{Schartmann}, M., {Meisenheimer}, K., {Camenzind}, M., {Wolf}, S., \&
  {Henning}, T. 2005,
  \begingroup\hypersetup{urlcolor=[rgb]{0.9,0.0,0.9}}\href{http://dx.doi.org/10.1051/0004-6361:20042363}{\aap}\endgroup,
   \href{http://cdsads.u-strasbg.fr/abs/2005A\%26A...437..861S}{437, 861}

\bibitem[{{Schartmann} {et~al.}(2008){Schartmann}, {Meisenheimer}, {Camenzind},
  {Wolf}, {Tristram}, \& {Henning}}]{2008Schartmann}
{Schartmann}, M., {Meisenheimer}, K., {Camenzind}, M., {et~al.} 2008,
  \begingroup\hypersetup{urlcolor=[rgb]{0.9,0.0,0.9}}\href{http://dx.doi.org/10.1051/0004-6361:20078907}{\aap}\endgroup,
   \href{http://cdsads.u-strasbg.fr/abs/2008A%26A...482...67S}{482, 67}

\bibitem[{{Schartmann} {et~al.}(2009){Schartmann}, {Meisenheimer}, {Klahr},
  {Camenzind}, {Wolf}, \& {Henning}}]{2009Schartmann}
{Schartmann}, M., {Meisenheimer}, K., {Klahr}, H., {et~al.} 2009,
  \begingroup\hypersetup{urlcolor=[rgb]{0.9,0.0,0.9}}\href{http://dx.doi.org/10.1111/j.1365-2966.2008.14220.x}{\mnras}\endgroup,
   \href{http://adsabs.harvard.edu/abs/2009MNRAS.393..759S}{393, 759}

\bibitem[{{Schmitt} {et~al.}(2006){Schmitt}, {Pauls}, {Tycner}, {Armstrong},
  {Benson}, {Clark}, {Hindsley}, {Hutter}, {Peterson}, {Jorgensen},
  {Mozurkewich}, {Gilbreath}, \& {Zavala}}]{2006Schmitt}
{Schmitt}, H.~R., {Pauls}, T.~A., {Tycner}, C., {et~al.} 2006, in Proceedings
  of the
  SPIE,\begingroup\hypersetup{urlcolor=[rgb]{0.9,0.0,0.9}}\href{http://dx.doi.org/10.1117/12.672396}{
  Vol. 6268}\endgroup, Advances in Stellar Interferometry. Edited by Monnier,
  John D.; Sch{\"o}ller, Markus; Danchi, William C.,
  \href{http://cdsads.u-strasbg.fr/abs/2006SPIE.6268E.105S}{62683B}

\bibitem[{{Schweitzer} {et~al.}(2008){Schweitzer}, {Groves}, {Netzer}, {Lutz},
  {Sturm}, {Contursi}, {Genzel}, {Tacconi}, {Veilleux}, {Kim}, {Rupke}, \&
  {Baker}}]{2008Schweitzer}
{Schweitzer}, M., {Groves}, B., {Netzer}, H., {et~al.} 2008,
  \begingroup\hypersetup{urlcolor=[rgb]{0.9,0.0,0.9}}\href{http://dx.doi.org/10.1086/587097}{\apj}\endgroup,
   \href{http://cdsads.u-strasbg.fr/abs/2008ApJ...679..101S}{679, 101}

\bibitem[{{Siebenmorgen} \& {Efstathiou}(2001)}]{2001Siebenmorgen}
{Siebenmorgen}, R. \& {Efstathiou}, A. 2001,
  \begingroup\hypersetup{urlcolor=[rgb]{0.9,0.0,0.9}}\href{http://dx.doi.org/10.1051/0004-6361:20011085}{\aap}\endgroup,
   \href{http://cdsads.u-strasbg.fr/abs/2001A%26A...376L..35S}{376, L35}

\bibitem[{{Smith} {et~al.}(2000){Smith}, {Wright}, {Aitken}, {Roche}, \&
  {Hough}}]{2000Smith}
{Smith}, C.~H., {Wright}, C.~M., {Aitken}, D.~K., {Roche}, P.~F., \& {Hough},
  J.~H. 2000,
  \begingroup\hypersetup{urlcolor=[rgb]{0.9,0.0,0.9}}\href{http://dx.doi.org/10.1046/j.1365-8711.2000.03158.x}{\mnras}\endgroup,
   \href{http://cdsads.u-strasbg.fr/abs/2000MNRAS.312..327S}{312, 327}

\bibitem[{{Smith} \& {Wilson}(2001)}]{2001Smith}
{Smith}, D.~A. \& {Wilson}, A.~S. 2001,
  \begingroup\hypersetup{urlcolor=[rgb]{0.9,0.0,0.9}}\href{http://dx.doi.org/10.1086/321667}{\apj}\endgroup,
   \href{http://cdsads.u-strasbg.fr/abs/2001ApJ...557..180S}{557, 180}

\bibitem[{{Smith} {et~al.}(2007){Smith}, {Draine}, {Dale}, {Moustakas},
  {Kennicutt}, {Helou}, {Armus}, {Roussel}, {Sheth}, {Bendo}, {Buckalew},
  {Calzetti}, {Engelbracht}, {Gordon}, {Hollenbach}, {Li}, {Malhotra},
  {Murphy}, \& {Walter}}]{2007Smith}
{Smith}, J.~D.~T., {Draine}, B.~T., {Dale}, D.~A., {et~al.} 2007,
  \begingroup\hypersetup{urlcolor=[rgb]{0.9,0.0,0.9}}\href{http://dx.doi.org/10.1086/510549}{\apj}\endgroup,
   \href{http://cdsads.u-strasbg.fr/abs/2007ApJ...656..770S}{656, 770}

\bibitem[{{Soldi} {et~al.}(2005){Soldi}, {Beckmann}, {Bassani}, {Courvoisier},
  {Landi}, {Malizia}, {Dean}, {de Rosa}, {Fabian}, \& {Walter}}]{2005Soldi}
{Soldi}, S., {Beckmann}, V., {Bassani}, L., {et~al.} 2005,
  \begingroup\hypersetup{urlcolor=[rgb]{0.9,0.0,0.9}}\href{http://dx.doi.org/10.1051/0004-6361:20053875}{\aap}\endgroup,
   \href{http://adsabs.harvard.edu/abs/2005A%26A...444..431S}{444, 431}

\bibitem[{{Suganuma} {et~al.}(2006){Suganuma}, {Yoshii}, {Kobayashi},
  {Minezaki}, {Enya}, {Tomita}, {Aoki}, {Koshida}, \&
  {Peterson}}]{2006Suganuma}
{Suganuma}, M., {Yoshii}, Y., {Kobayashi}, Y., {et~al.} 2006,
  \begingroup\hypersetup{urlcolor=[rgb]{0.9,0.0,0.9}}\href{http://dx.doi.org/10.1086/499326}{\apj}\endgroup,
   \href{http://cdsads.u-strasbg.fr/abs/2006ApJ...639...46S}{639, 46}

\bibitem[{{Swain} {et~al.}(2003){Swain}, {Vasisht}, {Akeson}, {Monnier},
  {Millan-Gabet}, {Serabyn}, {Creech-Eakman}, {van Belle}, {Beletic},
  {Beichman}, {Boden}, {Booth}, {Colavita}, {Gathright}, {Hrynevych},
  {Koresko}, {Le Mignant}, {Ligon}, {Mennesson}, {Neyman}, {Sargent}, {Shao},
  {Thompson}, {Unwin}, \& {Wizinowich}}]{2003Swain}
{Swain}, M., {Vasisht}, G., {Akeson}, R., {et~al.} 2003,
  \begingroup\hypersetup{urlcolor=[rgb]{0.9,0.0,0.9}}\href{http://dx.doi.org/10.1086/379235}{\apjl}\endgroup,
   \href{http://cdsads.u-strasbg.fr/abs/2003ApJ...596L.163S}{596, L163}

\bibitem[{{Tristram}(2007)}]{2007Tristram1}
{Tristram}, K.~R.~W. 2007, PhD thesis, Max-Planck-Institut f{\"u}r Astronomie,
  K{\"o}nigstuhl 17, 69117 Heidelberg, Germany

\bibitem[{{Tristram}(2013)}]{2013Tristram1}
{Tristram}, K.~R.~W. 2013,
  \href{http://cdsads.u-strasbg.fr/abs/2013arXiv1312.3607T}{ArXiv e-prints}

\bibitem[{{Tristram} {et~al.}(2007){Tristram}, {Meisenheimer}, {Jaffe},
  {Schartmann}, {Rix}, {Leinert}, {Morel}, {Wittkowski}, {R{\"o}ttgering},
  {Perrin}, {Lopez}, {Raban}, {Cotton}, {Graser}, {Paresce}, \&
  {Henning}}]{2007Tristram2}
{Tristram}, K.~R.~W., {Meisenheimer}, K., {Jaffe}, W., {et~al.} 2007,
  \begingroup\hypersetup{urlcolor=[rgb]{0.9,0.0,0.9}}\href{http://dx.doi.org/10.1051/0004-6361:20078369}{\aap}\endgroup,
   \href{http://cdsads.u-strasbg.fr/abs/2007A%26A...474..837T}{474, 837}

\bibitem[{{Tristram} {et~al.}(2009){Tristram}, {Raban}, {Meisenheimer},
  {Jaffe}, {R{\"o}ttgering}, {Burtscher}, {Cotton}, {Graser}, {Henning},
  {Leinert}, {Lopez}, {Morel}, {Perrin}, \& {Wittkowski}}]{2009Tristram}
{Tristram}, K.~R.~W., {Raban}, D., {Meisenheimer}, K., {et~al.} 2009,
  \begingroup\hypersetup{urlcolor=[rgb]{0.9,0.0,0.9}}\href{http://dx.doi.org/10.1051/0004-6361/200811607}{\aap}\endgroup,
   \href{http://adsabs.harvard.edu/abs/2009A%26A...502...67T}{502, 67}

\bibitem[{{Tristram} \& {Schartmann}(2011)}]{2011Tristram}
{Tristram}, K.~R.~W. \& {Schartmann}, M. 2011,
  \begingroup\hypersetup{urlcolor=[rgb]{0.9,0.0,0.9}}\href{http://dx.doi.org/10.1051/0004-6361/201116867}{\aap}\endgroup,
   \href{http://cdsads.u-strasbg.fr/abs/2011A%26A...531A..99T}{531, A99}

\bibitem[{{Urry} \& {Padovani}(1995)}]{1995Urry}
{Urry}, C.~M. \& {Padovani}, P. 1995,
  \begingroup\hypersetup{urlcolor=[rgb]{0.9,0.0,0.9}}\href{http://dx.doi.org/10.1086/133630}{\pasp}\endgroup,
   \href{http://cdsads.u-strasbg.fr/abs/1995PASP..107..803U}{107, 803}

\bibitem[{{van Boekel}(2004)}]{2004vanBoekel}
{van Boekel}, R.~J.~H.~M. 2004, PhD thesis, University of Amsterdam

\bibitem[{{Veilleux} \& {Bland-Hawthorn}(1997)}]{1997Veilleux}
{Veilleux}, S. \& {Bland-Hawthorn}, J. 1997,
  \begingroup\hypersetup{urlcolor=[rgb]{0.9,0.0,0.9}}\href{http://dx.doi.org/10.1086/310588}{\apjl}\endgroup,
   \href{http://cdsads.u-strasbg.fr/abs/1997ApJ...479L.105V}{479, L105}

\bibitem[{{Verhoelst}(2005)}]{2005Verhoelst}
{Verhoelst}, T. 2005, PhD thesis, Institute of Astronomy, K.U.Leuven, Belgium

\bibitem[{{Wada}(2012)}]{2012Wada}
{Wada}, K. 2012,
  \begingroup\hypersetup{urlcolor=[rgb]{0.9,0.0,0.9}}\href{http://dx.doi.org/10.1088/0004-637X/758/1/66}{\apj}\endgroup,
   \href{http://cdsads.u-strasbg.fr/abs/2012ApJ...758...66W}{758, 66}

\bibitem[{{Wada} \& {Norman}(2002)}]{2002Wada}
{Wada}, K. \& {Norman}, C.~A. 2002,
  \begingroup\hypersetup{urlcolor=[rgb]{0.9,0.0,0.9}}\href{http://dx.doi.org/10.1086/339438}{\apjl}\endgroup,
   \href{http://adsabs.harvard.edu/abs/2002ApJ...566L..21W}{566, L21}

\bibitem[{{Wada} {et~al.}(2009){Wada}, {Papadopoulos}, \& {Spaans}}]{2009Wada}
{Wada}, K., {Papadopoulos}, P.~P., \& {Spaans}, M. 2009,
  \begingroup\hypersetup{urlcolor=[rgb]{0.9,0.0,0.9}}\href{http://dx.doi.org/10.1088/0004-637X/702/1/63}{\apj}\endgroup,
   \href{http://cdsads.u-strasbg.fr/abs/2009ApJ...702...63W}{702, 63}

\bibitem[{{Weigelt} {et~al.}(2004){Weigelt}, {Wittkowski}, {Balega}, {Beckert},
  {Duschl}, {Hofmann}, {Men'shchikov}, \& {Schertl}}]{2004Weigelt}
{Weigelt}, G., {Wittkowski}, M., {Balega}, Y.~Y., {et~al.} 2004,
  \begingroup\hypersetup{urlcolor=[rgb]{0.9,0.0,0.9}}\href{http://dx.doi.org/10.1051/0004-6361:20040362}{\aap}\endgroup,
   \href{http://cdsads.u-strasbg.fr/abs/2004A%26A...425...77W}{425, 77}

\bibitem[{{Wilson} {et~al.}(2000){Wilson}, {Shopbell}, {Simpson},
  {Storchi-Bergmann}, {Barbosa}, \& {Ward}}]{2000Wilson}
{Wilson}, A.~S., {Shopbell}, P.~L., {Simpson}, C., {et~al.} 2000,
  \begingroup\hypersetup{urlcolor=[rgb]{0.9,0.0,0.9}}\href{http://dx.doi.org/10.1086/301532}{\aj}\endgroup,
   \href{http://cdsads.u-strasbg.fr/abs/2000AJ....120.1325W}{120, 1325}

\bibitem[{{Wittkowski} {et~al.}(2004){Wittkowski}, {Kervella}, {Arsenault},
  {Paresce}, {Beckert}, \& {Weigelt}}]{2004Wittkowski}
{Wittkowski}, M., {Kervella}, P., {Arsenault}, R., {et~al.} 2004,
  \begingroup\hypersetup{urlcolor=[rgb]{0.9,0.0,0.9}}\href{http://dx.doi.org/10.1051/0004-6361:20040118}{\aap}\endgroup,
   \href{http://cdsads.u-strasbg.fr/abs/2004A\%26A...418L..39W}{418, L39}

\bibitem[{{Yang} {et~al.}(2009){Yang}, {Wilson}, {Matt}, {Terashima}, \&
  {Greenhill}}]{2009Yang}
{Yang}, Y., {Wilson}, A.~S., {Matt}, G., {Terashima}, Y., \& {Greenhill}, L.~J.
  2009,
  \begingroup\hypersetup{urlcolor=[rgb]{0.9,0.0,0.9}}\href{http://dx.doi.org/10.1088/0004-637X/691/1/131}{\apj}\endgroup,
   \href{http://adsabs.harvard.edu/abs/2009ApJ...691..131Y}{691, 131}

\end{thebibliography}

\Online \begin{appendix}

\section{Table and plots of all $uv$ points\label{sec:list-of-uvpoints}}

Table~\ref{tab:list-of-uvpoints} lists all measurements of the Circinus
galaxy obtained with MIDI, that is all $uv$ points. The columns are:
(1) a running number of the measurement; (2) the date and time at
the start of the fringe track, i.e. the interferometric observation;
(3) modified Julian date, MJD; (4) telescope combination used for
the observation; (5) projected baseline length; (6) position angle
of the projected baseline; (7) adaptive optics system used, either
STRAP or MACAO (``\_T'' indicates additional usage of the technical
CCD); (8) usage of IRIS with: ``n/a'' -- IRIS not available, ``off''
-- IRIS not used for field stabilisation, ``on'' -- IRIS used for
field stabilisation, a number -- IRIS used with the integration time
specified in seconds (before 2011 no IRIS integration times were stored
with the data); (9) mode of the interferometric observations with:
``0OPD'' -- tracking with a scan length of $84.2\,\mathrm{\mu m}$
at zero OPD, ``OFF'' -- tracking with a scan length of $84.2\,\mathrm{\mu m}$
at an offset of $40\,\mathrm{\mu m}$, ``SHRT'' -- tracking with
a reduced scan length of $53.0\,\mathrm{\mu m}$ at an offset of $40\,\mathrm{\mu m}$;
(10) detector integration time of MIDI; (11) chopping frequency for
the observations of the total flux spectra with the single dishes;
(12) airmass; (13) seeing; (14) coherence time; (15) name of the calibrator
star; (16) time of the calibrator observation; (17) correlated flux
and its error at $12\,\mathrm{\mu m}$, if the observation was successful
and is used for the further analysis. Measurements were considered
successful if a fringe signal was tracked and sufficient signal was
detected (mostly equivalent to more than 1500 good frames used for
the averaging). No entry in the chop frequency column indicates that
no photometry was observed.

All dispersed correlated fluxes are shown in Fig.~\ref{fig:alldata-correlated},
the differential phases are plotted in Fig.~\ref{fig:alldata-phases}.
The data is plotted in blue, the correlated fluxes and differential
phases of the model from fit 3 are plotted in red. As in Table~\ref{tab:list-of-uvpoints},
the measurements are ordered by observing date and time. The numbering
is also the same as in Table~\ref{tab:list-of-uvpoints}, however
unsuccessful measurements are not plotted.

\global\long\def\tabcolsep{0.5mm}
\addtocounter{table}{1}\longtab{\value{table}}{{\tiny

\begin{longtable}{>{\raggedleft}p{0.5cm}>{\centering}p{2.4cm}>{\centering}p{1.7cm}>{\centering}p{0.8cm}>{\raggedleft}p{0.7cm}>{\raggedleft}p{0.8cm}>{\centering}p{1.4cm}>{\centering}p{0.6cm}>{\centering}p{0.8cm}>{\centering}p{0.6cm}>{\centering}p{0.6cm}>{\centering}p{0.6cm}>{\centering}p{0.8cm}>{\centering}p{0.7cm}>{\centering}p{1.3cm}>{\centering}p{1.2cm}>{\centering}p{1.2cm}}
\caption{List of interferometric measurements of the Circinus galaxy obtained
with MIDI. For each UV point, the list includes parameters and conditions
of the observations. See App.~\ref{sec:list-of-uvpoints} for details.\label{tab:list-of-uvpoints}}
\tabularnewline
\hline 
\hline \centering{\tiny \# } & {\tiny Date \& time } & {\tiny MJD } & {\tiny baseline } & \centering{\tiny BL}{\tiny \par}

{\tiny {[}m{]} } & \centering{\tiny PA}{\tiny \par}

{\tiny {[}\textdegree{}{]} } & {\tiny AO}{\tiny \par}

{\tiny mode } & {\tiny IRIS } & {\tiny track}{\tiny \par}

{\tiny mode } & {\tiny DIT}{\tiny \par}

{\tiny {[}ms{]} } & {\tiny $f_{\mathrm{chop}}$}{\tiny \par}

{\tiny {[}Hz{]} } & {\tiny AM } & {\tiny seeing}{\tiny \par}

{\tiny {[}arcsec{]} } & {\tiny $\tau_{0}$}{\tiny \par}

{\tiny {[}ms{]} } & {\tiny calibrator}{\tiny \par}

{\tiny name } & {\tiny calibrator}{\tiny \par}

{\tiny obs. date } & {\tiny $F_{\mathrm{cor}}$}{\tiny \par}

{\tiny {[}Jy{]}}\tabularnewline
\centering{\tiny (1) } & {\tiny (2) } & {\tiny (3) } & {\tiny (4) } & \centering{\tiny (5) } & \centering{\tiny (6) } & {\tiny (7) } & {\tiny (8) } & {\tiny (9) } & {\tiny (10) } & {\tiny (11) } & {\tiny (12) } & {\tiny (13) } & {\tiny (14) } & {\tiny (15) } & {\tiny (16) } & {\tiny (17)}\tabularnewline
\hline 
\endfirsthead
\caption{continued.} \tabularnewline % &  &  &  &  &  &  &  &  &  &  &  &  &  &  &  & \tabularnewline
\hline 
\hline \centering{\tiny \# } & {\tiny Date \& time } & {\tiny MJD } & {\tiny baseline } & \centering{\tiny BL}{\tiny \par}

{\tiny {[}m{]} } & \centering{\tiny PA}{\tiny \par}

{\tiny {[}\textdegree{}{]} } & {\tiny AO}{\tiny \par}

{\tiny mode } & {\tiny IRIS } & {\tiny track}{\tiny \par}

{\tiny mode } & {\tiny DIT}{\tiny \par}

{\tiny {[}ms{]} } & {\tiny $f_{\mathrm{chop}}$}{\tiny \par}

{\tiny {[}Hz{]} } & {\tiny AM } & {\tiny seeing}{\tiny \par}

{\tiny {[}arcsec{]} } & {\tiny $\tau_{0}$}{\tiny \par}

{\tiny {[}ms{]} } & {\tiny calibrator}{\tiny \par}

{\tiny name } & {\tiny calibrator}{\tiny \par}

{\tiny obs. date } & {\tiny $F_{\mathrm{cor}}$}{\tiny \par}

{\tiny {[}Jy{]}}\tabularnewline
\centering{\tiny (1) } & {\tiny (2) } & {\tiny (3) } & {\tiny (4) } & \centering{\tiny (5) } & \centering{\tiny (6) } & {\tiny (7) } & {\tiny (8) } & {\tiny (9) } & {\tiny (10) } & {\tiny (11) } & {\tiny (12) } & {\tiny (13) } & {\tiny (14) } & {\tiny (15) } & {\tiny (16) } & {\tiny (17)}\tabularnewline
\hline 
\endhead
\hline 
% &  &  &  &  &  &  &  &  &  &  &  &  &  &  &  & \tabularnewline
\endfoot
{\tiny 1\phantom{\selectlanguage{english}%
{*}\selectlanguage{british}%
} } & {\tiny 2004-02-12 06:55:11 } & {\tiny 53047.28832 } & {\tiny U3-U2 } & {\tiny 43.5 } & {\tiny -160.7 } & {\tiny STRAP\_T } & {\tiny n/a } & {\tiny 0OPD } & {\tiny 20 } & {\tiny 2.0 } & {\tiny 1.47 } & {\tiny 0.96 } & {\tiny 3.4 } & {\tiny HD120404 } & {\tiny 07:36:34 } & {\tiny $0.72\pm0.12$ }\tabularnewline
{\tiny 2\phantom{\selectlanguage{english}%
{*}\selectlanguage{british}%
} } & {\tiny 2004-02-12 07:06:46 } & {\tiny 53047.29637 } & {\tiny U3-U2 } & {\tiny 43.4 } & {\tiny -158.5 } & {\tiny STRAP\_T } & {\tiny n/a } & {\tiny 0OPD } & {\tiny 20 } & {\tiny 2.0 } & {\tiny 1.45 } & {\tiny 0.91 } & {\tiny 3.6 } & {\tiny HD120404 } & {\tiny 07:36:34 } & {\tiny $0.93\pm0.15$ }\tabularnewline
\hline 
{\tiny 3\phantom{\selectlanguage{english}%
{*}\selectlanguage{british}%
} } & {\tiny 2004-06-03 05:50:45 } & {\tiny 53159.24358 } & {\tiny U3-U2 } & {\tiny 29.2 } & {\tiny -87.0 } & {\tiny STRAP\_T } & {\tiny n/a } & {\tiny 0OPD } & {\tiny 12 } & {\tiny 2.0 } & {\tiny 1.69 } & {\tiny 0.68 } & {\tiny 4.3 } & {\tiny HD120404 } & {\tiny 06:42:24 } & {\tiny $1.06\pm0.13$ }\tabularnewline
{\tiny 4\phantom{\selectlanguage{english}%
{*}\selectlanguage{british}%
} } & {\tiny 2004-06-03 07:58:04 } & {\tiny 53159.33199 } & {\tiny U3-U2 } & {\tiny 20.7 } & {\tiny -50.3 } & {\tiny STRAP\_T } & {\tiny n/a } & {\tiny 0OPD } & {\tiny 12 } & {\tiny 2.0 } & {\tiny 2.54 } & {\tiny 0.73 } & {\tiny 4.2 } & {\tiny HD120404 } & {\tiny 06:42:24 } & {\tiny $2.30\pm0.29$ }\tabularnewline
\hline 
{\tiny 5\phantom{\selectlanguage{english}%
{*}\selectlanguage{british}%
} } & {\tiny 2005-02-21 05:25:00 } & {\tiny 53422.22569 } & {\tiny U2-U4 } & {\tiny 87.4 } & {\tiny 35.7 } & {\tiny MACAO\_T } & {\tiny n/a } & {\tiny 0OPD } & {\tiny 12 } & {\tiny -- } & {\tiny 1.62 } & {\tiny 0.77 } & {\tiny 3.1 } & {\tiny HD107446 } & {\tiny 04:38:01 } & {\tiny -- }\tabularnewline
\end{longtable}}}

\begin{figure*}
\includegraphics[bb=28bp 52bp 566bp 788bp,clip,width=18cm]{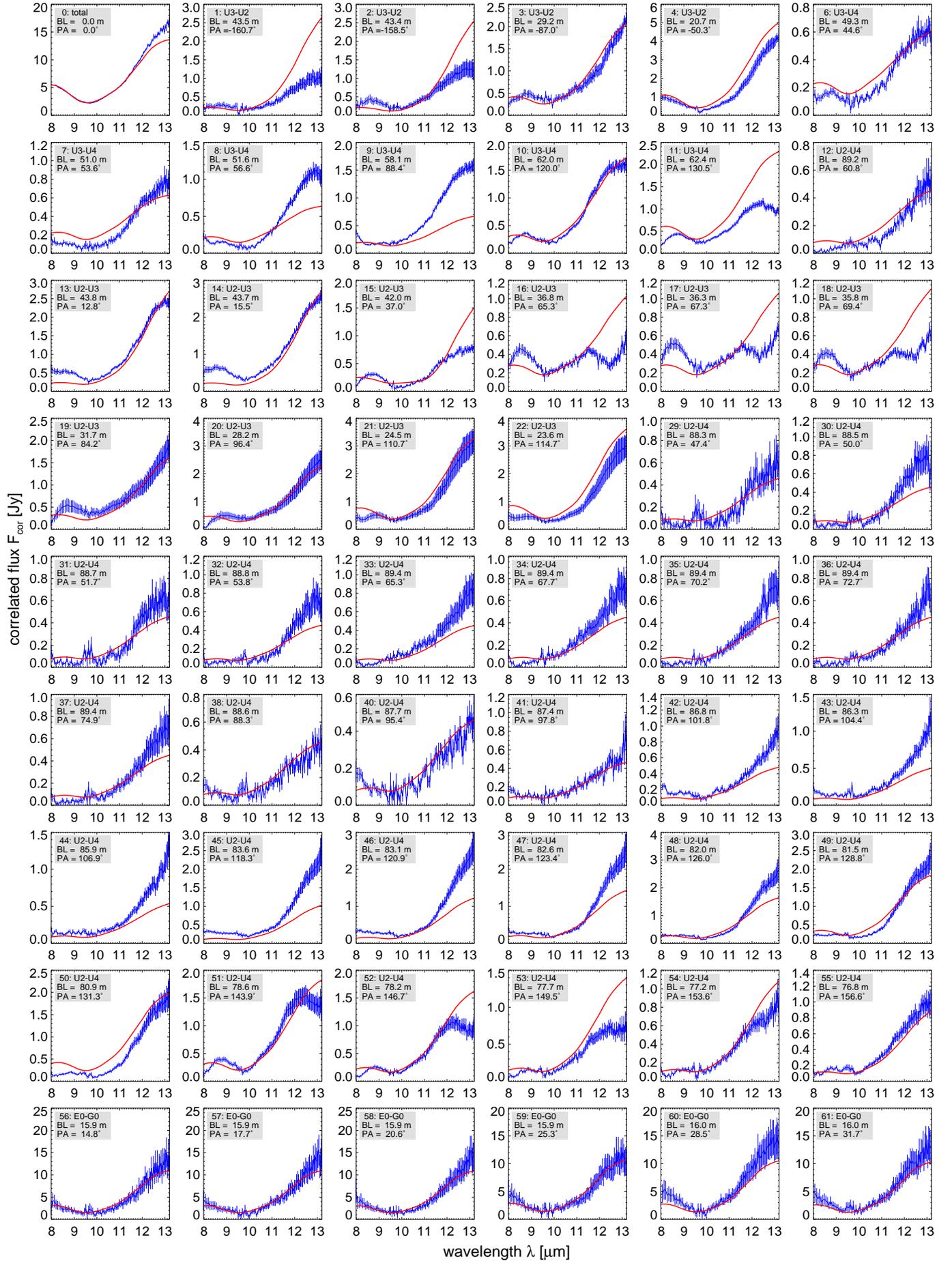}

\caption{Correlated fluxes of all $uv$ points (blue). The correlated fluxes
of our three-component model (fit 3) are plotted in red.\label{fig:alldata-correlated}}

\end{figure*}
\addtocounter{figure}{-1}

\begin{figure*}
\includegraphics[bb=28bp 52bp 566bp 788bp,clip,width=18cm]{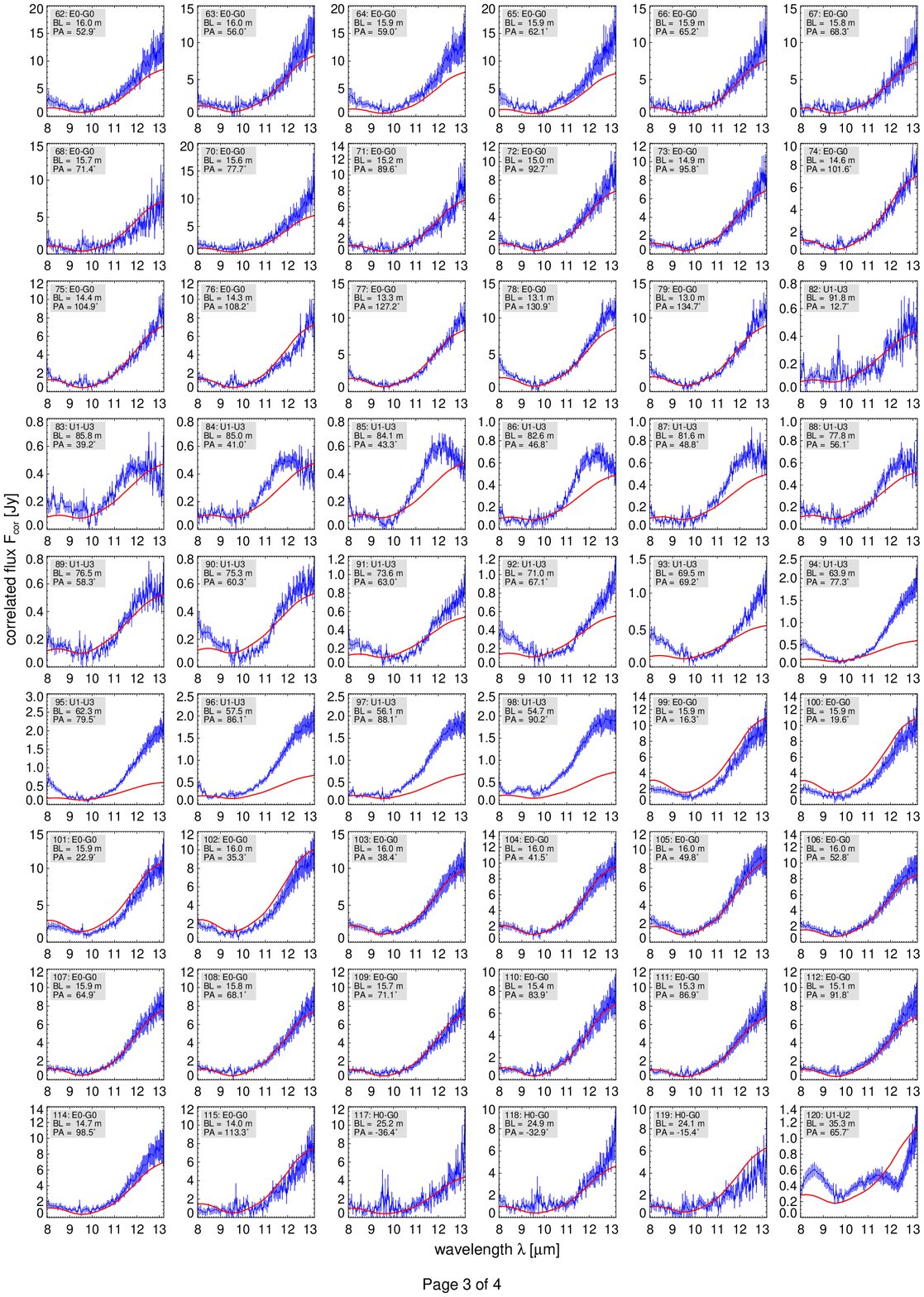}

\caption{continued.}
\end{figure*}
\addtocounter{figure}{-1}

\begin{figure*}
\includegraphics[bb=28bp 52bp 566bp 788bp,clip,width=18cm]{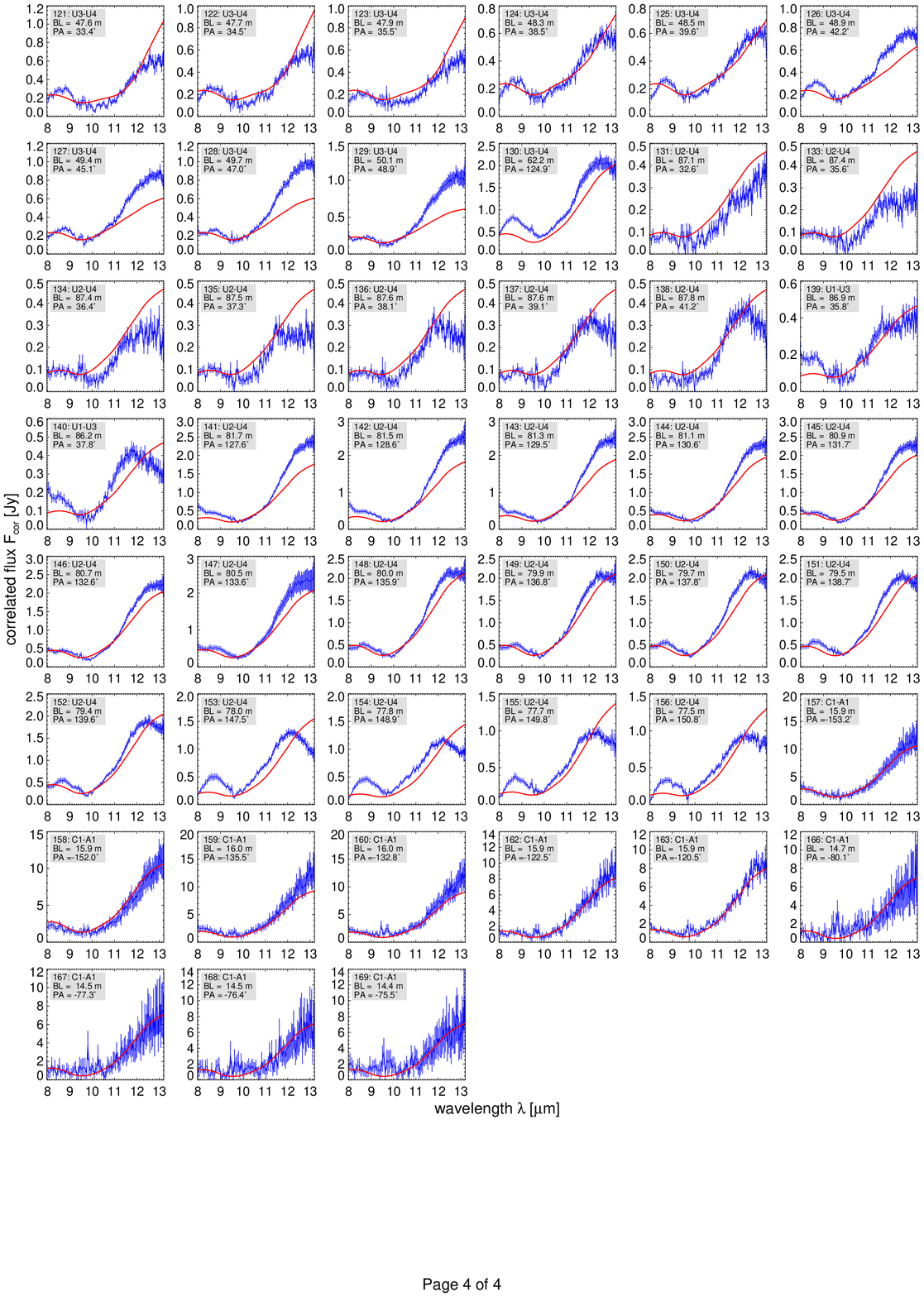}

\caption{continued.}
\end{figure*}

\begin{figure*}
\includegraphics[bb=28bp 52bp 566bp 788bp,clip,width=18cm]{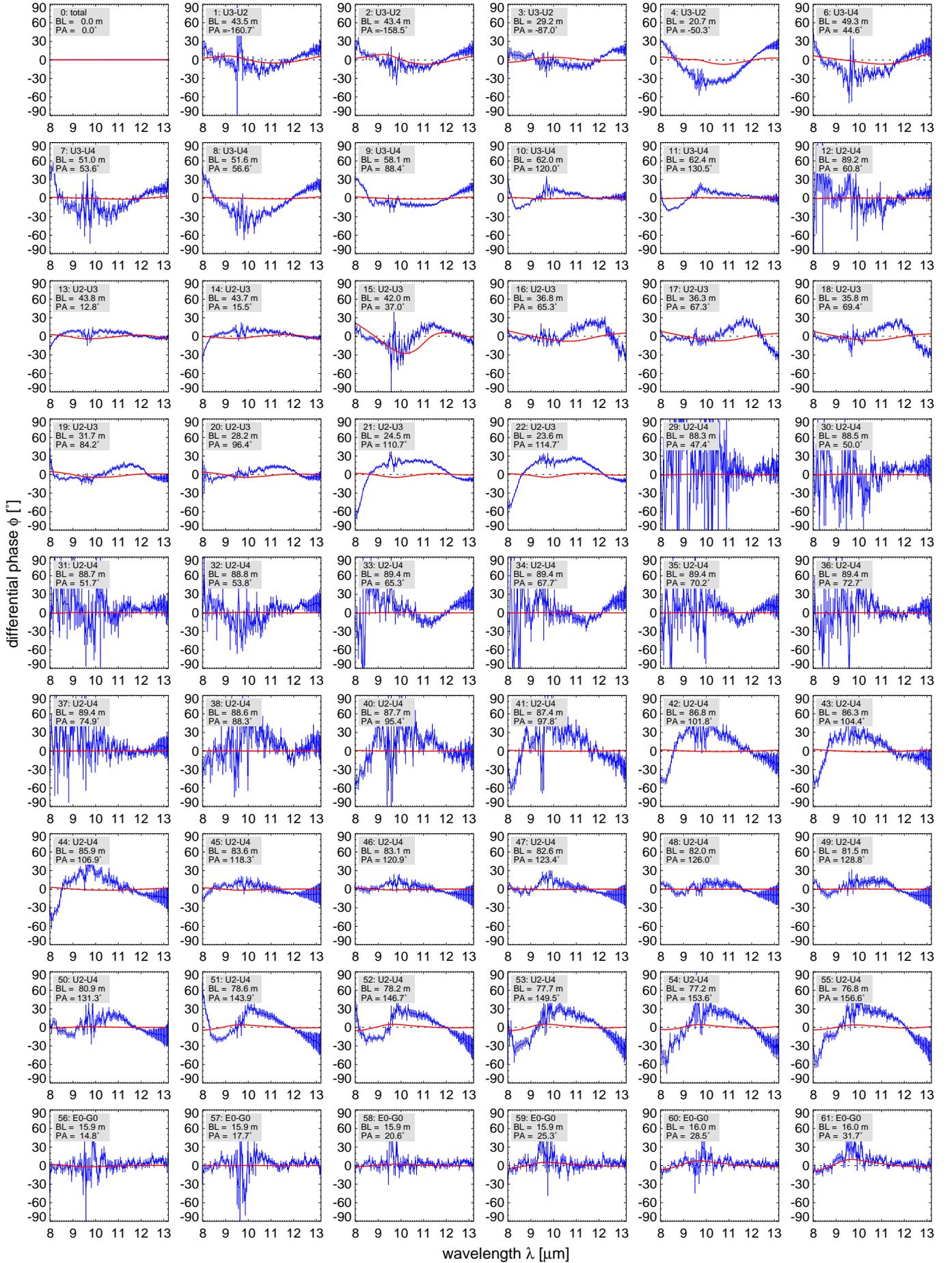}

\caption{Differential phases of all $uv$ points (blue). The differential phases
of our model (fit 3) are plotted in red.\label{fig:alldata-phases}}
\end{figure*}
\addtocounter{figure}{-1}

\begin{figure*}
\includegraphics[bb=28bp 52bp 566bp 788bp,clip,width=18cm]{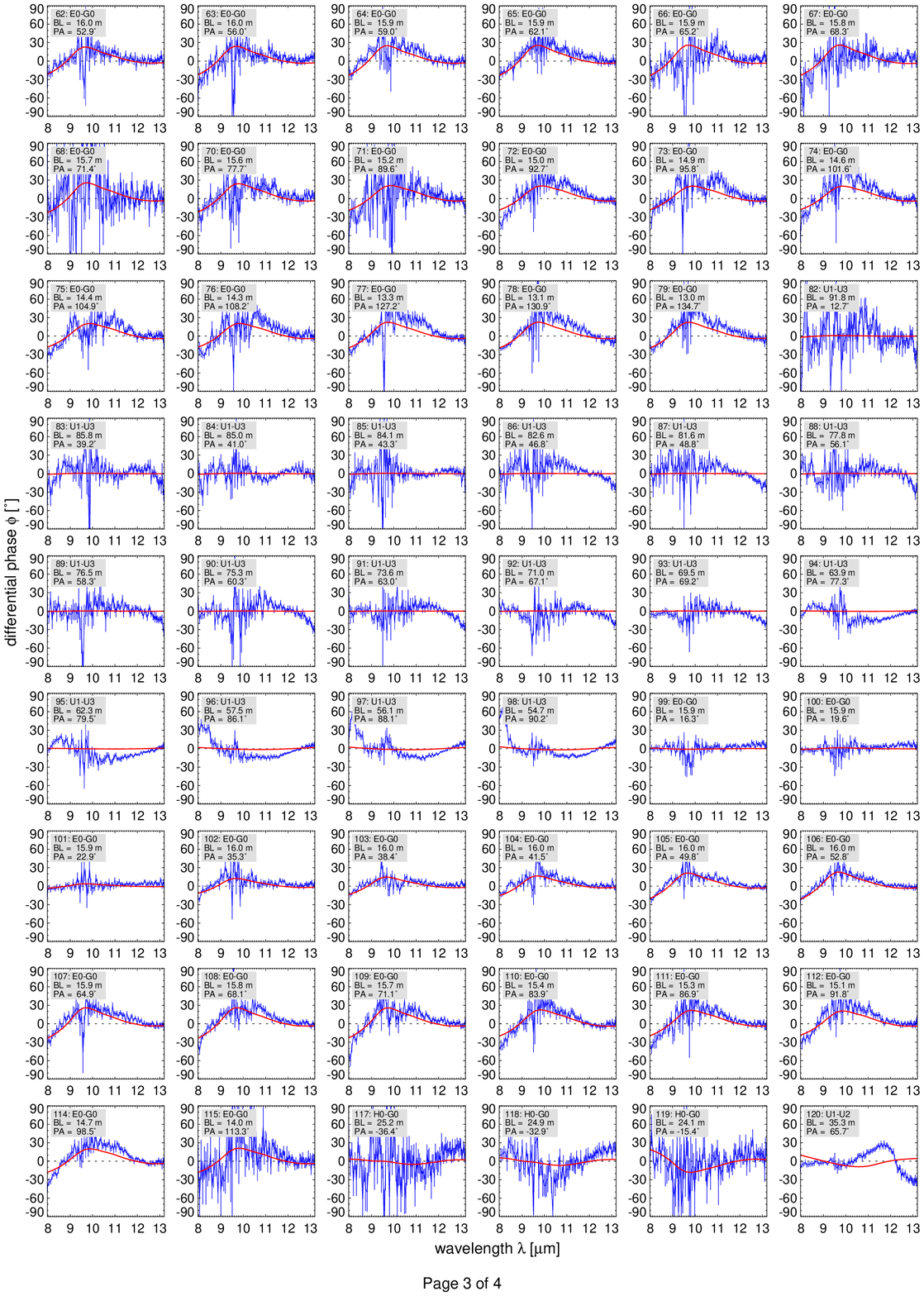}

\caption{continued.}
\end{figure*}
\addtocounter{figure}{-1}

\begin{figure*}
\includegraphics[bb=28bp 52bp 566bp 788bp,clip,width=18cm]{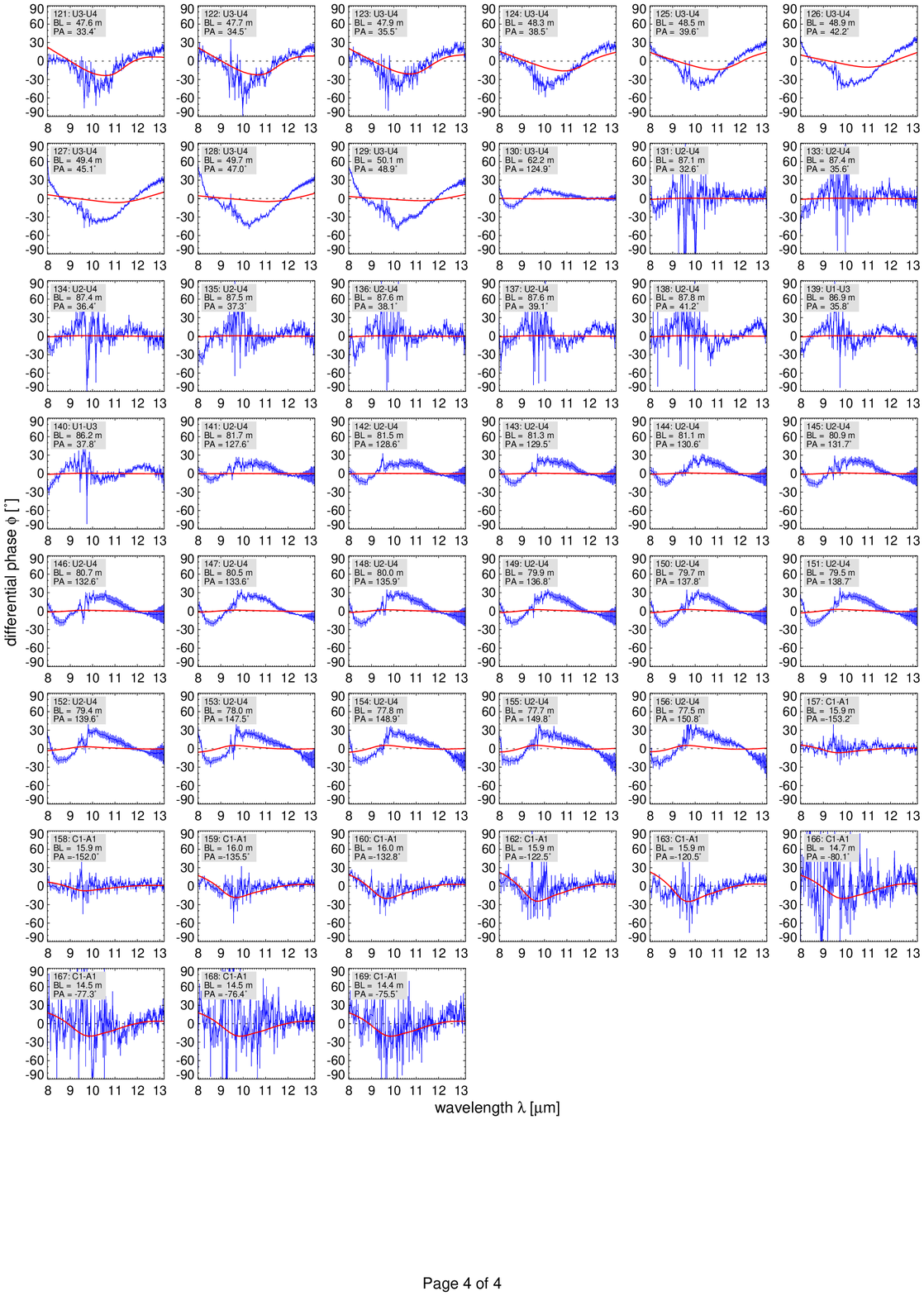}

\caption{continued.}
\end{figure*}

\end{appendix}
\end{document}